\documentclass[12pt]{article}
\pdfoutput=1
\usepackage[]{hyperref}
\usepackage{xcolor}
\usepackage[]{putex}
\usepackage[utf8]{inputenc}
\usepackage{cite}

\usepackage{amsmath,amsthm,amssymb}
\usepackage{tikz}
\usepackage{tikz-cd}
\usetikzlibrary{decorations.markings}
\usetikzlibrary{calc}
\usepackage{subcaption}
\usepackage{enumerate}
\usepackage{enumitem}

\numberwithin{equation}{section}

\makeatletter
\DeclareRobustCommand*{\bfseries}{%
  \not@math@alphabet\bfseries\mathbf
  \fontseries\bfdefault\selectfont
  \boldmath
}
\makeatother
\hypersetup{
  colorlinks=true,
  linkcolor=blue,
  citecolor=green!60!black,
  urlcolor=cyan!70!black,
  linktoc=all
  }

\input{glyphtounicode}
\def\musepic#1{\vcenter{\hbox{\usebox{#1}}}}

\newcommand{\DrawVertexPostMellin}[6]{
\sbox{\figVertexPostMellin}{%
    \begin{tikzpicture}[scale=.8]
    \tikzstyle{vint}=[draw,scale=0.3,color=blue,fill=blue,circle]
    \coordinate (y4) at (1,1/4);
    \coordinate (x5) at (1,1);
    \coordinate (w3) at (1/2,7/4);
    \coordinate (w4) at (3/2,7/4);
	\ifthenelse{\isempty{#6}}{\draw[thick] (y4)--(x5);}{\draw[thick,red] (y4)--(x5);}
	\ifthenelse{\isempty{#2}}{\draw[thick] (w3)--(x5);}{\draw[thick,red] (w3)--(x5);}
	\ifthenelse{\isempty{#4}}{\draw[thick] (w4)--(x5);}{\draw[thick,red] (w4)--(x5);}
	\ifthenelse{\isempty{#6}}{\draw (y4) node[anchor=north] {$#5$};}{\draw[red] (y4) node[anchor=north] {$#5$};}
	\ifthenelse{\isempty{#2}}{\draw (w3) node[anchor=south] {$#1$};}{\draw[red] (w3) node[anchor=south] {$#1$};}	
	\ifthenelse{\isempty{#4}}{\draw (w4) node[anchor=south] {$#3$};}{\draw[red] (w4) node[anchor=south] {$#3$};}
    \draw (x5) node[vint] {};
\end{tikzpicture}
    }
\musepic{\figVertexPostMellin}
}

\newsavebox{\figNptAdS}
\savebox{\figNptAdS}{%
\begin{tikzpicture}[scale=1.1]
    \tikzstyle{vint}=[draw,scale=0.55,color=teal,fill=green,circle]
    \tikzstyle{blob}=[draw,scale=2,color=black,fill=lightgray,circle]
    \draw[thick, color=black!30] (0,0) circle[radius=2];
    \coordinate (x1) at (240:2);
    \coordinate (x2) at (200:2);
    \coordinate (x4) at (-20:2);
    \coordinate (x5) at (-60:2);
    \coordinate (x3) at (110:2);
    \coordinate (x6) at (70:2);
    \coordinate (y) at (220:1.2);
    \coordinate (yp) at (-40:1.2);
    \coordinate (ypp) at (90:1.2);
    \coordinate (z) at (90:0.5);
	\draw[thick] (x1)--(y)--(x2);
	\draw[thick] (x4)--(yp)--(x5);
	\draw[thick] (y)--(z)--(yp);
	\draw[thick] (x6)--(ypp)--(x3);
	\draw[thick] (ypp)--(z);
	\draw (x1) node[anchor=north] {\footnotesize $\Delta_a$};
	\draw (x2) node[anchor= east] {\footnotesize $\Delta_3$};
	\draw (x3) node[anchor=south east] {\footnotesize $\Delta_2$};
	\draw (x4) node[anchor= west] {\footnotesize $\Delta_M$};
	\draw (x5) node[anchor=north] {\footnotesize $\Delta_{a+1}$};
	\draw (x6) node[anchor=south west] {\footnotesize $\Delta_1$};
 	\draw  ($(y)!0.75!(z)$) node[anchor= east] {\footnotesize $\Delta_{k_{\alpha_1}}$};
	\draw  ($(yp)!0.75!(z)$) node[anchor= west] {\footnotesize $\Delta_{k_{\alpha_2}}$};
	\draw  ($(ypp)!0.5!(z)$) node[anchor=west] {\footnotesize $\Delta_{k_1}$};
	\draw[very thick, line cap=round, dash pattern=on 0 off 15] (210:2.3) arc (210:240:2.3);
	\draw[very thick, line cap=round, dash pattern=on 0 off 15] (310:2.3) arc (310:340:2.3);
	\draw (y) node[blob] {\footnotesize $3$};
	\draw (yp) node[blob] {\scriptsize $M$};
	\draw (ypp) node[vint] {};
	\draw (z) node[vint] {};
\end{tikzpicture}}
\newsavebox{\figNMinusOneptAdS}
\savebox{\figNMinusOneptAdS}{%
\begin{tikzpicture}[scale=1.1]
    \tikzstyle{vint}=[draw,scale=0.55,color=teal,fill=green,circle]
    \tikzstyle{blob}=[draw,scale=1.7,color=black,fill=lightgray,circle]
    \draw[thick, color=black!30] (0,0) circle[radius=2];
    \coordinate (x1) at (240:2);
    \coordinate (x2) at (200:2);
    \coordinate (x4) at (-20:2);
    \coordinate (x5) at (-60:2);
    \coordinate (x3) at (110:2);
    \coordinate (y) at (220:1.2);
    \coordinate (yp) at (-40:1.2);
    \coordinate (z) at (90:0.5);
	\draw[thick] (x1)--(y)--(x2);
	\draw[thick] (x4)--(yp)--(x5);
	\draw[thick] (y)--(z)--(yp);
	\draw[thick] (x3)--(z);
	\draw (x1) node[anchor=north] {\footnotesize $\Delta_a^\prime$};
	\draw (x2) node[anchor= east] {\footnotesize $\Delta_3^\prime$};
	\draw (x3) node[anchor=south east] {\footnotesize $\Delta_2^\prime$};
	\draw (x4) node[anchor= west] {\footnotesize $\Delta_M^\prime$};
	\draw (x5) node[anchor=north] {\footnotesize $\Delta_{a+1}^\prime$};
 	\draw  ($(y)!0.75!(z)$) node[anchor= east] {\footnotesize $\Delta_{k_{\alpha_1}}^\prime$};
	\draw  ($(yp)!0.75!(z)$) node[anchor= west] {\footnotesize $\Delta_{k_{\alpha_2}}^\prime$};
	\draw[very thick, line cap=round, dash pattern=on 0 off 15] (210:2.3) arc (210:240:2.3);
	\draw[very thick, line cap=round, dash pattern=on 0 off 15] (310:2.3) arc (310:340:2.3);
	\draw (y) node[blob] {\footnotesize $3^\prime$};
	\draw (yp) node[blob] {\scriptsize $M^\prime$};
	\draw (z) node[vint] {};
\end{tikzpicture}}
\newsavebox{\figFiveptAdS}
\savebox{\figFiveptAdS}{%
\begin{tikzpicture}[scale=1.1]
    \tikzstyle{vint}=[draw,scale=0.55,color=teal,fill=green,circle];
    \draw[thick, color=black!30] (0,0) circle[radius=2];
    \coordinate (x1) at (240:2);
    \coordinate (x4) at (-20:2);
    \coordinate (x5) at (-60:2);
    \coordinate (x3) at (110:2);
    \coordinate (x6) at (70:2);
    \coordinate (y) at (220:2);
    \coordinate (yp) at (-40:1.2);
    \coordinate (ypp) at (90:1.2);
    \coordinate (z) at (90:0);
	\draw[thick] (x4)--(yp)--(x5);
	\draw[thick] (y)--(z)--(yp);
	\draw[thick] (x6)--(ypp)--(x3);
	\draw[thick] (ypp)--(z);
	\draw (y) node[anchor=north east] {\footnotesize $\Delta_3$};
	\draw (x3) node[anchor=south east] {\footnotesize $\Delta_2$};
	\draw (x4) node[anchor=  west] {\footnotesize $\Delta_5$};
	\draw (x5) node[anchor=north] {\footnotesize $\Delta_{4}$};
	\draw (x6) node[anchor=south west] {\footnotesize $\Delta_1$};
	\draw  ($(yp)!0.75!(z)$) node[anchor= west] {\footnotesize $\Delta_{k_2}$};
	\draw  ($(ypp)!0.5!(z)$) node[anchor=west] {\footnotesize $\Delta_{k_1}$};
    \draw (yp) node[vint] {};
	\draw (ypp) node[vint] {};
	\draw (z) node[vint] {};
\end{tikzpicture}}
\newsavebox{\figFourptAdS}
\savebox{\figFourptAdS}{%
\begin{tikzpicture}[scale=1.1]
    \tikzstyle{vint}=[draw,scale=0.55,color=teal,fill=green,circle];
    \draw[thick, color=black!30] (0,0) circle[radius=2];
    \coordinate (x1) at (240:2);
    \coordinate (x4) at (-20:2);
    \coordinate (x5) at (-60:2);
    \coordinate (x3) at (110:2);
    \coordinate (y) at (220:2);
    \coordinate (yp) at (-40:1.2);
    \coordinate (ypp) at (90:1.2);
    \coordinate (z) at (90:0);
	\draw[thick] (x4)--(yp)--(x5);
	\draw[thick] (y)--(z)--(yp);
	\draw[thick] (z)--(x3);
	\draw (y) node[anchor=north east] {\footnotesize $\Delta_3^\prime$};
	\draw (x3) node[anchor=south east] {\footnotesize $\Delta_2^\prime$};
	\draw (x4) node[anchor=  west] {\footnotesize $\Delta_5^\prime$};
	\draw (x5) node[anchor=north] {\footnotesize $\Delta_{4}^\prime$};
	\draw  ($(yp)!0.75!(z)$) node[anchor= west] {\footnotesize $\Delta_{k_2}^\prime$};
    \draw (yp) node[vint] {};
	\draw (z) node[vint] {};
\end{tikzpicture}}
\newsavebox{\figNineMixedCB}
\savebox{\figNineMixedCB}{%
   \begin{tikzpicture}[scale=2.0]
    \coordinate (y1) at (-3/2,0);
    \coordinate (y2) at (-1,0);
    \coordinate (x3) at (-1,1);
    \coordinate (y3) at (0,0);
    \coordinate (z3) at (0,1/2);
    \coordinate (x4) at (0,1);
    \coordinate (y4) at (1,0);
    \coordinate (z4) at (1,1/2);
    \coordinate (x5) at (1,1);
    \coordinate (y5) at (2,0);
    \coordinate (x6) at (2,1);
    \coordinate (y6) at (9/2,0);
    \coordinate (x7) at (3,1);
    \coordinate (y7) at (3,0);
    \coordinate (x8) at (4,1);
    \coordinate (y8) at (4,0);
    \coordinate (w3) at (1/2,3/2);
    \coordinate (w4) at (3/2,3/2);
	\draw[thick] (y1)--(y2)--(x3);
	\draw[thick] (y3)--(x4);
	\draw[thick] (y5)--(x6);	
	\draw[thick] (y5)--(y6);
	\draw[thick] (x7)--(y7);
	\draw[thick] (x8)--(y8);
	\draw[thick] (w3)--(x5)--(w4);
	\draw (y1) node[anchor=east] {\scriptsize $\mathcal{O}_{1}(x_{1})$};
	\draw (x3) node[anchor=south] {\scriptsize $\mathcal{O}_{2}(x_{2})$};
	\draw (x4) node[anchor=south] {\scriptsize $\mathcal{O}_{3}(x_{3})$};
	\draw (w3) node[anchor=south] {\scriptsize $\mathcal{O}_{4}(x_{4})$};
	\draw (w4) node[anchor=south] {\scriptsize $\mathcal{O}_{5}(x_{5})$};
	\draw (x6) node[anchor=south] {\scriptsize $\mathcal{O}_{6}(x_{6})$};
	\draw (x7) node[anchor=south] {\scriptsize $\mathcal{O}_{7}(x_{7})$};
	\draw (x8) node[anchor=south] {\scriptsize $\mathcal{O}_{8}(x_{8})$};
	\draw (y6) node[anchor=west] {\scriptsize $\mathcal{O}_{9}(x_{9})$};
	\draw[thick] (y2)--(y3); 
	\draw ($(y2)!0.5!(y3)$) node[anchor=north] {\scriptsize $\mathcal{O}_{k_1}$};
	\draw[thick] (y3)--(y4); 
	\draw ($(y3)!0.5!(y4)$) node[anchor=north] {\scriptsize $\mathcal{O}_{k_2}$};
	\draw[thick] (x5)--(y4); 
	\draw ($(x5)!0.5!(y4)$) node[anchor=west] {\scriptsize $\mathcal{O}_{k_3}$};
	\draw[thick] (y5)--(y4); 
	\draw ($(y5)!0.5!(y4)$) node[anchor=north] {\scriptsize $\mathcal{O}_{k_4}$};
	\draw ($(y8)!0.5!(y4)$) node[anchor=north] {\scriptsize $\mathcal{O}_{k_5}$};
	\draw ($(y6)!0.5!(y7)$) node[anchor=north] {\scriptsize $\mathcal{O}_{k_6}$};
\end{tikzpicture} 
    }
\newsavebox{\figfiveCB}
\savebox{\figfiveCB}{%
      \begin{tikzpicture}[scale=2.0]
    \coordinate (y1) at (-3/2,0);
    \coordinate (y2) at (-1,0);
    \coordinate (x3) at (-1,1);
    \coordinate (y3) at (0,0);
    \coordinate (z3) at (0,1/2);
    \coordinate (x4) at (0,1);
    \coordinate (y4) at (1,0);
    \coordinate (z4) at (1,1/2);
    \coordinate (x5) at (1,1);
    \coordinate (y5) at (3/2,0);
    \coordinate (x6) at (2,1);
    \coordinate (y6) at (9/2,0);
	\draw[thick] (y1)--(y2)--(x3);
	\draw[thick] (y3)--(x4);
	\draw (y1) node[anchor=east] {\scriptsize $\mathcal{O}_{1}(x_{1})$};
	\draw (x3) node[anchor=south] {\scriptsize $\mathcal{O}_{2}(x_{2})$};
	\draw (x4) node[anchor=south] {\scriptsize $\mathcal{O}_{3}(x_{3})$};
	\draw[thick] (y2)--(y3); 
	\draw ($(y2)!0.5!(y3)$) node[anchor=north] {\scriptsize $\mathcal{O}_{k_1}$};
	\draw[thick] (y3)--(y4); 
	\draw ($(y3)!0.5!(y4)$) node[anchor=north] {\scriptsize $\mathcal{O}_{k_2}$};
	\draw[thick] (x5)--(y4); 
	\draw (x5) node[anchor=south] {\scriptsize $\mathcal{O}_{4}(x_4)$};
	\draw[thick] (y5)--(y4); 
	\draw (y5) node[anchor=west] {\scriptsize $\mathcal{O}_{5}(x_5)$};
\end{tikzpicture} 
    }

\newsavebox{\figFourptAdSUnprime}
\savebox{\figFourptAdSUnprime}{%
\begin{tikzpicture}[scale=1.1]
    \tikzstyle{vint}=[draw,scale=0.55,color=teal,fill=green,circle];
    \draw[thick, color=black!30] (0,0) circle[radius=2];
    \coordinate (x1) at (240:2);
    \coordinate (x4) at (-20:2);
    \coordinate (x5) at (-60:2);
    \coordinate (x3) at (160:2);
    \coordinate (y) at (120:2);
    \coordinate (yp) at (-40:1);
    \coordinate (z) at (140:1);
	\draw[thick] (x4)--(yp)--(x5);
	\draw[thick] (y)--(z)--(yp);
	\draw[thick] (z)--(x3);
	\draw (y) node[anchor=south] {\footnotesize $\Delta_1$};
	\draw (x3) node[anchor= east] {\footnotesize $\Delta_2$};
	\draw (x4) node[anchor=  west] {\footnotesize $\Delta_4$};
	\draw (x5) node[anchor=north] {\footnotesize $\Delta_{3}$};
	\draw  ($(yp)!0.55!(z)$) node[anchor= west] {\footnotesize $\Delta_{k_1}$};
    \draw (yp) node[vint] {};
	\draw (z) node[vint] {};
\end{tikzpicture}}

\begin{document}

\title{Feynman Rules for Scalar Conformal Blocks}
\authors{Jean-Fran\c cois Fortin,$^1$\footnote{\tt jean-francois.fortin@phy.ulaval.ca} Sarah Hoback,$^2$\footnote{\tt sarahhoback98@gmail.com} Wen-Jie Ma,$^{3,4}$\footnote{\tt wenjia.ma@bimsa.cn}\\Sarthak Parikh$^{5,6}$\footnote{\tt sarthak@physics.iitd.ac.in} $\&$ Witold Skiba$^7$\footnote{\tt witold.skiba@yale.edu}}
\institution{Laval}{$^1$D\'epartement de Physique, de G\'enie Physique et d'Optique,\cr\hskip0.06in Universit\'e Laval, Qu\'ebec, QC G1V 0A6, Canada
}
\institution{}{$^2$Department of Physics and Astronomy, Pomona College, Claremont, CA 91711, USA}
\institution{}{$^3$Beijing Institute of Mathematical Sciences and Applications (BIMSA),\cr Huairou District, Beijing 101408, P. R. China}
\institution{}{$^4$Yau Mathematical Sciences Center, Tsinghua University, Beijing 100084, China}
\institution{}{$^5$Division of Physics, Mathematics and Astronomy, California Institute of Technology,\cr Pasadena, CA 91125, USA}
\institution{}{$^6$Department of Physics, Indian Institute of Technology Delhi, \cr Hauz Khas, New Delhi 110016, India}
\institution{Yale}{$^7$Department of Physics, Yale University, New Haven, CT 06520, USA}

\abstract{
We complete the proof of ``Feynman rules'' for constructing $M$-point conformal blocks with external and internal scalars in any topology for arbitrary $M$ in any spacetime dimension by combining the rules for the blocks (based on their Witten diagram interpretation) with the rules for the construction of conformal cross ratios (based on the OPE and ``flow diagrams'').  The full set of Feynman rules leads to blocks as power series of the hypergeometric type in the conformal cross ratios.  We then provide a proof by recursion of the Feynman rules which relies heavily on the first Barnes lemma and the decomposition of the topology of interest in comb structures.  Finally, we provide a nine-point example to illustrate the rules.
}

\date{October 2022}

\maketitle

{\hypersetup{linkcolor=black}
\tableofcontents
}

\section{Introduction}
\label{INTRO}

The interest in conformal field theories (CFTs) is multifaceted. It ranges from applications to phase transitions in condensed matter systems, through application of conformal theories to extensions of the electroweak Standard Model, to describing one side of the AdS-CFT duality. In part, this interest is driven by the many successes of the conformal bootstrap program that allowed non-perturbative treatment of CFTs~\cite{Ferrara:1973yt,Polyakov:1974gs,Rattazzi:2008pe,Poland:2018epd}. The bootstrap program relies on the decomposition of correlators in a CFT into conformal blocks and pure numbers: operator dimensions and the operator product expansion (OPE) coefficients. Another important motivation for studying CFTs  is the AdS-CFT duality, in which the CFT correlation functions describe bulk interactions that can be encoded in terms of Witten diagrams~\cite{Maldacena:1997re,Gubser:1998bc,Witten:1998qj}.

Conformal blocks are crucial for computing correlation functions as they encode kinematic constraints on the form of correlation functions imposed by the conformal symmetry, while the information about the dynamics is encoded in terms of the ``conformal data:" operator dimensions and OPE coefficients. Given how fundamental conformal blocks are for CFTs significant effort has been devoted to computing such blocks and several different methods have been presented in the literature. However, the majority of this effort has been focused on four-point correlation functions and associated four-point blocks, going back to the seminal works~\cite{Ferrara:1971vh,Ferrara:1973vz,Ferrara:1974ny,Dolan:2000ut,Dolan:2003hv,Dolan:2011dv}. There are only a handful of results on higher-point global conformal blocks that are applicable for obtaining $M$-point correlation functions, where $M\geq5$~\cite{Alkalaev:2015fbw,Rosenhaus:2018zqn,Parikh:2019ygo,Fortin:2019dnq,Goncalves:2019znr,Jepsen:2019svc,Parikh:2019dvm,Fortin:2019zkm,Irges:2020lgp,Fortin:2020ncr,Fortin:2020yjz,Anous:2020vtw,Pal:2020dqf,Fortin:2020bfq,Hoback:2020pgj,Fortin:2020zxw,Buric:2020dyz, Hoback:2020syd, Poland:2021xjs, Buric:2021ywo,  Buric:2021ttm, Buric:2021kgy}. One reason is that a complete set of crossing equations, which is one of the basic principles behind conformal bootstrap, can be formulated in terms of four-point functions alone. One could utilize crossing equations for higher-point functions as well, but they would be automatically satisfied if all four-point crossing equations are satisfied. For operators with spin, higher-point functions might provide a practical advantage for bootstrap, but for now it is not clear if this will turn out to be the case.

Another reason for the scarcity of results on higher-point blocks is that they are notoriously difficult to compute. Blocks with as few as four points require rather involved approaches, while higher-point blocks are even more complicated. The complication certainly grows with the number of points, partly because conformal blocks are functions of the invariant cross ratios and there are $\frac{M(M-3)}{2}$ independent cross rations with $M$ points. This count provides the maximum number of independent cross ratios, while in smaller number of spacetime dimensions $d$ there are dependencies that reduce the set of independent cross ratios. As a consequence, the problem of finding global conformal blocks in $d=1$ and $d=2$  is simpler and there are comparatively more results for conformal blocks in those dimensions~\cite{Alkalaev:2015fbw,Rosenhaus:2018zqn,Fortin:2019zkm,Anous:2020vtw,Fortin:2020zxw}. For a general number of dimensions there are only a few results for specific number of points and selected topologies\footnote{Blocks with $M\geq 6$ can have several topologies. Any block can be thought of as arising from decomposition of the correlation function using the OPE and visualized as a tree diagram with $M$ external vertices and $M-2$ internal vertices with three lines joining at each internal vertex. There are $(2 M-5)!!$ different OPE decompositions of an $M$-point function with fixed external vertices~\cite{Fortin:2020yjz}. Topology of the block refers to the topology of the diagram. There is no closed formula for the number of topologies. }~\cite{Rosenhaus:2018zqn,Parikh:2019ygo,Jepsen:2019svc,Parikh:2019dvm,Fortin:2019zkm,Fortin:2020ncr,Fortin:2020yjz,Fortin:2020bfq,Hoback:2020pgj,Poland:2021xjs}.

Recently, a conjecture was made on the form of an arbitrary scalar conformal block, that is a block with both external and internal scalar operators, for any $M$ and any topology with no restriction on the number of dimensions~\cite{Hoback:2020pgj}. The conjecture provides a set of rules for writing a conformal block based on its diagrammatic representation that is reminiscent of the well-known rules that convert a Feynman diagram to a scattering amplitude. A crucial ingredient for this construction was the use of Mellin amplitudes and their close connection to amplitudes in the AdS space~\cite{Mack:2009gy,Mack:2009mi,Penedones:2010ue,Fitzpatrick:2011ia, Paulos:2011ie,Nandan:2011wc}.

The conjecture was originally verified in specific examples, by comparison with other calculations in the literature and some novel higher-point results. Here, our goal is a proof of the conjecture.  The proof is inductive: we obtain an $M$-point block from an $(M-1)$-point block. We borrow methods both from work that uses the OPE directly to construct blocks~\cite{Fortin:2019fvx,Fortin:2019dnq} and from work that relies on Mellin amplitudes~\cite{Hoback:2020pgj}.

One of the key ingredients that made the proof possible was a convenient choice of the invariant cross ratios. The choice of cross ratios is clearly not unique as any function of cross ratios is invariant as well. The arbitrariness is not there because one might consider some complicated functions of the invariant cross ratios. All cross ratios we work with are plain ratios of  $x^2_{ij}=(x_i-x_j)^2$, but with $M$ external variables there is no unique choice. Already for $M=4$, one could choose the familiar $u$ and $v$ cross ratios, or choose instead $\frac{u}{v}$ and $\frac{1}{v}$. Both choices are just ratios of $x^2_{ij}$'s and there are clearly more choices in addition to the two we just mentioned. What turned out important for the proof was tailoring cross ratios for an $M$-point function such that they reduce straightforwardly to cross ratios for an $(M-1)$-point function. The original conjecture divided the Mellin variables, and in turn associated cross ratios, into three sets: $\cal{U}$, $\cal{V}$, and  $\cal{D}$. Set $\cal{U}$ corresponds to, what we call, $u$-type cross ratios, set $\cal{V}$ to $v$-type, while set $\cal{D}$ contains dependent variables that are eliminated by integrating over a set of Dirac delta functions that incorporate dependencies. The particular choice of cross ratios that is described in detail later on is amenable to proof by induction. Choosing cross ratios might seem mundane, and yet it was significant for our proof.

We derive suitable cross ratios using a diagrammatic method that we termed flow diagrams~\cite{Fortin:2020zxw}. The flow refers to a choice of internal coordinates in a diagram. Using the convergence property of the OPE in CFTs, one can systematically reduce an $M$-point function to an $(M-1)$-point function by combining two operators and replacing them with a single operator on the right hand side of the OPE, and so on until there is a nothing left but a simple two- or  three-point function.  Schematically, the OPE can be written as ${\cal O}_i(x_i) {\cal O}_j(x_j) \sim {\cal O}_k$, where we neglected  functions of coordinates and derivatives on the right-hand side of the OPE. What flow diagrams represent is the position at which ${\cal O}_k$ appears. One could choose ${\cal O}_k(x_i)$ or ${\cal O}_k(x_j)$ and such a selection needs to be made with every OPE that is used to reduce a correlator to a conformal block and the conformal data. When working in the $d+2$-dimensional embedding space, no other possibilities exists for the operator position on the right-hand side of the OPE as every coordinate needs to be placed on the light cone\cite{Ferrara:1971vh,Ferrara:1973yt}. We do not use the embedding space here, but simply pick one coordinate or the other for the placement of ${\cal O}_k$. The intermediate steps depend on such choices, but the validity of the proof holds for any choice.  Following how a coordinate ``flows" in a diagram leads to a prescription for the cross ratios. We introduce flow diagrams in Section~\ref{CROSSFEYN}. Several concrete examples are worked out in Sections~\ref{FOURCROSSRATIOS} and \ref{LOWPOINT}.

With the cross ratios at hand we turn to the inductive proof in Section~\ref{PROOF}.  The conjecture for the $M$-point blocks is given in terms of Mellin-Barnes integrals that convert a Mellin amplitude to a conformal block. We organized our calculation such that 
\begin{displaymath}
\text{CB}_M \sim \left(\prod_{j=2}^{M} \int\frac{d\gamma_{1j}}{2\pi i} \right) 
 {2\pi i}\delta(\sum_{j=2}^M \gamma_{1j}-\Delta_1)
 \frac{1}{\frac{\Delta_{k_{1}}-s_{1}}{2} +m_{1}} 
  \left(\prod_{j=2}^{M} \frac{\Gamma(\gamma_{1j})}{(x_{1j}^2)^{\gamma_{1j}}} \right) 
  {\text{CB}}_{M-1},
\end{displaymath}
where $\text{CB}_M$ stands for an $M$-point conformal block.  The $(M-1)$-point block, $ {\text{CB}}_{M-1}$, is obtained from the conjectured form of such a block. We show that once the necessary integrals are performed we obtain $\text{CB}_M$ of exactly the form predicted by the conjecture. An inductive proof has to start somewhere, by verifying that the initial step holds. It turns out, the initial step ($M=3$) holds quite trivially. While only one starting point is needed, the $M=3$ correlators are special as there are no invariant cross ratios with three points. However, the conjecture and the proof do apply to the $3\longrightarrow 4$ case. In Appendix \ref{BASECASE} we inspect that everything checks out when going from $M=3$ to $M=4$, as well as going from $M=4$ to $M=5$. We spelled out the $4\longrightarrow 5$ case to display a more generic procedure.

Admittedly, the proof is rather technical with quite a few steps. Therefore, we showcase in detail an example of a $9$-point function in Section~\ref{9-point}. We display how to choose the cross ratios in that case and how to construct the corresponding conformal block. We also discuss certain discrete symmetries that a block must possess. In the simplest case, any two external operators that are connected together to a vertex can be interchanged leading to a $\mathbb{Z}_2$ symmetry. The block must be invariant under such a $\mathbb{Z}_2$, which provides a non-trivial consistency check as many cross ratios need to be rearranged under operator exchange. The $9$-point example in Section~\ref{9-point} posses a $\mathbb{Z}_2\times \mathbb{Z}_2\times \mathbb{Z}_2$ symmetry although analogous discrete symmetries can be significantly larger and more complicated for more symmetric diagrams. Afterwards, we conclude in Section~\ref{DISCUSSION}. We note that with this paper the dimensional reduction of higher-point blocks discussed in~\cite{Hoback:2020syd}, which assumed the Feynman rules conjecture, is now proven.

\section{Feynman-like Rules}
\label{CROSSFEYN}

This paper uses a method of ``flow diagrams''~\cite{Fortin:2020zxw} in order to find $u$-type and $v$-type cross ratios for a conformal block. We can always use the OPE to reduce an $M$-point conformal block to an $(M-1)$-point block. Iterating such a reduction we decompose a block to a diagram consisting of 3-point vertices. When specific position space coordinates are chosen for a given OPE we will call the associated 3-point vertex an OPE vertex. Not all coordinate choices are consistent when the entire diagram with $M$ external legs is concerned. Flow diagrams provide a way of consistently assigning position space coordinates for all 3-point vertices of a given diagram. A consistent coordinate assignment depends on the order in which one decomposes the block into OPE vertices. Flow diagrams are pictorial representations of position coordinate assignments for each OPE throughout a block. We will illustrate these concepts in \ref{INDEXCROSSRATIOS} while specific examples are shown in Sections \ref{FOURCROSSRATIOS} and \ref{LOWPOINT}.

The term ``flow" refers to how coordinates are shared among OPE vertices or how they flow from one vertex to another. We say that a leg ``flows'' in a diagram if its position space coordinate is shared in a neighboring OPE vertex. The collection of OPE vertices that have the same flowing leg, naturally leads to ``comb'' structures within the flow diagram. Reduction of an $M$-point block can lead to several different overall topologies of diagrams. The ``comb" topology, illustrated in Figure~\ref{FigVComb}, is the simplest one. Coordinate flows single out sub-diagrams picking out those vertices in a diagram that share the same coordinate. Naturally, the vertices sharing one coordinate form a comb topology when the rest of diagram is ignored to concentrate on one particular coordinate flow.  The comb structures within a flow diagram provide an extremely useful method for figuring out in what order to integrate the Mellin-Barnes integral representation of conformal blocks to obtain explicit series representations for conformal blocks in arbitrary topologies. 

In this section we first introduce flow diagrams and explain how to flow coordinates along a given topology, effectively decomposing it in its comb structures.  We stress that this decomposition depends on the choice of flows. Using the flow diagrams prescription, we then define rules to compute all conformal cross ratios.  Finally, we combine all of the above with the known Feynman-like rules, introducing an explicit recipe for constructing scalar conformal blocks in arbitrary topologies.

\subsection{OPE Vertices and Cross Ratios for Arbitrary Topologies}\label{INDEXCROSSRATIOS}

Any scalar contribution to a $M$-point correlation function $\langle\mathcal{O}_1(x_1)\ldots\mathcal{O}_M(x_M)\rangle$ is of the form
\eqn{EqW}{
    W_M(x_i) = L(\Delta_1,\ldots,\Delta_M) \left(\prod_{i=1}^{M-3} u_i^{\Delta_{k_i}/2} \right) g(u,1-v)\,,
}
where $L(\Delta_1,\ldots,\Delta_M)$ is the leg factor which ensures that the correlation function transforms covariantly under the conformal group and $g(u,1-v)$ is the conformal block for that particular topology and set of exchanged operators.  Indeed, an OPE decomposition of any correlation function into its OPE vertices leads to a specific topology, with conformal blocks summed over internal exchanged operators.  Moreover, the appropriate $u$-type and $v$-type cross ratios appearing in the conformal blocks also depend on the corresponding topology and the chosen flow of coordinates, which can be found via flow diagrams presented in this paper.  We now introduce the rules to build the leg factor and the cross ratios of any conformal block with arbitrary topology. The construction of the conformal block is described in the next subsection.

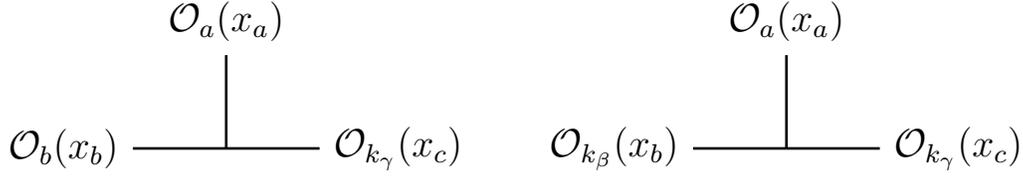
\begin{figure}[t]
\centering
\resizebox{14cm}{!}{%
\begin{tikzpicture}[thick]
\begin{scope}
\node at (0,2.5) {$1I$ OPE vertex};
\draw[-] (0,0)--+(90:1) node[above]{$\mathcal{O}_{a}(x_{a})$};
\draw[-] (0,0)--+(180:1) node[left]{$\mathcal{O}_{b}(x_{b})$};
\draw[-] (0,0)--+(0:1) node[right]{$\mathcal{O}_{k_{\gamma}}(x_{c})$};
\end{scope}
\begin{scope}[xshift=6cm]
\node at (0,2.5) {$2I$ OPE vertex};
\draw[-] (0,0)--+(90:1) node[above]{$\mathcal{O}_{a}(x_{a})$};
\draw[-] (0,0)--+(180:1) node[left]{$\mathcal{O}_{k_{\beta}}(x_{b})$};
\draw[-] (0,0)--+(0:1) node[right]{$\mathcal{O}_{k_{\gamma}}(x_{c})$};
\end{scope}
\end{tikzpicture}
}
\caption{$1I$ and $2I$ OPE vertices. Here, $\mathcal{O}_{a}$ ($\mathcal{O}_{k_a}$) represent external (internal, or exchanged) quasi-primary operators.}
\label{Fig1&2IOPE}
\end{figure}

For the OPE decomposition of the correlation function leading to the topology of interest, we first define $nI$ OPE vertices as OPE vertices with $n$ internal (or exchanged) operators.  The leg factor $L(\Delta_1,\ldots,\Delta_M)$ can then be determined by looking at the $1I$ and $2I$ OPE vertices. Specifically, after defining $X_{ab;c}$ as
\eqn{X}{
    X_{ab;c}=X_{ba;c}=\frac{x^2_{ab}}{x^2_{ac}x^2_{bc}},
}
one can associate factors
\eqn{1ILegs}{
L_{a}(\Delta_a)L_{b}(\Delta_b)=X^{\frac{\Delta_{a}}{2}}_{bc;a}X^{\frac{\Delta_{b}}{2}}_{ac;b},
}
and
\eqn{2ILegs}{
L_{a}(\Delta_a)=X^{\frac{\Delta_{a}}{2}}_{bc;a}  
}
to the $1I$ and $2I$ OPE vertices depicted in Figure \ref{Fig1&2IOPE}, respectively.  Then, the full leg factor $L(\Delta_1,\ldots,\Delta_M)$ can be built by multiplying all $L_{a}(\Delta_a)$ for $1\leq a\leq M$ together.  We note that in Figure \ref{Fig1&2IOPE}, external operators are indexed by $a$ or $b$ while internal (or exchanged) operators are indexed by $k_\beta$ or $k_\gamma$. 

To construct the cross ratios, it is important to first define a proper prescription for the flow of the coordinates in the OPE decomposition of the correlation function.  This is accomplished by ensuring that every operator on a given OPE vertex is at a different coordinate.  Then the flow of coordinates between adjacent OPE vertices is defined by choosing the coordinate that flows, in the OPE limit, from one OPE vertex to the other.  This prescription is most easily explained with concrete examples.

For example, the $u$-type conformal cross ratios are obtained from pairs of OPE vertices connected by a shared internal line yielding $M-3$ cross ratios. Equivalently, one can think of the $u$-type cross ratios as arising from $M-3$ four-point structures inside the $M$-point conformal blocks. There is one $u$-type cross ratio for each unique choice of a four-point structure.  For the four-point structure depicted in Figure \ref{FigU4pt} and its OPE decomposition shown in Figure \ref{FigU4ptOPE} (where $\mathcal{O}_{\sigma}$ can be an external or an internal operator),\footnote{Additional explicit examples can be found in \eno{fourpointudiagrams}, \eno{ucross5}, and \eno{ucross9} with their associated OPE vertices in \eno{fourpointflow}, \eno{fivepointflow}, and Figure \ref{flow diagram 9p}, respectively.}  we can build a $u$-type conformal cross ratio as
\begin{align}
\label{uCR}u_{\alpha}=\frac{x^2_{a_1a_2}x^2_{a_3a_4}}{x_{a_1a_4}^2x_{a_2a_3}^2},
\end{align}
following the flow of coordinates.

\begin{figure}[!t]
\centering
\resizebox{8cm}{!}{%
\begin{tikzpicture}[thick]
\begin{scope}
\node at (0,0) {$\mathcal{O}_{\sigma_1}(x_{a_1})$};
\draw[-] (1,0)--(4.5,0);
\draw[-] (2,0)--(2,1) node at (2,1.5) {$\mathcal{O}_{\sigma_2}(x_{a_2})$};
\draw[-] (3.5,0)--(3.5,1) node at (3.5,1.5) {$\mathcal{O}_{\sigma_3}(x_{a_3})$};
\draw[-] (3.5,0)--(4.5,0) node at (5.5,0)  {$\mathcal{O}_{\sigma_4}(x_{a_4})$};
\node at (2.8,-0.5) {$\mathcal{O}_{k_{\alpha}}$};
\end{scope}
\end{tikzpicture}
}
\caption{The four-point structure in which $\mathcal{O}_{\sigma_a}$ can be either external or internal operators.}
\label{FigU4pt}
\end{figure}
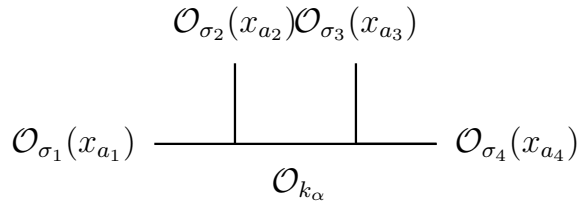
\begin{figure}[!t]
\centering
\resizebox{14cm}{!}{%
\begin{tikzpicture}[thick]
\begin{scope}
\node at (0.0,0) {$\mathcal{O}_{\sigma_1}(x_{a_1})$};
\draw[-] (1,0)--(3.0,0);
\draw[-] (2,0)--(2,1) node at (2,1.5) {$\mathcal{O}_{\sigma_2}(x_{a_2})$};
\node at (4,0) {$\mathcal{O}_{k_{\alpha}}(x_{a_4})$};
\end{scope}
\begin{scope}[xshift=7cm]
\node at (0.5,0) {$\mathcal{O}_{k_{\alpha}}(x_{a_2})$};
\draw[-] (1.5,0)--(3.5,0);
\draw[-] (2.5,0)--(2.5,1) node at (2.5,1.5) {$\mathcal{O}_{\sigma_3}(x_{a_3})$};
\draw[-] (2.5,0)--(3.5,0) node at (4.5,0)  {$\mathcal{O}_{\sigma_4}(x_{a_4})$};
\end{scope}
\end{tikzpicture}
}
\caption{The choice of OPE vertices which are equivalent to the OPEs $\mathcal{O}_{\sigma_1}(x_{a_1})\mathcal{O}_{\sigma_2}(x_{a_2})\sim\mathcal{O}_{k_{\alpha}}(x_{a_2})$ and $\mathcal{O}_{\sigma_3}(x_{a_3})\mathcal{O}_{\sigma_4}(x_{a_4})\sim\mathcal{O}_{k_{\alpha}}(x_{a_4})$.}
\label{FigU4ptOPE}
\end{figure}
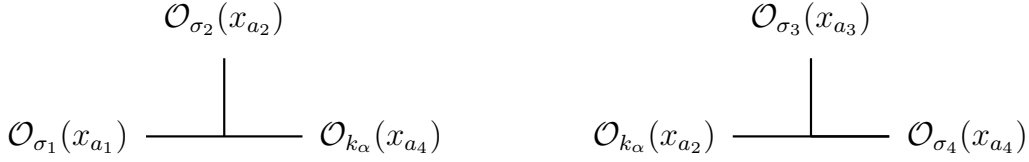
In Figure \ref{FigU4pt}, the coordinates on the operators are fixed (obtained consistently from the flow) if the corresponding operator is external (internal).  To build the flow of coordinates, it is necessary to ensure that adjacent OPE vertices are consistent, as in Figure \ref{FigU4ptOPE}.  Indeed, in Figure \ref{FigU4ptOPE}, the internal operator $\mathcal{O}_{k_\alpha}$ that pops out of the OPE decomposition appears in each OPE vertex, and its coordinate is defined by the OPE limit.  For Figure \ref{FigU4ptOPE}, the coordinate of the left OPE vertex is $x_{a_4}$ because the OPE of the right OPE vertex is chosen as $\mathcal{O}_{\sigma_3}(x_{a_3})\mathcal{O}_{\sigma_4}(x_{a_4})\sim\mathcal{O}_{k_\alpha}(x_{a_4})$.  Equivalently, the coordinate of the right OPE vertex is $x_{a_2}$ because the OPE of the left OPE vertex is chosen as $\mathcal{O}_{\sigma_1}(x_{a_1})\mathcal{O}_{\sigma_2}(x_{a_2})\sim\mathcal{O}_{k_\alpha}(x_{a_2})$.  The coordinate of any internal operator on any OPE vertex is obtained following this flow prescription.  There are obviously more than one choice of flow of coordinates per topology, and the corresponding $u$-type cross ratio depends on the choices made [the denominator in \eqref{uCR} clearly depends on the coordinates that flow], but every flow is an appropriate starting point for the Feynman-like rules described below.  We note here that the symbol $\sim$ in the OPE is a shortcut notation to indicate which operator (and its tower of descendants) is exchanged.

The $v$-type conformal cross ratios can be obtained similarly by selecting all pairs of OPE vertices yielding $\binom{M-2}{2}$ of such cross sections. For every pair of vertices, $v$ cross ratios also involve all the vertices needed to connect the selected pair via internal lines. Therefore, this prescription associates each $v$ with a comb structure inside the $M$-point conformal block. The comb structures can have anywhere between 2 and $M-2$ vertices that is between 4 and $M$ points. Note that the $u$-type ratios involved adjacent vertex pairs only.

Specifically, for the comb structure depicted in Figure \ref{FigVComb} and the chosen OPE decomposition depicted in Figure \ref{FigVCombOPE}, we can build a $v$-type conformal cross ratio
\begin{align}
\label{vCR}v_{a_1a_n}=\frac{x^2_{a_1a_n}x^2_{a_{n-1}q_{2n-6}}}{x^2_{a_1q_1}x^2_{a_nq_{2n-6}}}\prod_{i=1}^{n-4}\frac{x^2_{q_{2i-1}q_{2i}}}{x^2_{q_{2i}q_{2i+1}}},
\end{align}
with $q_2=a_2$ and $q_{2n-7}=a_{n-1}$.
\begin{figure}[!t]
\centering
\resizebox{10cm}{!}{%
\begin{tikzpicture}[thick]
\begin{scope}
\node at (0.0,0) {$\mathcal{O}_{\sigma_1}$};
\draw[-] (0.5,0)--(8.5,0);
\draw[-] (1.5,0)--(1.5,1) node at (1.5,1.5) {$\mathcal{O}_{\sigma_2}$};
\node at (2,-0.5) {$\mathcal{O}_{k_{\alpha_1}}$};
\draw[-] (2.5,0)--(2.5,1) node at (2.5,1.5) {$\mathcal{O}_{\sigma_3}$};
\node at (4.5,1) {$\dots$};
\draw[-] (6.5,0)--(6.5,1) node at (6.5,1.5) {$\mathcal{O}_{\sigma_{n-2}}$};
\node at (7.3,-0.5) {$\mathcal{O}_{k_{\alpha_{n-3}}}$};
\draw[-] (7.5,0)--(7.5,1) node at (7.5,1.5) {$\mathcal{O}_{\sigma_{n-1}}$};
\draw[-] (7.5,0)--(8.5,0) node at (9,0) {$\mathcal{O}_{\sigma_{n}}$};
\end{scope}
\end{tikzpicture}
}
\caption{The comb structure ($n\geq 4$) from which one $v$-type conformal cross ratio can be built. It should be understood that $\mathcal{O}_{\sigma_i}\equiv\mathcal{O}_{\sigma_i}(x_{a_i})$.}
\label{FigVComb}
\end{figure}
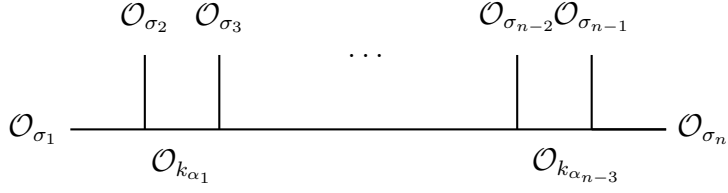
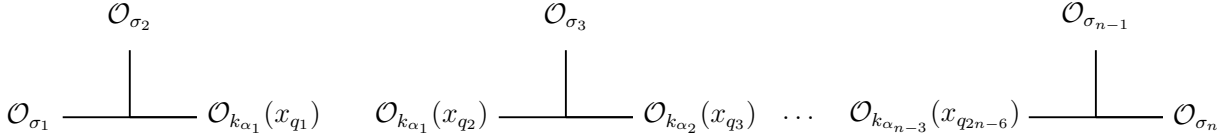
\begin{figure}[!t]
\centering
\resizebox{16.5cm}{!}{%
\begin{tikzpicture}[thick]
\begin{scope}
\node at (0,0) {$\mathcal{O}_{\sigma_1}$};
\draw[-] (0.5,0)--(2.5,0);
\draw[-] (1.5,0)--(1.5,1) node at (1.5,1.5) {$\mathcal{O}_{\sigma_2}$};
\draw[-] (1.5,0)--(2.5,0) node at (3.5,0) {$\mathcal{O}_{k_{\alpha_1}}(x_{q_1})$};
\end{scope}
\begin{scope}[xshift=6cm]
\node at (0.0,0) {$\mathcal{O}_{k_{\alpha_1}}(x_{q_2})$};
\draw[-] (1,0)--(3,0);
\draw[-] (2,0)--(2,1) node at (2,1.5) {$\mathcal{O}_{\sigma_3}$};
\draw[-] (2,0)--(3,0) node at (4,0) {$\mathcal{O}_{k_{\alpha_2}}(x_{q_3})$};
\end{scope}
\begin{scope}[xshift=11.5cm]
\node at (0,0) {$\dots$};
\end{scope}
\begin{scope}[xshift=13.5cm]
\node at (0.0,0) {$\mathcal{O}_{k_{\alpha_{n-3}}}(x_{q_{2n-6}})$};
\draw[-] (1.4,0)--(3.4,0);
\draw[-] (2.4,0)--(2.4,1) node at (2.4,1.5) {$\mathcal{O}_{\sigma_{n-1}}$};
\draw[-] (2.4,0)--(3.4,0) node at (3.9,0) {$\mathcal{O}_{\sigma_{n}}$};
\end{scope}
\end{tikzpicture}
}
\caption{The associated OPE vertices for the comb structure in Figure \ref{FigVComb}. The coordinates $x_{q_i}$ must be chosen in a way that is consistent with the OPE, or in other words, the flow of coordinates. Without loss of generality, we assume $q_2=a_2$ and $q_{2n-7}=a_{n-1}$. In other words, we take the OPEs $\mathcal{O}_{\sigma_{1}}(x_{a_1})\mathcal{O}_{\sigma_2}(x_{a_2})\sim\mathcal{O}_{k_{\alpha_1}}(x_{a_2})$ and $\mathcal{O}_{\sigma_{n}}(x_{a_n})\mathcal{O}_{\sigma_{n-1}}(x_{a_{n-1}})\sim\mathcal{O}_{k_{\alpha_{n-3}}}(x_{a_{n-1}})$.}
\label{FigVCombOPE}
\end{figure}
We stress that the OPEs $\mathcal{O}_{\sigma_{1}}(x_{a_1})\mathcal{O}_{\sigma_2}(x_{a_2})\sim\mathcal{O}_{k_{\alpha_1}}(x_{a_2})$ and $\mathcal{O}_{\sigma_{n}}(x_{a_n})\mathcal{O}_{\sigma_{n-1}}(x_{a_{n-1}})\sim\mathcal{O}_{k_{\alpha_{n-3}}}(x_{a_{n-1}})$ have been chosen to agree with \eqref{vCR}.  Moreover, the coordinates on the $\mathcal{O}_{k_i}$ must be chosen consistently following the OPE flow mentioned above.  For example, $x_{q_4}$ is either $x_{a_3}$ if the OPE is taken as $\mathcal{O}_{k_{\alpha_1}}(x_{q_2})\mathcal{O}_{\sigma_3}(x_{a_3})\sim\mathcal{O}_{k_{\alpha_2}}(x_{a_3})$, or $x_{q_2}$ if the OPE is taken as $\mathcal{O}_{k_{\alpha_1}}(x_{q_2})\mathcal{O}_{\sigma_3}(x_{a_3})\sim\mathcal{O}_{k_{\alpha_2}}(x_{q_2})$.

The rules for $v$-type conformal cross ratios \eqref{vCR} can be re-expressed in many different forms. To see this, we first assume that $x_{a_2}$  flows until the vertex containing $\mathcal{O}_{\sigma_{r_1}}$ where $x_{\sigma_{r_1}}$ starts flowing.  Concretely, this means that the coordinate of the right operator of all OPE vertices up to and including the OPE vertex with the top operator at $x_{\sigma_{r_1}}$, are $x_{a_2}$.  Then, $x_{\sigma_{r_1}}$ flows up to the vertex containing $\mathcal{O}_{\sigma_{r_2}}$ where $x_{\sigma_{r_2}}$ starts flowing.  We keep going on until the last vertex is reached.  As a result, we find that
\begin{align}
\label{q=a}\nonumber&q_{2i}=a_2,\hspace{0.5cm}1\leq i\leq r_1-2,\\
&q_{2i}=a_{r_1},\hspace{0.5cm}r_1-1\leq i\leq r_2-2,\\\nonumber
&\hspace{3cm}\vdots
\end{align}
where the odd-subscript $q$'s are not defined in terms of $a$'s by the flow as described above.  We note however that their explicit coordinates are not needed since they always cancel in \eqref{vCR} (for every numerator containing an odd-subscript $q$, there is an equal denominator with the same odd-subscript $q$).

To proceed further, we define the boundary vertices as vertices at which the flowing coordinates change. We use $\bar{T}^{a_1a_n}_{abc}$ to denote the boundary vertex where $x_a$ stops flowing while $x_b$ starts flowing and associate a factor
\eqn{wDef}{
w^{a_1a_n}_{abc}=\frac{x^2_{bc}}{x^2_{ac}},
}
to $\bar{T}^{a_1a_n}_{abc}$. In $\bar{T}^{a_1a_n}_{abc}$, we use the upper indices $a_1a_n$ to represent the fact that the boundary vertices are defined by focusing on the direction of the flow of coordinates from $a_1$ to $a_n$ in the comb structure, and the values of $a$, $b$ and $c$ depend on this choice.  In the specific case described here, the flow starts from the left OPE vertex in Figure \ref{FigVCombOPE} and goes toward the right vertex.  After substituting \eqref{q=a} into \eqref{vCR}, it is easy to check that \eqref{vCR} can be rewritten as
\begin{align}
\label{vCR1}v_{a_1a_n}=\frac{x^2_{a_1a_n}x^2_{a_2q_1}}{x^2_{a_nq_{2n-6}}x^2_{a_1q_1}}\prod_{t}w^{a_1a_n}_t,
\end{align}
where the product has been taken over all of boundary vertices. Similarly, focusing on the direction from $a_n$ to $a_1$, \eqref{vCR} can also be rewritten as
\begin{align}
\label{vCR2}v_{a_1a_n}=\frac{x^2_{a_1a_n}x^2_{a_{n-1}q_{2n-6}}}{x^2_{a_nq_{2n-6}}x^2_{a_1q_1}}\prod_{t}w^{a_na_1}_t.
\end{align}

Therefore, once the flow of coordinates for a given OPE decomposition of the correlation function in a corresponding topology has been chosen, it is straightforward to construct the $u$-type and $v$-type cross ratios.  The cross ratios can then be used as a starting point for the Feynman-like rules of the conformal block.

We end this discussion by constructing an important index set for the $M$-point topology associated with the choice of OPE vertex, denoted ${\cal V}$. For each $v$-type cross ratio there exists a corresponding element in ${\cal V}$. Thus the cardinality of ${\cal V}$ is $\binom{M-2}{2}$. Particularly, for the cross ratio in~\eno{vCR}, the corresponding element in ${\cal V}$ will be the index pairing $(a_1 a_n)$. 
This element corresponds to the indices of position coordinates of external operator insertions at the opposite extremes of the comb structure in Figure~\ref{FigVComb} which also feature in the numerator of the $v$-type cross-ratio~\eno{vCR} as $x_{a_1a_n}^2$, in other words the coordinates that do not flow.


\subsection{Feynman-like Rules for Conformal Blocks}
\label{FEYNMAN}

In this section we will briefly state the Feynman-like rules for conformal blocks~\cite{Hoback:2020pgj}. 
The complete set of rules are as follows:
\begin{itemize}
\item For a $M$-point correlation function, choose an OPE decomposition (in other words, a topology), and assign a consistent flow of coordinates between every OPE vertex.
\item From the OPE flow, determine the cross ratios with the help of \eqref{uCR} and \eqref{vCR},
\item Given the associated $M$-point conformal block, let the dimensions and insertion coordinates of the external operators be respectively, $\Delta_i$  and $x_i$ for $i=1,\ldots, M$. 
Let the dimensions of the exchanged operators be enumerated $\Delta_{k_i}$ for $i=1,\ldots,M-3$.
\item Assign each internal edge with an index $i$ running from $1$ to $M-3$. Associate to each such edge an integer parameter $m_i$, which we refer to as the ``single-trace parameter,'' and a factor of
    \eqn{EdgeDef}{
    { E_i := \frac{ (\Delta_{k_i} - h +1)_{m_i}}{ (\Delta_{k_i})_{2m_i + \ell_{k_i}} }\,, }
    }
    where $h := d/2$ and $\ell_{k_i}$ is an integer parameter, which we call the ``post-Mellin parameter,'' associated with the conformal dimension $\Delta_{k_i}$.

    \item Label each (cubic) vertex with an index $i$ running from $1$ to $M-2$. Assuming each leg of the cubic vertex has incident conformal dimensions, $\Delta_{a}, \Delta_{b}$, and $\Delta_{c}$,  write the factor associated to this vertex as
    \eqn{VertexDef}{
V_i &:= (\Delta_{ab,c})_{m_{ab,c}+{1\over 2}\ell_{ab,c}} (\Delta_{ac,b})_{m_{ac,b}+{1\over 2}\ell_{ac,b}} (\Delta_{bc,a})_{m_{bc,a}+{1\over 2}\ell_{bc,a}} \cr 
     &  \times F_A^{(3)}\!\left[\Delta_{abc,}- h; \{-m_a, -m_b, -m_c \}; \left\{\Delta_a -h+1, \Delta_b -h+1,\Delta_c -h+1 \right\}; 1,1,1 \right],}
where $F_A^{(3)}$ is the Lauricella function given by~\cite{Lauricella1893,Srivastava1985book,AartsMathWorld} (see also ref.~\cite{Paulos:2011ie})
\eqn{LauricellaDef}{
    F_A^{(\ell)}\Big[g;\{a_1,\ldots,a_\ell\};\{b_1,\ldots,b_\ell\};x_1,\ldots,x_\ell\Big]
    := \sum_{n_1,\ldots,n_\ell=0}^\infty(g)_{n_1+\cdots+n_\ell}\prod_{i=1}^\ell \frac{(a_i)_{n_i}}{(b_i)_{n_i}}\frac{x_i^{n_i}}{n_i!}\,.
}
\noindent Here $m_a, m_b$, and $m_c$ ($\ell_a, \ell_b$, and $\ell_c$) are the respective single-trace parameters (post-Mellin parameters) associated with each edge as mentioned in the previous point, and for conformal dimensions $\Delta_i$ we use the notation
\eqn{}{    \Delta_{i_1\ldots i_m,i_{m+1} \ldots i_n} := {1\over 2} \left( \Delta_{i_1} + \cdots + \Delta_{i_m} - \Delta_{i_{m+1}} - \cdots - \Delta_{i_n} \right),
}
while for single-trace parameters and post-Mellin parameters we use 
\begin{align}
m_{i_1 \ldots i_n, i_{n+1} \ldots i_{k}} := m_{i_1} + \cdots + m_{i_n} - m_{i_{n+1}} - \cdots - m_{i_k}\, ,\\
\ell_{i_1 \ldots i_n, i_{n+1}\ldots i_{k}} := \ell_{i_1} + \cdots + \ell_{i_n} - \ell_{i_{n+1}} - \cdots - \ell_{i_k} \,.
\end{align}

Note that the single-trace parameter associated to an external leg is identically zero. Thus depending on the type of vertex ($1I, 2I$ or $3I$; see the discussion around Figure~\ref{Fig1&2IOPE}) the vertex factor will include a Lauricella function of 1, 2, or 3 variables. 

\item Then the full conformal block is given by~\eno{EqW} where the leg factor is given by the discussion around equations~\eno{X}-\eno{2ILegs} with the conformally invariant function of cross-ratios $g(u,1-v)$ constructed by multiplying together all edge- and vertex-factors, including appropriate powers of the cross-ratios, and summing over all integer parameters, as follows:
\eqn{Feynman}{
g(u,1-v) =  \sum_{m_i, j_{rs}=0}^\infty \left[\left( \prod_{i=1}^{M-3} {u_i^{m_i} \over m_i!} \right) \left(\prod_{(rs) \in {\cal V}}^{} {\left(1-v_{rs}\right)^{j_{rs}} \over j_{rs}!} \right) \left( \prod_{i=1}^{M-3} E_i \right) \left(\prod_{i=1}^{M-2} V_i \right) \right],
}
where ${\cal V}$ is the index set associated to the $v$-type cross ratios.\footnote{We assume throughout the paper that $d>M-3$ so that the number of independent cross ratios is $M(M-3)/2$.  For $d\leq M-3$, by rewriting dependent cross ratios in terms of independent cross ratios, we have checked in several cases that the final answer~\eno{Feynman} is correct and we expect that it is valid for all $d$ and $M$.} The post-Mellin parameters $\ell_i$ and $\ell_{k_1}$ are linear combinations of the dummy variables $j_{rs}$, as we describe next.

\end{itemize}

\subsubsection{Determining the post-Mellin Parameters}
\label{POSTMELLIN}

The Feynman-like rules collected above leave the edge- and vertex-factors in terms of post-Mellin parameters $\ell_i$, one associated to each conformal dimension. We will now complete the Feynman-like prescription by specifying how these parameters are to be calculated.

For an external operator with conformal dimension $\Delta_i$ inserted at position $x_i$,the associated post-Mellin parameter is defined to be
\eqn{PostMellinExternal}{
 \ell_i : = \sum_{\substack{(rs) \in {\cal V}_{\rm} \\  r=i {\rm \ or\ } s=i}} j_{rs}\,.
}

For exchanged operators of conformal dimensions $\Delta_{k_i}$, we compute the post-Mellin parameters by repeating the method below until all post-Mellin parameters have been computed. Begin by identifying all internal vertices with two external edges and only one internal edge (which we refer to as $1I$ vertices). In this case one simply adds post-Mellin parameters of the external edges together and subtract all even multiples of $j_{rs}$ variables appearing in the sum.
That is, for the following $1I$ vertex, 
\eqn{PostMellinInternal1}{
&
\begin{tikzpicture}[thick][r]
\begin{scope}
\node at (-.3,0) {$\Delta_{1}$};
\node at (.5,-.3) {};
\draw[-] (0.0,0)--(1,0);
\draw[-] (.5,0)--(.5,1) node at (.5,1.2) {$\Delta_{2}$};
\draw[-] (.5,0)--(1,0) node at (1.4,0) {$\Delta_{k_3}$} ;
\end{scope}
\end{tikzpicture}
\qquad   \ell_{k_3} \stackrel{2{\cal J}}{=} \ell_1 + \ell_2 \,, }
where the symbol $\stackrel{2{\cal J}}{=}$ means equality holds once all even multiples of $j_{rs}$ variables have been dropped. If there are any post-Mellin parameters that have still yet to be computed, solve for them iteratively using the procedure above by adding together the known post-Mellin parameters of two legs of a cubic vertex. That is, if two post-Mellin parameters are known at a shared vertex, the unknown post-Mellin parameter is computed as
\eqn{PostMellinInternal2}{
\begin{tikzpicture}[thick][r]
\begin{scope}
\node at (-.3,0) {$\Delta_{k_1}$};
\node at (.5,-.3) {};
\draw[-] (0.0,0)--(1,0);
\draw[-] (.5,0)--(.5,1) node at (.5,1.2) {$\Delta_{k_2}$};
\draw[-] (.5,0)--(1,0) node at (1.4,0) {$\Delta_{k_3}$} ;
\end{scope}
\end{tikzpicture}
\qquad
    \ell_{k_3} \stackrel{2{\cal J}}{=} \ell_{k_1} + \ell_{k_2} \,. }
Repeat this process until all post-Mellin parameters have been determined.

\section{Examples of Cross Ratios}
\label{FOURCROSSRATIOS}

We first show how to use \eqref{uCR} for an arbitrary four-point structure and \eqref{vCR} for a specific comb structure inside a nine-point topology.

\subsection{\texorpdfstring{$u$}{u}-type Cross Ratios}
In this section we will demonstrate how to use \eno{uCR} to solve for the $u$-type cross ratios given a flow diagram. The construction of these ratios is sketched in four steps. 

\begin{enumerate}
\item

Identify the pairs of neighboring OPE vertices in a flow diagram. The pairs should result in two OPE vertices that have one $\mathcal{O}_{k_{i}}$ between them\footnote{See \eno{fourpointudiagrams}, \eno{ucross5} and \eno{ucross9} for further examples.} so that any given structure should be of the form below. 
\eqn{ucrossstep1}{ 
&\begin{tikzpicture}[thick][r]
\begin{scope}
\node at (-.2,0) {$x_{a}$};
\node at (.5,-.3) {$V_{r}$};
\node at (1.6,-0.5) {$\mathcal{O}_{k_{i}}$};
\draw[-] (0.0,0)--(1,0);
\draw[-] (.5,0)--(.5,1) node at (.5,1.2) {$x_{b}$};
\draw[-] (.5,0)--(1,0) node at (1.2,0) {$x_c$} ;
\end{scope}
\begin{scope}[xshift=2cm]
\node at (-.2,0) {$x_{b}$};
\node at (.5,-.3) {$V_{s}$};
\draw[-] (0.0,0)--(1,0);
\draw[-] (.5,0)--(.5,1) node at (.5,1.2) {$x_{c}$};
\draw[-] (.5,0)--(1,0) node at (1.2,0) {$x_{d}$};
\end{scope}
\end{tikzpicture} 
}

\item Next, identify which legs do not flow in the pair of OPE vertices. We say a leg does not flow if we do not assign those factors in the flow diagram for that structure. In this paper we circled the legs that do not flow in the $u$-type diagrams. Also identify the lower script labeling of $\mathcal{O}_{k_{i}}$. We simply label the $u$-type cross ratios with the subscript corresponding to the $\mathcal{O}_{k_{i}}$ in the diagram that is denote them as $u_i$. 

\eqn{ucrossstep2}{ 
&\begin{tikzpicture}[thick][r]
\begin{scope}
\node at (-.2,0) {$x_{a}$};
\node at (.5,-.3) {$V_{r}$};
\node at (1.6,-0.5) {$\mathcal{O}_{k_{i}}$};
\draw[-] (0.0,0)--(1,0);
\draw[-] (.5,0)--(.5,1) node at (.5,1.2) {$x_{b}$};
\draw[-] (.5,0)--(1,0) node at (1.2,0) {$x_c$} ;
\draw[black] (-.25,-.05) circle (.3 cm);
\end{scope}
\begin{scope}[xshift=2cm]
\node at (-.2,0) {$x_{b}$};
\node at (.5,-.3) {$V_{s}$};
\draw[-] (0.0,0)--(1,0);
\draw[-] (.5,0)--(.5,1) node at (.5,1.2) {$x_{c}$};
\draw[-] (.5,0)--(1,0) node at (1.2,0) {$x_{d}$};
\draw[black] (1.25,-.05) circle (.3 cm);
\end{scope}
\end{tikzpicture} 
}

\item Once the legs which do not flow are identified, in this case $x_a$ and $x_d$, we draw a dashed green line to the leg that flows \textit{from} a vertex to its neighbor, in this case $x_b$ and $x_c$ respectively. The two $x_p$ and $x_q$ factors that are connected by the green dotted lines in a single vertex become subscripts on the \textit{numerator} of $u_i$ as an $x_{pq}^2$ factor. 
\eqn{ucrossstep3}{ 
&\begin{tikzpicture}[thick][r]
\begin{scope}
\node at (-.2,0) {$x_{a}$};
\node at (.5,-.3) {$V_{r}$};
\node at (1.6,-0.5) {$\mathcal{O}_{k_{i}}$};
\draw[-] (0.0,0)--(1,0);
\draw[-] (.5,0)--(.5,1) node at (.5,1.2) {$x_{b}$};
\draw[-] (.5,0)--(1,0) node at (1.2,0) {$x_c$} ;
\draw[black] (-.25,-.05) circle (.3 cm);
\draw[dashed,  teal]  (0,.08) to[bend right=5] (.45,1);
\end{scope}
\begin{scope}[xshift=2cm]
\node at (-.2,0) {$x_{b}$};
\node at (.5,-.3) {$V_{s}$};
\draw[-] (0.0,0)--(1,0);
\draw[-] (.5,0)--(.5,1) node at (.5,1.2) {$x_{c}$};
\draw[-] (.5,0)--(1,0) node at (1.2,0) {$x_{d}$};
\draw[black] (1.25,-.05) circle (.3 cm);
\draw[dashed,  teal]  (.55,1) to[bend right=5] (.8, .1);
\end{scope}
\end{tikzpicture} : u_i = \frac{x_{ab}^2x_{dc}^2}{}
}

\item Finally, the denominator of the expression for $u_i$ is obtained from the numerator by exchanging the labels of the vertices that do flow. Namely, 
\eqn{ucrossstep4}
 {u_i = \frac{x_{ab}^2x_{dc}^2}{x_{ac}^2x_{db}^2}.}

\end{enumerate}

For sufficiently complicated blocks, this prescription might occasionally lead to some confusion. A useful checkup on the expressions for the $u$-type cross ratios can be obtained by drawing a dashed red line to the leg that flows \textit{into} a vertex to its neighbor, in this case $x_c$ and $x_b$ respectively. The two $x_p$ and $x_q$ factors that are connected by the red dotted lines in a single vertex become subscripts on the \textit{denominator} of $u_i$ as an $x_{pq}^2$ factor 
\eqn{ucross-sanity}{ 
&\begin{tikzpicture}[thick][r]
\begin{scope}
\node at (-.2,0) {$x_{a}$};
\node at (.5,-.3) {$V_{r}$};
\node at (1.6,-0.5) {$\mathcal{O}_{k_{i}}$};
\draw[-] (0.0,0)--(1,0);
\draw[-] (.5,0)--(.5,1) node at (.5,1.2) {$x_{b}$};
\draw[-] (.5,0)--(1,0) node at (1.2,0) {$x_c$} ;
\draw[black] (-.25,-.05) circle (.3 cm);
\draw[dashed,  red]  (0,-.08) to[bend right=5] (.9,-.08);
\end{scope}
\begin{scope}[xshift=2cm]
\node at (-.2,0) {$x_{b}$};
\node at (.5,-.3) {$V_{s}$};
\draw[-] (0.0,0)--(1,0);
\draw[-] (.5,0)--(.5,1) node at (.5,1.2) {$x_{c}$};
\draw[-] (.5,0)--(1,0) node at (1.2,0) {$x_{d}$};
\draw[black] (1.25,-.05) circle (.3 cm);
\draw[dashed,  red]  (0,-.08) to[bend right=5] (.9,-.08);
\end{scope}
\end{tikzpicture} : u_i = \frac{}{x_{ac}^2x_{db}^2}
}
yielding again
\eqn{ucrossstep5}{ 
&\begin{tikzpicture}[thick][r]
\begin{scope}
\node at (-.2,0) {$x_{a}$};
\node at (.5,-.3) {$V_{r}$};
\node at (1.6,-0.5) {$\mathcal{O}_{k_{i}}$};
\draw[-] (0.0,0)--(1,0);
\draw[-] (.5,0)--(.5,1) node at (.5,1.2) {$x_{b}$};
\draw[-] (.5,0)--(1,0) node at (1.2,0) {$x_c$} ;
\draw[black] (-.25,-.05) circle (.3 cm);
\draw[dashed,  red]  (0,-.08) to[bend right=5] (.9,-.08);
\draw[dashed,  teal]  (0,.08) to[bend right=5] (.45,1);
\end{scope}
\begin{scope}[xshift=2cm]
\node at (-.2,0) {$x_{b}$};
\node at (.5,-.3) {$V_{s}$};
\draw[-] (0.0,0)--(1,0);
\draw[-] (.5,0)--(.5,1) node at (.5,1.2) {$x_{c}$};
\draw[-] (.5,0)--(1,0) node at (1.2,0) {$x_{d}$};
\draw[black] (1.25,-.05) circle (.3 cm);
\draw[dashed,  red]  (0,-.08) to[bend right=5] (.9,-.08);
\draw[dashed,  teal]  (.55,1) to[bend right=5] (.8, .1);
\end{scope}
\end{tikzpicture} : u_i = \frac{x_{ab}^2x_{dc}^2}{x_{ac}^2x_{db}^2}.
}

\subsection{\texorpdfstring{$v$}{v}-type Cross Ratios}
In this section we will demonstrate how to use \eno{vCR} to construct $v$-type cross ratios given an appropriate choice of flow diagram.

\begin{enumerate}
\item 
To solve for the $v$-type cross ratios, we need to find the comb structure that results for any vertex pair, $V_iV_j$. The subscripts of the $V_iV_j$ pair correspond to the labels of the two OPE vertices that are the end vertices of the corresponding comb structure. This will result in $\binom{M-2}{2}$ $v_{rs}$ cross ratios. For example if we consider the following flow diagram, 
\begin{figure}[h!]
\centering
\resizebox{16.5cm}{!}{%
\begin{tikzpicture}[thick]
\begin{scope}
\node at (-.2,0) {$x_{2}$};
\node at (.5,-.3) {$V_{7}$};
\draw[-] (0.0,0)--(1,0);
\draw[-] (.5,0)--(.5,1) node at (.5,1.2) {$x_{1}$};
\draw[-] (.5,0)--(1,0) node at (1.2,0) {$x_{3}$};
\end{scope}
\begin{scope}[xshift=2cm]
\node at (-.2,0) {$x_{2}$};
\node at (.5,-.3) {$V_{1}$};
\draw[-] (0.0,0)--(1,0);
\draw[-] (.5,0)--(.5,1) node at (.5,1.2) {$x_{3}$};
\draw[-] (.5,0)--(1,0) node at (1.2,0) {$x_{4}$};
\end{scope}
\begin{scope}[xshift=4cm]
\node at (-.2,0) {$x_{2}$};
\node at (.5,-.3) {$V_{2}$};
\draw[-] (0.0,0)--(1,0);
\draw[-] (.5,0)--(.5,1) node at (.5,1.2) {$x_{4}$};
\draw[-] (.5,0)--(1,0) node at (1.2,0) {$x_{9}$};
\end{scope}
\begin{scope}[xshift=4cm]
\node at (.5,1.5) {$x_{9}$};
\node at (1.3,1.7) {$V_{3}$};
\draw[-] (.8,1.5)--(1.4,2.5);
\draw[-] (1.1,2)--(.5,2.5) node at (.5,2.8) {$x_{4}$};
\node at (1.4,2.7) {$x_{5}$};
\end{scope}
\begin{scope}[xshift=6cm]
\node at (-.2,0) {$x_{2}$};
\node at (.5,-.3) {$V_{4}$};
\draw[-] (0.0,0)--(1,0);
\draw[-] (.5,0)--(.5,1) node at (.5,1.2) {$x_{6}$};
\draw[-] (.5,0)--(1,0) node at (1.2,0) {$x_{9}$};
\end{scope}
\begin{scope}[xshift=8cm]
\node at (-.2,0) {$x_{2}$};
\node at (.5,-.3) {$V_{5}$};
\draw[-] (0.0,0)--(1,0);
\draw[-] (.5,0)--(.5,1) node at (.5,1.2) {$x_{7}$};
\draw[-] (.5,0)--(1,0) node at (1.2,0) {$x_{9}$};
\end{scope}
\begin{scope}[xshift=10cm]
\node at (-.2,0) {$x_{7}$};
\node at (.5,-.3) {$V_{6}$};
\draw[-] (0.0,0)--(1,0);
\draw[-] (.5,0)--(.5,1) node at (.5,1.2) {$x_{8}$};
\draw[-] (.5,0)--(1,0) node at (1.2,0) {$x_{9}$};
\end{scope}
\end{tikzpicture}
}
\caption{A possible flow diagram for a 9-point asymmetrical block.}
\label{flow diagram 9p}
\end{figure}
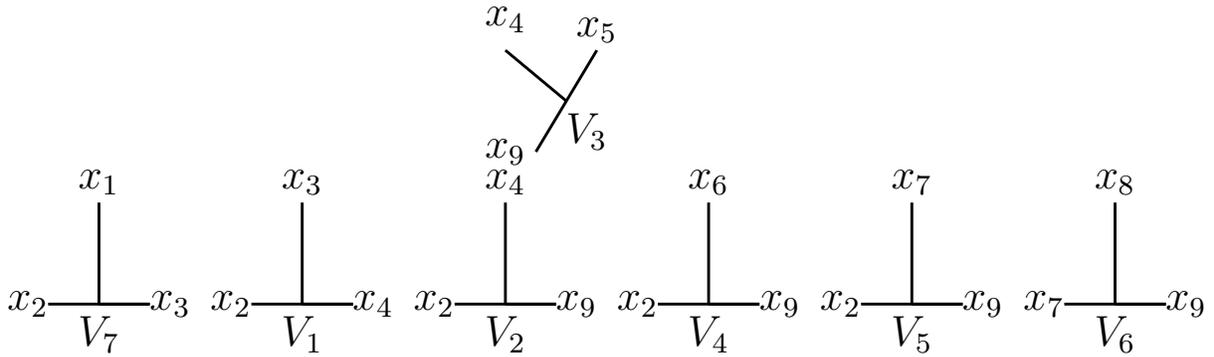

then we will have $V_iV_j$ pairs found in the upper right triangle of the matrix\footnote{Note that this matrix corresponds to $\binom{M-2}{2}$ in any given diagram.} below
\eqn{}{ 
\left(\begin{array}{cccccc}
   V_{1}V_{2} & V_{1}V_{3} & V_{1}V_{4} & V_{1}V_{5} & V_{1}V_{6}& V_{1}V_{7} \\
    0 & V_{2}V_{3} & V_{2}V_{4} & V_{2}V_{5} & V_{2}V_{6}& V_{2}V_{7}  \\
    0 & 0 & V_{3}V_{4} & V_{3}V_{5} & V_{3}V_{6}& V_{3}V_{7}  \\
      0 & 0 & 0 & V_{4}V_{5} & V_{4}V_{6}& V_{4}V_{7}  \\
        0 & 0 & 0 & 0 & V_{5}V_{6}& V_{5}V_{7} \\
        0 & 0 & 0 & 0 & 0& V_{6}V_{7} \\
\end{array}
\right).
}

\item  Now that we have identified all $V_iV_j$ pairs, draw the associated comb structure between these vertices. For reference, for the rest of this example we will consider the $V_{3}V_{7}$ pair. The vertex pair comb structure for $V_{3}V_{7}$ is
\eqn{}{ 
\begin{tikzpicture}[thick][r]
\begin{scope}
\node at (-.2,0) {$x_{2}$};
\node at (.5,-.3) {$V_{7}$};
\draw[-] (0.0,0)--(1,0);
\draw[-] (.5,0)--(.5,1) node at (.5,1.2) {$x_{1}$};
\draw[-] (.5,0)--(1,0) node at (1.2,0) {$x_{3}$};
\end{scope}
\begin{scope}[xshift=2cm]
\node at (-.2,0) {$x_{2}$};
\node at (.5,-.3) {$V_{1}$};
\draw[-] (0.0,0)--(1,0);
\draw[-] (.5,0)--(.5,1) node at (.5,1.2) {$x_{3}$};
\draw[-] (.5,0)--(1,0) node at (1.2,0) {$x_{4}$};
\end{scope}
\begin{scope}[xshift=4cm]
\node at (-.2,0) {$x_{2}$};
\node at (.5,-.3) {$V_{2}$};
\draw[-] (0.0,0)--(1,0);
\draw[-] (.5,0)--(.5,1) node at (.5,1.2) {$x_{4}$};
\draw[-] (.5,0)--(1,0) node at (1.2,0) {$x_{9}$};
\end{scope}
\begin{scope}[xshift=4cm]
\node at (.5,1.5) {$x_{9}$};
\node at (1.3,1.7) {$V_{3}$};
\draw[-] (.8,1.5)--(1.4,2.5);
\draw[-] (1.1,2)--(.5,2.5) node at (.5,2.8) {$x_{4}$};
\node at (1.4,2.7) {$x_{5}$};
\end{scope}
\end{tikzpicture}
: 
V_{3}V_{7}:
}

\item Next, in each structure associated to the vertex pair, identify which legs do not flow. We say a leg does not flow if we do not assign those factors in the flow diagram for that structure. The two legs that do not flow will become the subscripts for a $x_{rs}^2$ term in the numerator of the $v$-type cross ratio. Moreover, we now circle the two $x_i$ $x_j$ pairs that are between vertices. The $x_i$ and $x_j$ subscripts from within the blue circle will become $x_{ij}^2$ factors in the numerator.

In the case of the $V_{3}V_{7}$ comb structure $x_{1}$ and $x_{5}$ do not flow. Thus, the numerator contains $x_{15}^2$ and the $x_{i}$ $x_{j}$ pairs circled in blue, $x_{23}, x_{24}, x_{49}$.
\eqn{}{ 
\begin{tikzpicture}[thick][r]
\begin{scope}
\node at (-.2,0) {$x_{2}$};
\node at (.5,-.3) {$V_{7}$};
\draw[-] (0.0,0)--(1,0);
\draw[-] (.5,0)--(.5,1) node at (.5,1.2) {$x_{1}$};
\draw[-] (.5,0)--(1,0) node at (1.2,0) {$x_{3}$};
\draw[cyan] (1.5,.1) circle (.6 cm);
\end{scope}
\begin{scope}[xshift=2cm]
\node at (-.2,0) {$x_{2}$};
\node at (.5,-.3) {$V_{1}$};
\draw[-] (0.0,0)--(1,0);
\draw[-] (.5,0)--(.5,1) node at (.5,1.2) {$x_{3}$};
\draw[-] (.5,0)--(1,0) node at (1.2,0) {$x_{4}$};
\draw[cyan] (1.5,.1) circle (.6 cm);
\end{scope}
\begin{scope}[xshift=4cm]
\node at (-.2,0) {$x_{2}$};
\node at (.5,-.3) {$V_{2}$};
\draw[-] (0.0,0)--(1,0);
\draw[-] (.5,0)--(.5,1) node at (.5,1.2) {$x_{4}$};
\draw[-] (.5,0)--(1,0) node at (1.2,0) {$x_{9}$};
\draw[cyan] (.5,1.3) circle (.5 cm);
\end{scope}
\begin{scope}[xshift=4cm]
\node at (.5,1.5) {$x_{9}$};
\node at (1.3,1.7) {$V_{3}$};
\draw[-] (.8,1.5)--(1.4,2.5);
\draw[-] (1.1,2)--(.5,2.5) node at (.5,2.8) {$x_{4}$};
\node at (1.4,2.7) {$x_{5}$};
\end{scope}
\end{tikzpicture}
: 
V_{3}V_{7}: v_{15}= {x_{15}^2 x_{32}^2 x_{24}^2 x_{49}^2 \over } }

\item  Once we have identified which $x_r$ and $x_s$ legs do not flow, we draw a dashed orange line to show the shortest path from $x_r$ to $x_s$ assuming it must go through the other points labeled in the diagram, a geodesic between $x_r$ to $x_s$. The two $x_i$ and $x_j$ factors that are connected by the orange dotted lines in a single vertex become subscripts on an $x_{ij}^2$ factor in the denominator.  

Using our example, $x_{1}$ and $x_{5}$ do not flow, thus the denominator contains the pairs of the geodesic between $x_{1}$ and $x_{5}$, which are $x_{13}^2,x_{24}^2,x_{24}^2, x_{95}^2 $.
\eqn{}{ 
\begin{tikzpicture}[thick][r]
\begin{scope}
\node at (-.2,0) {$x_{2}$};
\node at (.5,-.3) {$V_{7}$};
\draw[-] (0.0,0)--(1,0);
\draw[-] (.5,0)--(.5,1) node at (.5,1.2) {$x_{1}$};
\draw[-] (.5,0)--(1,0) node at (1.2,0) {$x_{3}$};
\draw[dashed,  orange]  (.5,1.1) to[bend right=30] (1,.15);
\end{scope}
\begin{scope}[xshift=2cm]
\node at (-.2,0) {$x_{2}$};
\node at (.5,-.3) {$V_{1}$};
\draw[-] (0.0,0)--(1,0);
\draw[-] (.5,0)--(.5,1) node at (.5,1.2) {$x_{3}$};
\draw[-] (.5,0)--(1,0) node at (1.2,0) {$x_{4}$};
\draw[dashed,  orange]  (0,-.08) to[bend right=5] (.9,-.08);
\end{scope}
\begin{scope}[xshift=4cm]
\node at (-.2,0) {$x_{2}$};
\node at (.5,-.3) {$V_{2}$};
\draw[-] (0.0,0)--(1,0);
\draw[-] (.5,0)--(.5,1) node at (.5,1.2) {$x_{4}$};
\draw[-] (.5,0)--(1,0) node at (1.2,0) {$x_{9}$};
\draw[dashed,  orange]  (.1,.15) to[bend right=30] (.5,1.1);
\end{scope}
\begin{scope}[xshift=4cm]
\node at (.5,1.5) {$x_{9}$};
\node at (1.3,1.7) {$V_{3}$};
\draw[-] (.8,1.5)--(1.4,2.5);
\draw[-] (1.1,2)--(.5,2.5) node at (.5,2.8) {$x_{4}$};
\node at (1.4,2.7) {$x_{5}$};
\draw[dashed,  orange]  (.9,1.5) to[bend right=5] (1.2,2.4);
\end{scope}
\end{tikzpicture}
: 
V_{7}V_{3}: v_{15}= { \over x_{13}^2 x_{24}^2 x_{24}^2 x_{95}^2 }
}

\item  Put the method all together to get the full $v$-type cross ratio.\footnote{As can be seen in Section~\ref{9-point} in step 3 of the $v$-type cross ratios this method works regardless of the ordering of $V_iV_j$. In this example it is clear that $V_3V_7 = V_7V_3$.}
\eqn{}{ 
\begin{tikzpicture}[thick][r]
\begin{scope}
\node at (-.2,0) {$x_{2}$};
\node at (.5,-.3) {$V_{7}$};
\draw[-] (0.0,0)--(1,0);
\draw[-] (.5,0)--(.5,1) node at (.5,1.2) {$x_{1}$};
\draw[-] (.5,0)--(1,0) node at (1.2,0) {$x_{3}$};
\draw[cyan] (1.5,.1) circle (.6 cm);
\draw[dashed,  orange]  (.5,1.1) to[bend right=30] (1,.15);
\end{scope}
\begin{scope}[xshift=2cm]
\node at (-.2,0) {$x_{2}$};
\node at (.5,-.3) {$V_{1}$};
\draw[-] (0.0,0)--(1,0);
\draw[-] (.5,0)--(.5,1) node at (.5,1.2) {$x_{3}$};
\draw[-] (.5,0)--(1,0) node at (1.2,0) {$x_{4}$};
\draw[cyan] (1.5,.1) circle (.6 cm);
\draw[dashed,  orange]  (0,-.08) to[bend right=5] (.9,-.08);
\end{scope}
\begin{scope}[xshift=4cm]
\node at (-.2,0) {$x_{2}$};
\node at (.5,-.3) {$V_{2}$};
\draw[-] (0.0,0)--(1,0);
\draw[-] (.5,0)--(.5,1) node at (.5,1.2) {$x_{4}$};
\draw[-] (.5,0)--(1,0) node at (1.2,0) {$x_{9}$};
\draw[cyan] (.5,1.3) circle (.5 cm);
\draw[dashed,  orange]  (.1,.15) to[bend right=30] (.5,1.1);
\end{scope}
\begin{scope}[xshift=4cm]
\node at (.5,1.5) {$x_{9}$};
\node at (1.3,1.7) {$V_{3}$};
\draw[-] (.8,1.5)--(1.4,2.5);
\draw[-] (1.1,2)--(.5,2.5) node at (.5,2.8) {$x_{4}$};
\node at (1.4,2.7) {$x_{5}$};
\draw[dashed,  orange]  (.9,1.5) to[bend right=5] (1.2,2.4);
\end{scope}
\end{tikzpicture}
: 
V_{3}V_{7}: v_{15}= {x_{15}^2 x_{32}^2 x_{24}^2 x_{49}^2 \over x_{13}^2 x_{24}^2 x_{24}^2 x_{95}^2 }= {x_{15}^2 x_{32}^2  x_{49}^2 \over x_{13}^2 x_{24}^2 x_{59}^2 }
}

\end{enumerate}

We now turn to the construction of the $u$-type and $v$-type cross ratios for 4-point and 5-point conformal blocks.

\section{Low-point examples}
\label{LOWPOINT}

In this section we demonstrate how to find the cross ratios for a conformal block using the $4$-point and $5$-point blocks respectively. This section directly corresponds to Section~\ref{INDEXCROSSRATIOS} and is used to complement the equations there through illustrative examples. For a non-trivial example of how to find $u$-type and $v$-type cross ratios in a higher-point block see Section~\ref{9-point}. 

\subsection{4-point Block}
\label{4POINTFLOWS}

Consider a 4-point block with the following labeling
\eqn{4pointblock}{ 
&
\begin{tikzpicture}[thick]
\begin{scope}[xshift=2cm]
\node at (0,0) {$\mathcal{O}_{1}(x_{1})$};
\draw[-] (1,0)--(4.5,0);
\draw[-] (2,0)--(2,1) node at (2,1.5) {$\mathcal{O}_{2}(x_{2})$};
\draw[-] (3.5,0)--(3.5,1) node at (3.5,1.5) {$\mathcal{O}_{3}(x_{3})$};
\draw[-] (3.5,0)--(4.5,0) node at (5.5,0)  {$\mathcal{O}_{4}(x_{4})$};
\node at (2.8,-0.5) {$\mathcal{O}_{k_{1}}$};
\end{scope}
\end{tikzpicture}.
}

\subsubsection{Flow Diagrams}
In order to create an associated flow diagram for \eno{4pointblock}, it is convenient to expand out the conformal block into a set of OPE vertices. In the case of \eno{4pointblock} we see the following 3-point vertices,

\eqn{4baseOPE}{ 
\begin{tikzpicture}[thick][r]
\begin{scope}
\node at (-.2,0) {$x_{1}$};
\node at (.5,-.3) {$V_{1}$};
\draw[-] (0.0,0)--(1,0);
\draw[-] (.5,0)--(.5,1) node at (.5,1.2) {$x_{2}$};
\draw[-] (.5,0)--(1,0) node at (1.2,0) {} ;
\end{scope}
\begin{scope}[xshift=2cm]
\node at (-.2,0) {};
\node at (.5,-.3) {$V_{2}$};
\draw[-] (0.0,0)--(1,0);
\draw[-] (.5,0)--(.5,1) node at (.5,1.2) {$x_{3}$};
\draw[-] (.5,0)--(1,0) node at (1.2,0) {$x_{4}$};
\end{scope}
\end{tikzpicture}
}

In \eno{4baseOPE} we assigned a label of $V_1$, and $V_2$ in order to keep track of individual vertices in the diagram. To complete a flow diagram we have to assign position coordinates which satisfy the OPE vertices of the block. For the 4-point conformal block, we see four possible flow diagrams 

\eqn{fourpointflow}{ 
&\begin{tikzpicture}[thick][r]
\begin{scope}
\node at (-.2,0) {$x_{1}$};
\node at (.5,-.3) {$V_{1}$};
\draw[-] (0.0,0)--(1,0);
\draw[-] (.5,0)--(.5,1) node at (.5,1.2) {$x_{2}$};
\draw[-] (.5,0)--(1,0) node at (1.2,0) {$x_3$} ;
\end{scope}
\begin{scope}[xshift=2cm]
\node at (-.2,0) {$x_{2}$};
\node at (.5,-.3) {$V_{2}$};
\draw[-] (0.0,0)--(1,0);
\draw[-] (.5,0)--(.5,1) node at (.5,1.2) {$x_{3}$};
\draw[-] (.5,0)--(1,0) node at (1.2,0) {$x_{4}$};
\end{scope}
\end{tikzpicture}
\qquad 
\begin{tikzpicture}[thick][r]
\begin{scope}
\node at (-.2,0) {$x_{1}$};
\node at (.5,-.3) {$V_{1}$};
\draw[-] (0.0,0)--(1,0);
\draw[-] (.5,0)--(.5,1) node at (.5,1.2) {$x_{2}$};
\draw[-] (.5,0)--(1,0) node at (1.2,0) {$x_4$} ;
\end{scope}
\begin{scope}[xshift=2cm]
\node at (-.2,0) {$x_{2}$};
\node at (.5,-.3) {$V_{2}$};
\draw[-] (0.0,0)--(1,0);
\draw[-] (.5,0)--(.5,1) node at (.5,1.2) {$x_{3}$};
\draw[-] (.5,0)--(1,0) node at (1.2,0) {$x_{4}$};
\end{scope}
\end{tikzpicture}
\cr 
& \begin{tikzpicture}[thick][r]
\begin{scope}
\node at (-.2,0) {$x_{1}$};
\node at (.5,-.3) {$V_{1}$};
\draw[-] (0.0,0)--(1,0);
\draw[-] (.5,0)--(.5,1) node at (.5,1.2) {$x_{2}$};
\draw[-] (.5,0)--(1,0) node at (1.2,0) {$x_{3}$} ;
\end{scope}
\begin{scope}[xshift=2cm]
\node at (-.2,0) {$x_{1}$};
\node at (.5,-.3) {$V_{2}$};
\draw[-] (0.0,0)--(1,0);
\draw[-] (.5,0)--(.5,1) node at (.5,1.2) {$x_{3}$};
\draw[-] (.5,0)--(1,0) node at (1.2,0) {$x_{4}$};
\end{scope}
\end{tikzpicture}
\qquad 
\begin{tikzpicture}[thick][r]
\begin{scope}
\node at (-.2,0) {$x_{1}$};
\node at (.5,-.3) {$V_{1}$};
\draw[-] (0.0,0)--(1,0);
\draw[-] (.5,0)--(.5,1) node at (.5,1.2) {$x_{2}$};
\draw[-] (.5,0)--(1,0) node at (1.2,0) {$x_{4}$} ;
\end{scope}
\begin{scope}[xshift=2cm]
\node at (-.2,0) {$x_1$};
\node at (.5,-.3) {$V_{2}$};
\draw[-] (0.0,0)--(1,0);
\draw[-] (.5,0)--(.5,1) node at (.5,1.2) {$x_{3}$};
\draw[-] (.5,0)--(1,0) node at (1.2,0) {$x_{4}$};
\end{scope}
\end{tikzpicture}
}

Flow diagrams are not unique. Each flow diagram generates a corresponding unique set of cross ratios. For example, using \eno{uCR} and \eno{vCR2} we can recreate all $u$-type and $v$-type cross ratios that are possible for a 4-point block with our chosen labeling. 

In the case of the 4-point block there is simply only one $u$-type cross ratio per any flow diagram because there are only two vertices.
%
\eqn{fourpointudiagrams}{ 
&\begin{tikzpicture}[thick][r]
\begin{scope}
\node at (-.2,0) {$x_{1}$};
\node at (.5,-.3) {$V_{1}$};
\draw[-] (0.0,0)--(1,0);
\draw[-] (.5,0)--(.5,1) node at (.5,1.2) {$x_{2}$};
\draw[-] (.5,0)--(1,0) node at (1.2,0) {$x_3$} ;
\end{scope}
\begin{scope}[xshift=2cm]
\node at (-.2,0) {$x_{2}$};
\node at (.5,-.3) {$V_{2}$};
\draw[-] (0.0,0)--(1,0);
\draw[-] (.5,0)--(.5,1) node at (.5,1.2) {$x_{3}$};
\draw[-] (.5,0)--(1,0) node at (1.2,0) {$x_{4}$};
\end{scope}
\end{tikzpicture}:
\qquad
\begin{tikzpicture}[thick][r]
\begin{scope}
\node at (-.2,0) {$x_{1}$};
\node at (.5,-.3) {$V_{1}$};
\node at (1.6,-0.5) {$\mathcal{O}_{k_{1}}$};
\draw[-] (0.0,0)--(1,0);
\draw[-] (.5,0)--(.5,1) node at (.5,1.2) {$x_{2}$};
\draw[-] (.5,0)--(1,0) node at (1.2,0) {$x_3$} ;
\draw[black] (-.25,-.05) circle (.3 cm);
\draw[dashed,  red]  (0,-.08) to[bend right=5] (.9,-.08);
\draw[dashed,  teal]  (0,.08) to[bend right=5] (.45,1);
\end{scope}
\begin{scope}[xshift=2cm]
\node at (-.2,0) {$x_{2}$};
\node at (.5,-.3) {$V_{2}$};
\draw[-] (0.0,0)--(1,0);
\draw[-] (.5,0)--(.5,1) node at (.5,1.2) {$x_{3}$};
\draw[-] (.5,0)--(1,0) node at (1.2,0) {$x_{4}$};
\draw[black] (1.25,-.05) circle (.3 cm);
\draw[dashed,  red]  (0,-.08) to[bend right=5] (.9,-.08);
\draw[dashed,  teal]  (.55,1) to[bend right=5] (.8, .1);
\end{scope}
\end{tikzpicture} 
\qquad : u_{1}=\frac{x^2_{12}x^2_{34}}{x_{13}^2x_{24}^2}
\cr &  
\begin{tikzpicture}[thick][r]
\begin{scope}
\node at (-.2,0) {$x_{1}$};
\node at (.5,-.3) {$V_{1}$};
\draw[-] (0.0,0)--(1,0);
\draw[-] (.5,0)--(.5,1) node at (.5,1.2) {$x_{2}$};
\draw[-] (.5,0)--(1,0) node at (1.2,0) {$x_4$} ;
\end{scope}
\begin{scope}[xshift=2cm]
\node at (-.2,0) {$x_{2}$};
\node at (.5,-.3) {$V_{2}$};
\draw[-] (0.0,0)--(1,0);
\draw[-] (.5,0)--(.5,1) node at (.5,1.2) {$x_{3}$};
\draw[-] (.5,0)--(1,0) node at (1.2,0) {$x_{4}$};
\end{scope}
\end{tikzpicture}
\qquad :
\begin{tikzpicture}[thick][r]
\begin{scope}
\node at (-.2,0) {$x_{1}$};
\node at (.5,-.3) {$V_{1}$};
\node at (1.6,-0.5) {$\mathcal{O}_{k_{1}}$};
\draw[-] (0.0,0)--(1,0);
\draw[-] (.5,0)--(.5,1) node at (.5,1.2) {$x_{2}$};
\draw[-] (.5,0)--(1,0) node at (1.2,0) {$x_4$} ;
\draw[black] (-.25,-.05) circle (.3 cm);
\draw[dashed,  red]  (0,-.08) to[bend right=5] (.9,-.08);
\draw[dashed,  teal]  (0,.08) to[bend right=5] (.45,1);
\end{scope}
\begin{scope}[xshift=2cm]
\node at (-.2,0) {$x_{2}$};
\node at (.5,-.3) {$V_{2}$};
\draw[-] (0.0,0)--(1,0);
\draw[-] (.5,0)--(.5,1) node at (.5,1.2) {$x_{3}$};
\draw[-] (.5,0)--(1,0) node at (1.2,0) {$x_{4}$};
\draw[black] (.45,1.15) circle (.3 cm);
\draw[dashed,  red]  (0,.08) to[bend right=5] (.45,1);
\draw[dashed,  teal]  (.55,1) to[bend right=5] (.8, .1);
\end{scope}
\end{tikzpicture}
: \qquad 
u_{1}=\frac{x^2_{12}x^2_{34}}{x_{14}^2x_{23}^2}
\cr
& \begin{tikzpicture}[thick][r]
\begin{scope}
\node at (-.2,0) {$x_{1}$};
\node at (.5,-.3) {$V_{1}$};
\draw[-] (0.0,0)--(1,0);
\draw[-] (.5,0)--(.5,1) node at (.5,1.2) {$x_{2}$};
\draw[-] (.5,0)--(1,0) node at (1.2,0) {$x_{3}$} ;
\end{scope}
\begin{scope}[xshift=2cm]
\node at (-.2,0) {$x_{1}$};
\node at (.5,-.3) {$V_{2}$};
\draw[-] (0.0,0)--(1,0);
\draw[-] (.5,0)--(.5,1) node at (.5,1.2) {$x_{3}$};
\draw[-] (.5,0)--(1,0) node at (1.2,0) {$x_{4}$};
\end{scope}
\end{tikzpicture}
\qquad :
\begin{tikzpicture}[thick][r]
\begin{scope}
\node at (-.2,0) {$x_{1}$};
\node at (.5,-.3) {$V_{1}$};
\node at (1.6,-0.5) {$\mathcal{O}_{k_{1}}$};
\draw[-] (0.0,0)--(1,0);
\draw[-] (.5,0)--(.5,1) node at (.5,1.2) {$x_{2}$};
\draw[-] (.5,0)--(1,0) node at (1.2,0) {$x_{3}$} ;
\draw[black] (.45,1.15) circle (.3 cm);
\draw[dashed,  red]  (.55,1) to[bend right=5] (.8, .1);
\draw[dashed,  teal]  (0,.08) to[bend right=5] (.45,1);
\end{scope}
\begin{scope}[xshift=2cm]
\node at (-.2,0) {$x_{1}$};
\node at (.5,-.3) {$V_{2}$};
\draw[-] (0.0,0)--(1,0);
\draw[-] (.5,0)--(.5,1) node at (.5,1.2) {$x_{3}$};
\draw[-] (.5,0)--(1,0) node at (1.2,0) {$x_{4}$};
\draw[black] (1.25,-.05) circle (.3 cm);
\draw[dashed,  red]  (0,-.08) to[bend right=5] (.9,-.08);
\draw[dashed,  teal]  (.55,1) to[bend right=5] (.8, .1);
\end{scope}
\end{tikzpicture} 
\qquad :
u_{1}=\frac{x^2_{12}x^2_{34}}{x_{23}^2x_{14}^2}
\cr &
\begin{tikzpicture}[thick][r]
\begin{scope}
\node at (-.2,0) {$x_{1}$};
\node at (.5,-.3) {$V_{1}$};
\draw[-] (0.0,0)--(1,0);
\draw[-] (.5,0)--(.5,1) node at (.5,1.2) {$x_{2}$};
\draw[-] (.5,0)--(1,0) node at (1.2,0) {$x_{4}$} ;
\end{scope}
\begin{scope}[xshift=2cm]
\node at (-.2,0) {$x_1$};
\node at (.5,-.3) {$V_{2}$};
\draw[-] (0.0,0)--(1,0);
\draw[-] (.5,0)--(.5,1) node at (.5,1.2) {$x_{3}$};
\draw[-] (.5,0)--(1,0) node at (1.2,0) {$x_{4}$};
\end{scope}
\end{tikzpicture}
\qquad :
\begin{tikzpicture}[thick][r]
\begin{scope}
\node at (-.2,0) {$x_{1}$};
\node at (.5,-.3) {$V_{1}$};
\node at (1.6,-0.5) {$\mathcal{O}_{k_{1}}$};
\draw[-] (0.0,0)--(1,0);
\draw[-] (.5,0)--(.5,1) node at (.5,1.2) {$x_{2}$};
\draw[-] (.5,0)--(1,0) node at (1.2,0) {$x_{4}$} ;
\draw[black] (.45,1.15) circle (.3 cm);
\draw[dashed,  red]  (.55,1) to[bend right=5] (.8, .1);
\draw[dashed,  teal]  (0,.08) to[bend right=5] (.45,1);
\end{scope}
\begin{scope}[xshift=2cm]
\node at (-.2,0) {$x_1$};
\node at (.5,-.3) {$V_{2}$};
\draw[-] (0.0,0)--(1,0);
\draw[-] (.5,0)--(.5,1) node at (.5,1.2) {$x_{3}$};
\draw[-] (.5,0)--(1,0) node at (1.2,0) {$x_{4}$};
\draw[black] (.45,1.15) circle (.3 cm);
\draw[dashed,  red]  (0,.08) to[bend right=5] (.45,1);
\draw[dashed,  teal]  (.55,1) to[bend right=5] (.8, .1);
\end{scope}
\end{tikzpicture}
\qquad : 
u_{1}=\frac{x^2_{12}x^2_{34}}{x_{24}^2x_{13}^2}
}

Next, we demonstrate how to use flow diagrams to solve the $v$-type cross ratios. In the below diagrams we show the resulting $v$-type cross diagrams and their cross ratios based on every possible choice of 4-point flow diagrams.  

\eqn{fourVtypecross}{
&\begin{tikzpicture}[thick]
\begin{scope}[xshift=0cm]
\node at (-.2,0) {$x_{1}$};
\node at (.5,-.3) {$V_{1}$};
\draw[-] (0.0,0)--(1,0);
\draw[-] (.5,0)--(.5,1) node at (.5,1.2) {$x_{2}$};
\draw[-] (.5,0)--(1,0) node at (1.2,0) {$x_{3}$};
\end{scope}
\begin{scope}[xshift=2cm]
\node at (-.2,0) {$x_{2}$};
\node at (.5,-.3) {$V_{2}$};
\draw[-] (0.0,0)--(1,0);
\draw[-] (.5,0)--(.5,1) node at (.5,1.2) {$x_{3}$};
\draw[-] (.5,0)--(1,0) node at (1.2,0) {$x_{4}$};
\end{scope}
\end{tikzpicture}
\qquad :
\begin{tikzpicture}[thick]
\begin{scope}[xshift=0cm]
\node at (-.2,0) {$x_{1}$};
\node at (.5,-.3) {$V_{1}$};
\draw[-] (0.0,0)--(1,0);
\draw[-] (.5,0)--(.5,1) node at (.5,1.2) {$x_{2}$};
\draw[-] (.5,0)--(1,0) node at (1.2,0) {$x_{3}$};
\draw[cyan] (1.5,.1) circle (.6 cm);
\draw[dashed,  orange]  (0,-.08) to[bend right=5] (.9,-.08);
\end{scope}
\begin{scope}[xshift=2cm]
\node at (-.2,0) {$x_{2}$};
\node at (.5,-.3) {$V_{2}$};
\draw[-] (0.0,0)--(1,0);
\draw[-] (.5,0)--(.5,1) node at (.5,1.2) {$x_{3}$};
\draw[-] (.5,0)--(1,0) node at (1.2,0) {$x_{4}$};
\draw[dashed,  orange]  (0,-.08) to[bend right=5] (.9,-.08);
\end{scope}
\end{tikzpicture}
: V_{1}V_{2}:v_{14}= {x_{14}^2 x_{32}^2 \over x_{13}^2 x_{24}^2 }
\cr &
\begin{tikzpicture}[thick]
\begin{scope}[xshift=0cm]
\node at (-.2,0) {$x_{1}$};
\node at (.5,-.3) {$V_{1}$};
\draw[-] (0.0,0)--(1,0);
\draw[-] (.5,0)--(.5,1) node at (.5,1.2) {$x_{2}$};
\draw[-] (.5,0)--(1,0) node at (1.2,0) {$x_{4}$};
\end{scope}
\begin{scope}[xshift=2cm]
\node at (-.2,0) {$x_{2}$};
\node at (.5,-.3) {$V_{2}$};
\draw[-] (0.0,0)--(1,0);
\draw[-] (.5,0)--(.5,1) node at (.5,1.2) {$x_{3}$};
\draw[-] (.5,0)--(1,0) node at (1.2,0) {$x_{4}$};
\end{scope}
\end{tikzpicture}
\qquad :
\begin{tikzpicture}[thick]
\begin{scope}[xshift=0cm]
\node at (-.2,0) {$x_{1}$};
\node at (.5,-.3) {$V_{1}$};
\draw[-] (0.0,0)--(1,0);
\draw[-] (.5,0)--(.5,1) node at (.5,1.2) {$x_{2}$};
\draw[-] (.5,0)--(1,0) node at (1.2,0) {$x_{4}$};
\draw[cyan] (1.5,.1) circle (.6 cm);
\draw[dashed,  orange]  (0,-.08) to[bend right=5] (.9,-.08);
\end{scope}
\begin{scope}[xshift=2cm]
\node at (-.2,0) {$x_{2}$};
\node at (.5,-.3) {$V_{2}$};
\draw[-] (0.0,0)--(1,0);
\draw[-] (.5,0)--(.5,1) node at (.5,1.2) {$x_{3}$};
\draw[-] (.5,0)--(1,0) node at (1.2,0) {$x_{4}$};
\draw[dashed,  orange]  (.1,.15) to[bend right=30] (.5,1.1);
\end{scope}
\end{tikzpicture}
: V_{1}V_{2}:v_{13}= {x_{13}^2 x_{42}^2 \over x_{14}^2 x_{23}^2 }
\cr 
&
\begin{tikzpicture}[thick]
\begin{scope}[xshift=0cm]
\node at (-.2,0) {$x_{1}$};
\node at (.5,-.3) {$V_{1}$};
\draw[-] (0.0,0)--(1,0);
\draw[-] (.5,0)--(.5,1) node at (.5,1.2) {$x_{2}$};
\draw[-] (.5,0)--(1,0) node at (1.2,0) {$x_{3}$};
\end{scope}
\begin{scope}[xshift=2cm]
\node at (-.2,0) {$x_{1}$};
\node at (.5,-.3) {$V_{2}$};
\draw[-] (0.0,0)--(1,0);
\draw[-] (.5,0)--(.5,1) node at (.5,1.2) {$x_{3}$};
\draw[-] (.5,0)--(1,0) node at (1.2,0) {$x_{4}$};
\end{scope}
\end{tikzpicture}
\qquad : 
\begin{tikzpicture}[thick]
\begin{scope}[xshift=0cm]
\node at (-.2,0) {$x_{1}$};
\node at (.5,-.3) {$V_{1}$};
\draw[-] (0.0,0)--(1,0);
\draw[-] (.5,0)--(.5,1) node at (.5,1.2) {$x_{2}$};
\draw[-] (.5,0)--(1,0) node at (1.2,0) {$x_{3}$};
\draw[cyan] (1.5,.1) circle (.6 cm);
\draw[dashed,  orange]  (.5,1.1) to[bend right=30] (1,.15);
\end{scope}
\begin{scope}[xshift=2cm]
\node at (-.2,0) {$x_{1}$};
\node at (.5,-.3) {$V_{2}$};
\draw[-] (0.0,0)--(1,0);
\draw[-] (.5,0)--(.5,1) node at (.5,1.2) {$x_{3}$};
\draw[-] (.5,0)--(1,0) node at (1.2,0) {$x_{4}$};
\draw[dashed,  orange]  (0,-.08) to[bend right=5] (.9,-.08);
\end{scope}
\end{tikzpicture}
: V_{1}V_{2}:v_{24}= {x_{24}^2 x_{31}^2 \over x_{23}^2 x_{14}^2 }
\cr & 
\begin{tikzpicture}[thick]
\begin{scope}[xshift=0cm]
\node at (-.2,0) {$x_{1}$};
\node at (.5,-.3) {$V_{1}$};
\draw[-] (0.0,0)--(1,0);
\draw[-] (.5,0)--(.5,1) node at (.5,1.2) {$x_{2}$};
\draw[-] (.5,0)--(1,0) node at (1.2,0) {$x_{4}$};
\end{scope}
\begin{scope}[xshift=2cm]
\node at (-.2,0) {$x_{1}$};
\node at (.5,-.3) {$V_{2}$};
\draw[-] (0.0,0)--(1,0);
\draw[-] (.5,0)--(.5,1) node at (.5,1.2) {$x_{3}$};
\draw[-] (.5,0)--(1,0) node at (1.2,0) {$x_{4}$};
\end{scope}
\end{tikzpicture}
\qquad : 
\begin{tikzpicture}[thick]
\begin{scope}[xshift=0cm]
\node at (-.2,0) {$x_{1}$};
\node at (.5,-.3) {$V_{1}$};
\draw[-] (0.0,0)--(1,0);
\draw[-] (.5,0)--(.5,1) node at (.5,1.2) {$x_{2}$};
\draw[-] (.5,0)--(1,0) node at (1.2,0) {$x_{4}$};
\draw[cyan] (1.5,.1) circle (.6 cm);
\draw[dashed,  orange]  (.5,1.1) to[bend right=30] (1,.15);
\end{scope}
\begin{scope}[xshift=2cm]
\node at (-.2,0) {$x_{1}$};
\node at (.5,-.3) {$V_{2}$};
\draw[-] (0.0,0)--(1,0);
\draw[-] (.5,0)--(.5,1) node at (.5,1.2) {$x_{3}$};
\draw[-] (.5,0)--(1,0) node at (1.2,0) {$x_{4}$};
\draw[dashed,  orange]  (.1,.15) to[bend right=30] (.5,1.1);
\end{scope}
\end{tikzpicture}
: V_{1}V_{2}:v_{23}= {x_{23}^2 x_{41}^2 \over x_{24}^2 x_{13}^2 }
}

\subsection{5-point Block}
\label{5POINTFLOWS}

Consider a 5-point block with the following labeling
\begin{figure}[h!]
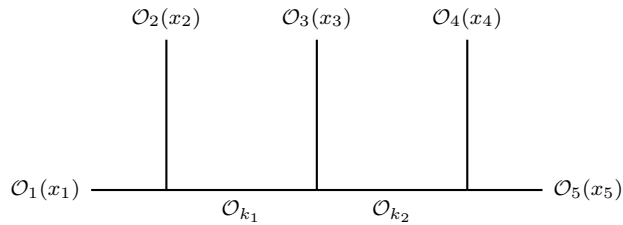

    \centering
     \[  \musepic{\figfiveCB} \]
    \caption{5-point conformal block. }
    \label{fig:fiveCB}
\end{figure}

\subsubsection{Flow Diagrams}
\begin{figure}[h!]
\centering
\begin{tikzpicture}[thick]
\begin{scope}
\node at (-.2,0) {$x_{1}$};
\node at (.5,-.3) {$V_{1}$};
\draw[-] (0.0,0)--(1,0);
\draw[-] (.5,0)--(.5,1) node at (.5,1.2) {$x_{2}$};
\draw[-] (.5,0)--(1,0) node at (1.2,0) {};
\end{scope}
\begin{scope}[xshift=2cm]
\node at (-.2,0) {};
\node at (.5,-.3) {$V_{2}$};
\draw[-] (0.0,0)--(1,0);
\draw[-] (.5,0)--(.5,1) node at (.5,1.2) {$x_{3}$};
\draw[-] (.5,0)--(1,0) node at (1.2,0) {};
\end{scope}
\begin{scope}[xshift=4cm]
\node at (-.2,0) {};
\node at (.5,-.3) {$V_{3}$};
\draw[-] (0.0,0)--(1,0);
\draw[-] (.5,0)--(.5,1) node at (.5,1.2) {$x_{4}$};
\draw[-] (.5,0)--(1,0) node at (1.2,0) {$x_{5}$};
\end{scope}
\end{tikzpicture}
\caption{OPE vertices for the 5-point block of Figure~\ref{fig:fiveCB}.}
\label{OPE Vertex 5}
\end{figure}
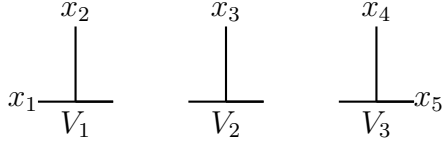

In order to create an associated flow diagram for Figure \ref{fig:fiveCB}, it is convenient to expand out the conformal block into a set of OPE vertices. In the case of Figure~\ref{fig:fiveCB} we see the 3-point vertices of Figure~\ref{OPE Vertex 5}.

We can complete our flow diagram by assigning position coordinates to any unassigned legs of the 3-point vertices. A few of the possible choices of flow diagrams for the 5-point conformal block are as follows\footnote{For sake of brevity we did not generate a list of all possible flow diagrams for the 5-point block. Instead, we list a few significantly different valid choices.}$^{,}$\footnote{If we had found the $u$-type and $v$-type cross ratios of each flow diagram, then we would have obtained different sets of cross ratios, including the sets found in \cite{Fortin:2020bfq,Hoback:2020pgj, Rosenhaus:2018zqn} as in \eno{fivepointflow}.}
\eqn{fivepointflow}{ 
&\begin{tikzpicture}[thick]
\begin{scope}
\node at (-.2,0) {$x_{1}$};
\node at (.5,-.3) {$V_{1}$};
\draw[-] (0.0,0)--(1,0);
\draw[-] (.5,0)--(.5,1) node at (.5,1.2) {$x_{2}$};
\draw[-] (.5,0)--(1,0) node at (1.2,0) {$x_3$};
\end{scope}
\begin{scope}[xshift=2cm]
\node at (-.2,0) {$x_2$};
\node at (.5,-.3) {$V_{2}$};
\draw[-] (0.0,0)--(1,0);
\draw[-] (.5,0)--(.5,1) node at (.5,1.2) {$x_{3}$};
\draw[-] (.5,0)--(1,0) node at (1.2,0) {$x_4$};
\end{scope}
\begin{scope}[xshift=4cm]
\node at (-.2,0) {$x_3$};
\node at (.5,-.3) {$V_{3}$};
\draw[-] (0.0,0)--(1,0);
\draw[-] (.5,0)--(.5,1) node at (.5,1.2) {$x_{4}$};
\draw[-] (.5,0)--(1,0) node at (1.2,0) {$x_{5}$};
\end{scope}
\end{tikzpicture}
\qquad 
\begin{tikzpicture}[thick]
\begin{scope}
\node at (-.2,0) {$x_{1}$};
\node at (.5,-.3) {$V_{1}$};
\draw[-] (0.0,0)--(1,0);
\draw[-] (.5,0)--(.5,1) node at (.5,1.2) {$x_{2}$};
\draw[-] (.5,0)--(1,0) node at (1.2,0) {$x_3$};
\end{scope}
\begin{scope}[xshift=2cm]
\node at (-.2,0) {$x_2$};
\node at (.5,-.3) {$V_{2}$};
\draw[-] (0.0,0)--(1,0);
\draw[-] (.5,0)--(.5,1) node at (.5,1.2) {$x_{3}$};
\draw[-] (.5,0)--(1,0) node at (1.2,0) {$x_5$};
\end{scope}
\begin{scope}[xshift=4cm]
\node at (-.2,0) {$x_2$};
\node at (.5,-.3) {$V_{3}$};
\draw[-] (0.0,0)--(1,0);
\draw[-] (.5,0)--(.5,1) node at (.5,1.2) {$x_{4}$};
\draw[-] (.5,0)--(1,0) node at (1.2,0) {$x_{5}$};
\end{scope}
\end{tikzpicture}
\cr &
\begin{tikzpicture}[thick]
\begin{scope}
\node at (-.2,0) {$x_{1}$};
\node at (.5,-.3) {$V_{1}$};
\draw[-] (0.0,0)--(1,0);
\draw[-] (.5,0)--(.5,1) node at (.5,1.2) {$x_{2}$};
\draw[-] (.5,0)--(1,0) node at (1.2,0) {$x_4$};
\end{scope}
\begin{scope}[xshift=2cm]
\node at (-.2,0) {$x_1$};
\node at (.5,-.3) {$V_{2}$};
\draw[-] (0.0,0)--(1,0);
\draw[-] (.5,0)--(.5,1) node at (.5,1.2) {$x_{3}$};
\draw[-] (.5,0)--(1,0) node at (1.2,0) {$x_4$};
\end{scope}
\begin{scope}[xshift=4cm]
\node at (-.2,0) {$x_1$};
\node at (.5,-.3) {$V_{3}$};
\draw[-] (0.0,0)--(1,0);
\draw[-] (.5,0)--(.5,1) node at (.5,1.2) {$x_{4}$};
\draw[-] (.5,0)--(1,0) node at (1.2,0) {$x_{5}$};
\end{scope}
\end{tikzpicture}
\qquad 
\begin{tikzpicture}[thick]
\begin{scope}
\node at (-.2,0) {$x_{1}$};
\node at (.5,-.3) {$V_{1}$};
\draw[-] (0.0,0)--(1,0);
\draw[-] (.5,0)--(.5,1) node at (.5,1.2) {$x_{2}$};
\draw[-] (.5,0)--(1,0) node at (1.2,0) {$x_5$};
\end{scope}
\begin{scope}[xshift=2cm]
\node at (-.2,0) {$x_1$};
\node at (.5,-.3) {$V_{2}$};
\draw[-] (0.0,0)--(1,0);
\draw[-] (.5,0)--(.5,1) node at (.5,1.2) {$x_{3}$};
\draw[-] (.5,0)--(1,0) node at (1.2,0) {$x_5$};
\end{scope}
\begin{scope}[xshift=4cm]
\node at (-.2,0) {$x_3$};
\node at (.5,-.3) {$V_{3}$};
\draw[-] (0.0,0)--(1,0);
\draw[-] (.5,0)--(.5,1) node at (.5,1.2) {$x_{4}$};
\draw[-] (.5,0)--(1,0) node at (1.2,0) {$x_{5}$};
\end{scope}
\end{tikzpicture}
}

\subsubsection{$u$-type and $v$-type Cross Ratios}

To finish solving for the $u$-type and $v$-type cross ratios we choose to use the flow diagram in the top left of \eno{fivepointflow}. In this paper, the inductive proof in Section \ref{PROOF} makes the choice to assume the $\mathcal{O}_{1}(x_1)$ leg never flows. To enforce the choice that $\mathcal{O}_{1}(x_1)$ never flows, we restricted our labeling to require the $\mathcal{O}_{1}(x_1)$ leg to always connect to the $\mathcal{O}_{2}(x_2)$ leg.\footnote{We can always make the choice to label an arbitrary conformal block such that the $\mathcal{O}_{1}(x_1)$ and $\mathcal{O}_{2}(x_2)$ leg connect. Since block labeling is arbitrary we can send some $\mathcal{O}_{1}(x_1)$ labeling to $\mathcal{O}_{a}(x_a)$ and $\mathcal{O}_{2}(x_2)$ labeling to $\mathcal{O}_{b}(x_b)$ such that our choice in flow restriction is possible in any block orientation and labeling.} Solving for the $u$-type cross ratios
\eqn{ucross5}{ 
&
\begin{tikzpicture}[thick]
\begin{scope}
\node at (-.2,0) {$x_{1}$};
\node at (.5,-.3) {$V_{1}$};
\draw[-] (0.0,0)--(1,0);
\draw[-] (.5,0)--(.5,1) node at (.5,1.2) {$x_{2}$};
\draw[-] (.5,0)--(1,0) node at (1.2,0) {$x_3$};
\end{scope}
\begin{scope}[xshift=2cm]
\node at (-.2,0) {$x_2$};
\node at (.5,-.3) {$V_{2}$};
\draw[-] (0.0,0)--(1,0);
\draw[-] (.5,0)--(.5,1) node at (.5,1.2) {$x_{3}$};
\draw[-] (.5,0)--(1,0) node at (1.2,0) {$x_4$};
\end{scope}
\begin{scope}[xshift=4cm]
\node at (-.2,0) {$x_3$};
\node at (.5,-.3) {$V_{3}$};
\draw[-] (0.0,0)--(1,0);
\draw[-] (.5,0)--(.5,1) node at (.5,1.2) {$x_{4}$};
\draw[-] (.5,0)--(1,0) node at (1.2,0) {$x_{5}$};
\end{scope}
\end{tikzpicture} :
\cr 
&
\begin{tikzpicture}[thick]
\begin{scope}
\node at (-.2,0) {$x_{1}$};
\node at (.5,-.3) {$V_{1}$};
\node at (1.6,-0.5) {$\mathcal{O}_{k_{1}}$};
\draw[-] (0.0,0)--(1,0);
\draw[-] (.5,0)--(.5,1) node at (.5,1.2) {$x_{2}$};
\draw[-] (.5,0)--(1,0) node at (1.2,0) {$x_3$};
\draw[black] (-.25,-.05) circle (.3 cm);
\draw[dashed,  red]  (0,-.08) to[bend right=5] (.9,-.08);
\draw[dashed,  teal]  (0,.08) to[bend right=5] (.45,1);
\end{scope}
\begin{scope}[xshift=2cm]
\node at (-.2,0) {$x_2$};
\node at (.5,-.3) {$V_{2}$};
\draw[-] (0.0,0)--(1,0);
\draw[-] (.5,0)--(.5,1) node at (.5,1.2) {$x_{3}$};
\draw[-] (.5,0)--(1,0) node at (1.2,0) {$x_4$};
\draw[black] (1.25,-.05) circle (.3 cm);
\draw[dashed,  red]  (0,-.08) to[bend right=5] (.9,-.08);
\draw[dashed,  teal]  (.55,1) to[bend right=5] (.8, .1);
\end{scope}
\end{tikzpicture}: u_1 = \frac{x_{12}^2x_{43}^2}{x_{13}^2x_{42}^2},
\qquad 
\begin{tikzpicture}[thick]
\begin{scope}
\node at (-.2,0) {$x_{2}$};
\node at (.5,-.3) {$V_{2}$};
\node at (1.6,-0.5) {$\mathcal{O}_{k_{2}}$};
\draw[-] (0.0,0)--(1,0);
\draw[-] (.5,0)--(.5,1) node at (.5,1.2) {$x_{3}$};
\draw[-] (.5,0)--(1,0) node at (1.2,0) {$x_4$};
\draw[black] (-.25,-.05) circle (.3 cm);
\draw[dashed,  red]  (0,-.08) to[bend right=5] (.9,-.08);
\draw[dashed,  teal]  (0,.08) to[bend right=5] (.45,1);
\end{scope}
\begin{scope}[xshift=2cm]
\node at (-.2,0) {$x_3$};
\node at (.5,-.3) {$V_{3}$};
\draw[-] (0.0,0)--(1,0);
\draw[-] (.5,0)--(.5,1) node at (.5,1.2) {$x_{4}$};
\draw[-] (.5,0)--(1,0) node at (1.2,0) {$x_5$};
\draw[black] (1.25,-.05) circle (.3 cm);
\draw[dashed,  red]  (0,-.08) to[bend right=5] (.9,-.08);
\draw[dashed,  teal]  (.55,1) to[bend right=5] (.8, .1);
\end{scope}
\end{tikzpicture}
: u_{2}=\frac{x_{23}^2x_{54}^2}{x^2_{24}x^2_{53}}
}
and solving for the $v$-type cross ratios, 
\begingroup\makeatletter\def\f@size{8}\check@mathfonts 
\eqn{vcross5}{ 
&\begin{tikzpicture}[thick]
\begin{scope}[xshift=0cm]
\node at (-.2,0) {$x_{1}$};
\node at (.5,-.3) {$V_{1}$};
\draw[-] (0.0,0)--(1,0);
\draw[-] (.5,0)--(.5,1) node at (.5,1.2) {$x_{2}$};
\draw[-] (.5,0)--(1,0) node at (1.2,0) {$x_{3}$};
\draw[cyan] (1.5,.1) circle (.6 cm);
\draw[dashed,  orange]  (0,-.08) to[bend right=5] (.9,-.08);
\end{scope}
\begin{scope}[xshift=2cm]
\node at (-.2,0) {$x_{2}$};
\node at (.5,-.3) {$V_{2}$};
\draw[-] (0.0,0)--(1,0);
\draw[-] (.5,0)--(.5,1) node at (.5,1.2) {$x_{3}$};
\draw[-] (.5,0)--(1,0) node at (1.2,0) {$x_{4}$};
\draw[dashed,  orange]  (0,-.08) to[bend right=5] (.9,-.08);
\end{scope}
\end{tikzpicture}
:V_{1}V_{2} :v_{14}= {x_{14}^2 x_{23}^2 \over x_{13}^2 x_{24}^2  }
\qquad 
\begin{tikzpicture}[thick]
\begin{scope}[xshift=2cm]
\node at (-.2,0) {$x_{2}$};
\node at (.5,-.3) {$V_{2}$};
\draw[-] (0.0,0)--(1,0);
\draw[-] (.5,0)--(.5,1) node at (.5,1.2) {$x_{3}$};
\draw[-] (.5,0)--(1,0) node at (1.2,0) {$x_{4}$};
\draw[cyan] (1.5,.1) circle (.6 cm);
\draw[dashed,  orange]  (0,-.08) to[bend right=5] (.9,-.08);
\end{scope}
\begin{scope}[xshift=4cm]
\node at (-.2,0) {$x_{3}$};
\node at (.5,-.3) {$V_{3}$};
\draw[-] (0.0,0)--(1,0);
\draw[-] (.5,0)--(.5,1) node at (.5,1.2) {$x_{4}$};
\draw[-] (.5,0)--(1,0) node at (1.2,0) {$x_{5}$};
\draw[dashed,  orange]  (0,-.08) to[bend right=5] (.9,-.08);
\end{scope}
\end{tikzpicture}
:V_{2}V_{3} :v_{25}= {x_{25}^2 x_{34}^2 \over x_{24}^2 x_{35}^2  }
\cr 
&\begin{tikzpicture}[thick]
\begin{scope}[xshift=0cm]
\node at (-.2,0) {$x_{1}$};
\node at (.5,-.3) {$V_{1}$};
\draw[-] (0.0,0)--(1,0);
\draw[-] (.5,0)--(.5,1) node at (.5,1.2) {$x_{2}$};
\draw[-] (.5,0)--(1,0) node at (1.2,0) {$x_{3}$};
\draw[cyan] (1.5,.1) circle (.6 cm);
\draw[dashed,  orange]  (0,-.08) to[bend right=5] (.9,-.08);
\end{scope}
\begin{scope}[xshift=2cm]
\node at (-.2,0) {$x_{2}$};
\node at (.5,-.3) {$V_{2}$};
\draw[-] (0.0,0)--(1,0);
\draw[-] (.5,0)--(.5,1) node at (.5,1.2) {$x_{3}$};
\draw[-] (.5,0)--(1,0) node at (1.2,0) {$x_{4}$};
\draw[cyan] (1.5,.1) circle (.6 cm);
\draw[dashed,  orange]  (0,-.08) to[bend right=5] (.9,-.08);
\end{scope}
\begin{scope}[xshift=4cm]
\node at (-.2,0) {$x_{3}$};
\node at (.5,-.3) {$V_{3}$};
\draw[-] (0.0,0)--(1,0);
\draw[-] (.5,0)--(.5,1) node at (.5,1.2) {$x_{4}$};
\draw[-] (.5,0)--(1,0) node at (1.2,0) {$x_{5}$};
\draw[dashed,  orange]  (0,-.08) to[bend right=5] (.9,-.08);
\end{scope}
\end{tikzpicture}
:V_{1}V_{3} :v_{15}= {x_{15}^2 x_{23}^2 x_{34}^2 \over x_{13}^2 x_{24}^2x_{35}^2  }
}
\endgroup

It is worth re-noting that since $x_{ij}^2$ is symmetric that we tend to use the convention of writing $i<j$. In this section we outlined the methodology for obtaining $u$-type and $v$-type cross ratios for any conformal block using two examples, the 4-point and 5-point blocks.

\section{Proof of Feynman Rules}
\label{PROOF}

In this section we provide the full proof of the Feynman rules based on the OPE flow and the associated conformal cross ratios discussed above.  The proof proceeds by induction.

\begin{figure}[th]
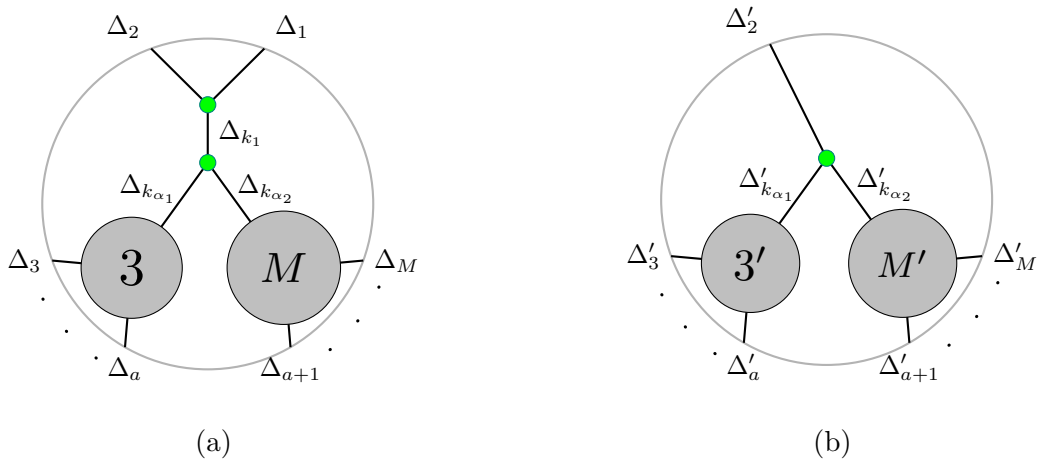

    \centering
    \begin{subfigure}[b]{0.49\textwidth}
    \[ \musepic{\figNptAdS}  \]
    \caption{}
    \label{fig:nptAdS}
    \end{subfigure}
    \begin{subfigure}[b]{0.49\textwidth}
    \[ \musepic{\figNMinusOneptAdS}  \]
    \caption{}
    \label{fig:nmin1ptAdS}
    \end{subfigure}
    \caption{(a) A canonical $M$-point AdS diagram, and (b) the associated $(M-1)$-point diagram. We have left the external insertion points implicit -- an operator of dimension $\Delta_i$ or $\Delta_i^\prime$ is inserted at position $x_i$. The shaded blobs labeled $3$ and $M$ represent arbitrary tree-level subdiagrams. The shaded blobs $3^\prime$ and $M^\prime$ have the same topology as their unprimed counterparts with all external and internal dimensions replaced by their primed counterparts. The precise relation between the primed and unprimed conformal dimensions will be explained later in this section.}
    \label{fig:AdS}
\end{figure}

Consider the $M$-point bulk diagram in AdS$_{d+1}$ shown in Figure~\ref{fig:nptAdS}. It is a canonical AdS diagram, which is constructed only with the help of cubic vertices, and its topology is the same as the topology of the associated conformal block we are interested in.  The only explicit cubic vertices, which correspond to OPEs in the CFT language, shown in Figure~\ref{fig:nptAdS} are denoted by green dots.  All the remaining cubic vertices are included inside the gray blobs.
It is well-known that the conformal block decomposition of this diagram contains a single-trace component where single-trace operators are exchanged at {\it every} internal leg, as well as contributions coming from the exchange of a mixture of single-trace, double-trace and higher multi-trace operators at internal legs. To obtain the $M$-point conformal block of the same topology as the AdS diagram corresponding to a scalar block built out of external and exchanged operators dual to the corresponding AdS scalar fields shown in Figure~\ref{fig:nptAdS}, we need only isolate the purely single-trace contribution to the diagram~\cite{Parikh:2019ygo,Jepsen:2019svc}. 
We will achieve this single-trace projection in Mellin space following the method utilized in Ref.~\cite{Hoback:2020pgj}.

This $M$-point bulk diagram position space amplitude $A_M$ can be expressed in Mellin space as, 
\eqn{MellinAmpDef}{
A_M &= {\cal N}_M \left( \prod_{(rs) \in {\cal U} \bigcup {\cal V}} \int {d\gamma_{rs} \over 2\pi i} \right)
{\cal M}_M(\gamma_{ij})  \left(\prod_{(ij) \in {\cal U} \bigcup {\cal V} \bigcup {\cal D}} {\Gamma(\gamma_{ij}) \over (x_{ij}^2)^{\gamma_{ij}} } \right) \cr 
&= {\cal N}_M \left( \prod_{(rs) \in {\cal U} \bigcup {\cal V} \bigcup {\cal D} } \int {d\gamma_{rs} \over 2\pi i} {\Gamma(\gamma_{rs}) \over (x_{rs}^2)^{\gamma_{rs}} } \right)
\left(\prod_{a=1}^{M} 2\pi i\, \delta(\sum_{b=1}^{M} \gamma_{ab}) \right)
{\cal M}_M(\gamma_{ab})   \,,
}
where the union of index sets ${\cal U} \cup {\cal V}$  labels the $M(M-3)/2$ independent Mellin variables. The index set ${\cal D}$ labeling the $M$ dependent Mellin variables is introduced as integration variables in the second line with the help of delta functions enforcing the constraints 
\eqn{MellinConstraints}{
\sum_{b=1}^M\gamma_{ab} = 0 \qquad \forall\; a=1,\ldots, M\,.
}
In writing~\eno{MellinConstraints}, we have defined the ``diagonal'' Mellin variables $\gamma_{aa} := -\Delta_a$.\footnote{Mellin variables are symmetric, complex variables so that the constraints~\eno{MellinConstraints} can be solved in terms of auxiliary momentum variables $p_a$, such that $\gamma_{ab}= p_a \cdot p_b$, along with momentum conservation $\sum_{a=1}^M p_a = 0$, where $p_a \cdot p_a= -\Delta_a$ on-shell. We can then define Mandelstam variables
\eqn{MandelstamDef}{
s_{a_1\ldots a_k} = -(p_{a_1} + \cdots + p_{a_k})^2 = \sum_{b=a_1}^{a_k} \Delta_{b}-2\sum_{a_1 \leq r < s \leq a_k} \gamma_{rs}\,,
}
which we will use in the proof later.
}
Together, the index sets constitute the off-diagonal upper-triangular matrix indices
\eqn{UVDunion}{
{\cal U} \cup {\cal V} \cup {\cal D} = \{ (rs) : 1 \leq r < s \leq M \}\,.
}
 The decomposition into dependent and independent sets of variables is not unique.
 For the set ${\cal V}$, we will be employing the choice presented in Section~\ref{INDEXCROSSRATIOS}.
 
In our convention, the normalization constant in~\eno{MellinAmpDef} is given by
\eqn{NDef}{
{\cal N}_M = \pi^{(M-2)h} \left( \prod_{a=1}^{M-3} {1 \over \Gamma(\Delta_{k_a}) } \right) \left( \prod_{a=1}^{M} {1\over \Gamma(\Delta_a)} \right)  ,
}
where $\Delta_a$ are the external conformal dimensions, $\Delta_{k_a}$ are the internal exchanged dimensions and $h$ was defined below~\eno{EdgeDef}. 
 Then, according to the Feynman rules for Mellin amplitudes of tree-level AdS diagrams~\cite{Paulos:2011ie,Fitzpatrick:2011ia,Nandan:2011wc}, the Mellin amplitude ${\cal M}_M$ in~\eno{MellinAmpDef} is given by
\eqn{MellinAmp}{
{\cal M}_M(\gamma_{ab}) = \left( \prod_{a=1}^{M-3} \sum_{m_a=0}^{\infty} \right) \left( \prod_{a=1}^{M-3} E_a^{\rm Mellin} \right)  \left( \prod_{a=1}^{M-2} V_a^{\rm Mellin} \right) .
}
The edge factors take the form
\eqn{EdgeMellin}{
    E_a^{\rm Mellin} = {1\over m_a!} {(\Delta_{k_a} - h+1)_{m_a} \over    {\Delta_{k_a}-s_a\over 2} + m_a} \qquad (1\leq a \leq M-3)\,,
}
where $s_a$ is the Mandelstam invariant associated to the internal leg $\Delta_{k_a}$, and $m_a$ is an integral parameter associated to the same leg to be summed over as shown in~\eno{MellinAmp}.
All vertices of the canonical $M$-point AdS diagram are cubic vertices; the associated vertex factors are
\eqn{VertexMellin}{
    V_a^{\rm Mellin} &= {1 \over 2} \Gamma(\Delta_{\alpha\beta\gamma,}-h)  \cr 
    & \hspace{-1em} \times F_A^{(3)}\!\left[\Delta_{\alpha\beta\gamma,}- h; \{-m_\alpha, -m_\beta, -m_\gamma \}; \left\{\Delta_\alpha -h+1, \Delta_\beta -h+1,\Delta_\gamma -h+1 \right\}; 1,1,1 \right],
}
for $1 \leq a \leq M-2$, where $\Delta_\alpha,\Delta_\beta,\Delta_\gamma$ correspond to the conformal dimensions incident at the associated cubic vertex in the AdS diagram and $m_\alpha$, $m_\beta$, $m_\gamma$ are the dummy integral parameters associated to the legs.  Moreover, $F_A^{(3)}$ is the Lauricella function of type $A$ of three variables,\footnote{If one (respectively, two) leg of the cubic vertex is external, the corresponding  $m_\alpha$ parameter is set to zero, and the Lauricella function of three variables reduces to the Lauricella function of two (respectively, one) variable(s).} given by \eqref{LauricellaDef}.

Mellin amplitudes are merophormic functions of Mandelstam invariants which have simple poles corresponding precisely to the exchange of single-trace operators~\cite{Mack:2009gy,Mack:2009mi,Penedones:2010ue}. 
These simple poles occur when the Mandelstam invariants in the edge-factors~\eno{EdgeMellin} go ``on-shell.'' 
Thus the task of isolating the conformal block reduces to evaluating the residue at all simple poles of the Mellin amplitude of the associated canonical AdS diagram.
We will achieve this inductively for a generic canonical AdS diagram in the remainder of this section.

\subsection{Setting Up Induction}
\label{SETINDUCTION}

To proceed with the induction, we first write:
\eqn{ASplit}{
A_M &= {\cal N}_M \left(\prod_{a=1}^{M-3} \sum_{m_a=0}^\infty \right) 
\left(\prod_{a=1}^{M-2} V_a^{\rm Mellin} \right)  \left( \prod_{a=1}^{M-3} {(\Delta_{k_a} - h + 1)_{m_a} \over m_a!} \right) B_M \,,
}
with
\eqn{BDef}{
B_M &:= \left(\prod_{(rs) \in {\cal U} \bigcup {\cal V} \bigcup {\cal D}} \int{ d \gamma_{rs} \over 2\pi i} {\Gamma(\gamma_{rs}) \over (x_{rs}^2)^{\gamma_{rs}} } \right)
\left(\prod_{a=1}^M 2\pi i\delta(\sum_{\substack{b=1\\b\neq a}}^M \gamma_{ab} - \Delta_a) \right) 
\left(\prod_{a=1}^{M-3}{1 \over {\Delta_{k_a}-s_a \over 2} +m_a} \right)\cr
&=: \left(\prod_{a=2}^M \int {d\gamma_{1a} \over 2\pi i} {\Gamma(\gamma_{1a}) \over (x_{1a}^2)^{\gamma_{1a}} } \right) 
2\pi i\delta(\sum_{a=2}^M \gamma_{1a} - \Delta_1) 
{1 \over {\Delta_{k_1}-s_1 \over 2} + m_1}B_{M-1} \,,
}
where in the second equality we defined $B_{M-1}$ as 
\eqn{BMins1}{
B_{M-1} := \left(\prod_{(rs) \in {\cal U}^\prime \bigcup {\cal V}^\prime \bigcup {\cal D}^\prime} \int{ d \gamma_{rs} \over 2\pi i} {\Gamma(\gamma_{rs}) \over (x_{rs}^2)^{\gamma_{rs}} } \right)
\left(\prod_{a=2}^M 2\pi i\delta(\sum_{\substack{b=1\\b\neq a}}^M \gamma_{ab} - \Delta_a) \right) 
\left(\prod_{a=2}^{M-3} {1 \over {\Delta_{k_a}-s_a \over 2} +m_a} \right)\!.
}
In writing~\eno{ASplit}-\eno{BDef} we have suggestively rearranged the integrals in~\eno{MellinAmpDef}.
Note that we interchanged the order of integration and summation to pull out the $(M-3)$-dimensional infinite sum to the front in~\eno{ASplit}, and we used the fact that the vertex factors $V_a^{\rm Mellin}$ are Mellin variable-independent to move them outside the contour integrals.
The edge factors $E_a^{\rm Mellin}$ have been collected in $B_{M-1}$ in~\eno{BMins1} except for the edge factor $E_1$ which corresponds to the internal edge labeled $\Delta_{k_1}$ in Figure~\ref{fig:nptAdS}.
We could pull $E_1$ out of the integral in~\eno{BMins1} over the primed Mellin index sets ${\cal U}^\prime \cup {\cal V}^\prime \cup {\cal D}^\prime$, defined via
\eqn{}{ 
  {\cal U}^\prime \cup {\cal V}^\prime \cup {\cal D}^\prime := {\cal U} \cup {\cal V} \cup {\cal D} \smallsetminus \{ (1a): 2 \leq a \leq M\} \,,
}
because we have chosen to express the associated Mandelstam invariant $s_1$ solely in terms of Mellin variables from the set $\{ \gamma_{1a}: 2 \leq a \leq M\}$.
The contour integral in the second equality of~\eno{BDef} is precisely over the Mellin variables whose indices belong to the set $\{ \gamma_{1a}: 2 \leq a \leq M\}$.
Specifically,
\eqn{s1Choice}{
s_1 = \Delta_1 + \Delta_2 -2\gamma_{12}\,.
}

A convenient choice for the individual primed subsets ${\cal U}^\prime, {\cal V}^\prime, {\cal D}^\prime$ will eventually allow us to identify them as the Mellin index subsets of an $(M-1)$-point Witten diagram obtained by removing the  external leg labeled $\Delta_1$ from the original Witten diagram (see Figure~\ref{fig:nmin1ptAdS}).\footnote{As we will demonstrate shortly, the precise values of external and exchanged conformal dimensions in the $(M-1)$-point diagram will differ from those of the original $M$-point diagram. However, the Mellin index sets do not depend on conformal dimensions but only on the position space insertion points, which are preserved upon going from the $M$-point diagram to the $(M-1)$-point diagram.} 
The set ${\cal D}^\prime$ will play the role of indexing the dependent Mellin variables of the $(M-1)$-point diagram, while the set ${\cal U}^\prime \cup {\cal V}^\prime$ will index the independent Mellin variables.
Just like for the $M$-point diagram the choice of the primed sets is non-unique, but  we will employ the algorithm of Section~\ref{INDEXCROSSRATIOS} to construct the set ${\cal V}^\prime$.
This choice turns out to be convenient from the point of view of setting up an inductive procedure as by construction it guarantees the following strict containment,
\eqn{PrimeSubsets}{
{\cal V}^\prime \subset {\cal V} \,.
}
As will become clear later in the proof, the choice of $s_1$ in~\eno{s1Choice} allows us to set
\eqn{UUprime}{
{\cal U} = {\cal U}^\prime \cup \{(12) \}\,.
}
Furthermore, we choose
\eqn{DDprime}{
{\cal D} = {\cal D}^\prime \cup \{(13)\} \,,
}
which immediately yields
\eqn{VVprime}{
{\cal V} = {\cal V}^\prime  \cup \{ (1a) : 4 \leq a \leq M\}\,.
}

The reason it is useful to rewrite the $M$-point position space amplitude as in~\eno{ASplit} is that it can now be expressed in terms of the position space amplitude of the closely related $(M-1)$-point diagram shown in Figure~\ref{fig:nmin1ptAdS}. 
The aforementioned $M$- and $(M-1)$-point diagrams share the same topology, except for the fact that the external leg labeled $\Delta_1$ has been removed in the $(M-1)$-point diagram and all conformal dimensions have been shifted in a way we will describe shortly. 
To distinguish between the original and shifted conformal dimensions, we will be using primed labels to refer to the external and exchanged dimensions of the $(M-1)$-point diagram, as shown in Figure~\ref{fig:nmin1ptAdS}.

In \eqref{BDef} and $\eqref{BMins1}$, $s_a$ can be determined as follows: We cut the $\Delta_{k_a}$-internal line inside the $M$-point diagram, leading to two disconnected parts. One part contains the external operator $\mathcal{O}_{1}(x_1)$ while the other part does not. 
We denote the external operators in the part which does not contain $\mathcal{O}_1(x_1)$ by $\mathcal{O}_b(x_b)$, and we define the corresponding set $J_{k_a} \ni b$. As shown in Figure \ref{fig:nptAdS}, it is clear that $J_{k_a}$ is a subset of $\{3,4,\ldots,M\}$. Indeed our choice in cutting the $\Delta_{k_a}$-internal line is such that one part contains the external operator $\mathcal{O}_1(x_1)$ [remember $\mathcal{O}_1(x_1)$ will connect to $\mathcal{O}_2(x_2)$], while the remaining internal operators from the other part are contained in $J_{k_a}$. As a result, $s_a$ can be computed from\footnote{The Mandelstam variable $s_a$ can also be obtained from the part which contains $\mathcal{O}_{1}(x_1)$. Let us denote the external operators in the part with $\mathcal{O}_1(x_1)$ by $\mathcal{O}_b(x_b)$ for $b\in J_{k_a}^c$, where $J_{k_a}^c$ is the complement of $J_{k_a}$ in $\{1,2,\ldots,M\}$. Due to the conservation condition of the auxiliary momentum, \textit{i.e.} $\sum_{b=1}^{M}p_b=0$, $s_a$ is independent of this choice.}
\eqn{}{
s_a=&-(\sum_{b\in J_{k_a}}p_{b})^2=\sum_{b\in J_{k_a}}\Delta_{b}-2\sum_{\substack{b,c\in J_{k_a}\\b<c}}\gamma_{bc}.
}
To get the corresponding conformal blocks, we want to evaluate $B_M$ at the “single-trace poles,” by evaluating the residue at the simple
poles at which the Mandelstam invariant goes “on-shell.” Let us call that quantity $B_{M}^{\text{s.t.}}$. We will obtain $B_{M}^{\text{s.t.}}$ inductively starting with
\eqn{BMst}{
B^{\text{s.t.}}_M  &= \left[\left(\prod_{(rs) \in {\cal U} \bigcup {\cal V} \bigcup {\cal D}} \int{ d \gamma_{rs} \over 2\pi i} {\Gamma(\gamma_{rs}) \over (x_{rs}^2)^{\gamma_{rs}} } \right)
\left(\prod_{a=1}^M 2\pi i\delta(\sum_{\substack{b=1\\b\neq a}}^M \gamma_{ab} - \Delta_a) \right) 
\left(\prod_{a=1}^{M-3}{1 \over {\Delta_{k_a}-s_a \over 2} +m_a} \right) \right]_{\rm s.t.} \cr
&=\left[\left(\prod_{a=2}^M \int {d\gamma_{1a} \over 2\pi i} {\Gamma(\gamma_{1a}) \over (x_{1a}^2)^{\gamma_{1a}} } \right) 
2\pi i\delta(\sum_{a=2}^M \gamma_{1a} - \Delta_1) 
 {1 \over {\Delta_{k_1}-s_1 \over 2} + m_1}B^{\text{s.t.}}_{M-1}\right]_{\text{s.t.}},
}
where $B^{\text{s.t.}}_{M-1}$ is given by
\eqn{BMins1st}{
B^{\text{s.t.}}_{M-1} &= \left[\left(\prod_{(rs) \in {\cal U}^\prime \bigcup {\cal V}^\prime \bigcup {\cal D}^\prime} \int{ d \gamma_{rs} \over 2\pi i} {\Gamma(\gamma_{rs}) \over (x_{rs}^2)^{\gamma_{rs}} } \right) \!\!
\left(\prod_{a=2}^M 2\pi i\delta(\sum_{\substack{b=1\\b\neq a}}^M \gamma_{ab} - \Delta_a) \right) \!\!
\left(\prod_{a=2}^{M-3} {1 \over {\Delta_{k_a}-s_a \over 2} +m_a} \right)\right]_{\text{s.t.}}\!\!\!.
}

The Feynman-like rules for conformal blocks given in Section~\ref{FEYNMAN} are equivalent to the following expression for $B_M^{\rm s.t.}$: 
\eqn{RulesBMst}{
B^{\text{s.t.}}_M&=L(\Delta_1,\ldots,\Delta_M)\left(\prod_{a=1}^{M-3}u_a^{\frac{\Delta_{k_a}}{2}+m_a}\right)\left(\prod_{(rs)\in\mathcal{V}}\sum_{j_{rs}=0}^{\infty}\frac{(1-v_{rs})^{j_{rs}}}{j_{rs}!}\right)\left(\prod_{a=1}^{M-2}\hat{V}_a\right)\left(\prod_{a=1}^{M-3}\hat{E}_{a}\right),
}
where $\hat{V}_a$ and $\hat{E}_a$ are “Gamma-vertex” and “Gamma-edge” factors for the $M$-point conformal blocks. Per the rules, $\hat{V}_a$ and $\hat{E}_a$ are given by
\eqn{GammaVertexEdge}{
\widehat{V}_a &=\Gamma(\Delta_{\sigma_1\sigma_2,\sigma_3}+m_{\sigma_1\sigma_2,\sigma_3}+\frac{1}{2}\ell_{\sigma_1\sigma_2,\sigma_3})\Gamma(\Delta_{\sigma_2\sigma_3,\sigma_1}+m_{\sigma_2\sigma_3,\sigma_1}+\frac{1}{2}\ell_{\sigma_2\sigma_3,\sigma_1})\cr
&\qquad\times\Gamma(\Delta_{\sigma_3\sigma_1,\sigma_2}+m_{\sigma_3\sigma_1,\sigma_2}+\frac{1}{2}\ell_{\sigma_3\sigma_1,\sigma_2}),\cr
\widehat{E}_a&=\frac{1}{\Gamma(\Delta_{k_a}+2m_{a}+\ell_{k_a})},
}
where $\ell_a$ are the post-Mellin parameters for the $M$-point topology, and $\sigma_1$, $\sigma_2$, $\sigma_3$ label the operators incident on the internal vertex. 

Let us briefly describe why~\eno{RulesBMst}-\eno{GammaVertexEdge} is equivalent to the original Feynman-like rules~\eno{EdgeDef}-\eno{Feynman}. 
Looking at the definition of $B_M$ in~\eno{ASplit} and comparing~\eno{RulesBMst} with the original Feynman-like rules~\eno{Feynman}, it is clear the factor of $(m_a!)$ and the overall sum over integers $m_a$ should not appear in~\eno{RulesBMst}.
It is also clear that the numerator of the edge factor~\eno{EdgeDef} has been factored out of $B_M$ in~\eno{ASplit}, so it doesn't appear in the Gamma-edge factor $\widehat{E}_a$.
Similarly, the factors of Lauricella functions appearing in the Feynman vertex factor~\eno{VertexDef} get absorbed into $V_a^{\rm Mellin}$ in~\eno{ASplit}, thus they are left out of the Gamma-vertex factor $\widehat{V}_a$. 
Now, the denominator of the edge factor~\eno{EdgeDef} is the Pochhammer symbol $(\Delta_{k_a})_{2m_a+\ell_{k_a}} = \Gamma(\Delta_{k_a}+2m_a+\ell_{k_a})/\Gamma(\Delta_{k_a})$. The factor of $\Gamma(\Delta_{k_{a}})$ combines with the same factor in the normalization constant ${\cal N}_M$ given in~\eno{NDef}. 
Modulo this factor of $\prod_{k_i}\Gamma(\Delta_{k_i})^2$, what remains is precisely the Gamma-edge factor in~\eno{GammaVertexEdge}. We will return to what happens to this excess factor shortly. 
The explanation of the Gamma-vertex factor is slightly more involved. 
First, note that to obtain the theory-independent conformal block, after taking the single-trace projection of~\eno{ASplit} one must divide out by theory-specific OPE coefficients. The OPE coefficients associated with a scalar effective field theory in AdS space are the mean field theory OPE coefficients given in~\eno{OPEreal}, one for each internal vertex of the conformal block. 
For instance, the factor of ${1\over 2} \Gamma(\Delta_{ijk,}-h)$ in $V_a^{\rm Mellin}$ for each vertex cancels against the same factor appearing in the OPE coefficient. 
The factors of $\Gamma(\Delta_{ij,k})\Gamma(\Delta_{jk,i})\Gamma(\Delta_{ki,j})$ in the OPE coefficient cancel against the same factors appearing inside the Pochhammer symbols in the vertex factor~\eno{VertexDef}, leaving behind the Gamma function factors shown in the Gamma-vertex factor in~\eno{GammaVertexEdge}.
Furthermore it is easy to see that the $M-2$ factors of $\pi^h$ and $M$ factors of $\Gamma(\Delta_a)$ in the normalization constant~\eno{NDef} cancel out against the factors of $\pi^h$ and $\Gamma(\Delta_a)$ in the $M-2$ OPE coefficients.
What remains in the product over all OPE coefficients is the factor $\prod_{k_i}\Gamma(\Delta_{k_i})^2$, which precisely cancels with the excess factor mentioned above. This accounts for all factors and proves the equivalence between~\eno{Feynman} and~\eno{RulesBMst}.

Now, it is easily checked that the Feynman-like rules~\eno{RulesBMst} are satisfied for $M=3$. In that case, 
\eqn{B3Def}{
B_3^{\rm s.t.} = \left[\left( \prod_{(rs) \in {\cal U} \bigcup {\cal V} \bigcup {\cal D}} \int {d\gamma_{rs} \over 2\pi i} {\Gamma(\gamma_{rs}) \over (x_{rs}^2)^{\gamma_{rs}}} \right) \left( \prod_{a=1}^3 2\pi i\, \delta(\sum_{\substack{b=1 \\b\neq a}}^3 \gamma_{ab}-\Delta_a)  \right) \right]_{\rm s.t.},
}
where 
\eqn{}{
{\cal U}  = \emptyset  \qquad {\cal V} = \emptyset \qquad {\cal D} = \{ (12),(13),(23) \} \,.
}
The single-trace operation, denoted by $\left[ \,\cdot\, \right]_{\rm s.t.}$, acts trivially since there are no single-trace poles in $B_3^{\rm s.t.}$ to begin with.\footnote{Recall that the single-trace poles arise from the exchange of a single-trace quasi-primary operator in an intermediate channel; for $M=3$ there are no such exchanges.}
 We can thus easily evaluate $B_3^{\rm s.t.}$ since the three delta functions in~\eno{B3Def} eat up the three contour integrals, enforcing the following constraints
\eqn{}{
\gamma_{12} = \Delta_{12,3}  \qquad  
\gamma_{13} = \Delta_{13,2} \qquad 
\gamma_{23} = \Delta_{23,1}  \,.
}
Consequently,
\eqn{B3eval}{
B_3^{\rm s.t.} = {\Gamma(\Delta_{12,3}) \Gamma(\Delta_{13,2}) \Gamma(\Delta_{23,1}) \over (x_{12}^2)^{\Delta_{12,3} } (x_{13}^2)^{\Delta_{13,2} } (x_{23}^2)^{\Delta_{23,1}} } 
 = L(\Delta_1,\Delta_2,\Delta_3) \, \Gamma(\Delta_{12,3}) \Gamma(\Delta_{13,2} )  \Gamma(\Delta_{23,1}) \,,
}
where the leg factor is given by
\eqn{3LegFactors}{
L(\Delta_1,\Delta_2,\Delta_3) = \left({x_{23}^2 \over x_{12}^2 x_{13}^2} \right)^{\Delta_1 \over 2} 
\, \left({x_{13}^2 \over x_{12}^2 x_{23}^2} \right)^{\Delta_2 \over 2} 
\, \left({x_{12}^2 \over x_{13}^2 x_{23}^2} \right)^{\Delta_3 \over 2} \,.
}
It is straightforward to check that this is consistent with~\eno{RulesBMst} for $M=3$:
There are no cross ratios or Gamma-edge factors while the lone Gamma-vertex factor is given by \eno{GammaVertexEdge} with $\sigma_i=i$ for $i=1,2,3$, where $m_{i}=0$ since all incident edges on the vertex are external. Since the ${\cal V}$ set is empty (as there are no $v$-type cross ratios), the post-Mellin parameters $\ell_i$ also vanish identically. Thus the Gamma-vertex factor in~\eno{GammaVertexEdge} reproduces the Gamma functions in~\eno{B3eval}, completing the check.

\subsection{Proof by induction}

We will now prove \eqref{RulesBMst} by induction. We have already established the base case above (for $M=3$).  With the initial step verified, the proof by induction proceeds as follows:

\begin{enumerate}
    \item Sec.~\ref{sec:Mminus1ptShifts}: The Mandelstam invariants and conformal dimensions of the $M$-point diagram are re-expressed in terms of shifted Mandelstam invariants and conformal dimensions of the $(M-1)$-point diagram;
    \item Sec.~\ref{sec:Mminus1ptGamma}: After using the Feynman-like rules for the shifted $(M-1)$-point diagram, the edge- and vertex-factors of the corresponding $(M-1)$-point block are obtained;
    \item Sec.~\ref{sec:SingelTraceProj}: With the $(M-1)$-point diagram conveniently written, the $M$-point diagram can be represented in terms of shifted Mandelstam invariants and conformal dimensions, as well as appropriate edge- and vertex-factors, from which it is possible to project out the single-trace contributions to get the $M$-point conformal block.  The resulting answer is split in two parts, a coordinate dependent part and a Gamma function dependent part;
    \item Sec.~\ref{sec:RecovMptCrossRatios}: The coordinate dependent part is then used to recover the proper $M$-point cross ratios appearing in the leg factor and in the conformal block sums;
    \item Sec.~\ref{sec:SetUpMptGamma}:  One product of the resulting coordinate dependent part is thereafter manipulated by introducing extra contour integrals;
    \item Sec.~\ref{sec:FirstBarnes}: The original contour integrals, which combine the Gamma function dependent part and the manipulated product introduced before, are then performed by application of the first Barnes lemma;
    \item Sec.~\ref{sec:RecovEdgeVert}: Finally, the extra contour integrals are trivially performed such that the missing sums on the $v$ cross ratios and the correct $M$-point edge- and vertex-factors are recovered, completing the proof.
\end{enumerate}

We refer the reader to Appendix~\ref{BASECASE} where these steps are carried out for the first two non-trivial cases, corresponding to $M=4$ and $M=5$, for illustrative purposes. In the following, we will proceed in full generality for an arbitrary $M$ in the arbitrary topology associated with the canonical diagram shown in Figure~\ref{fig:nptAdS}.

\subsubsection{Shifted Conformal Dimensions and Mandelstam Invariants of the $(M-1)$-point Diagram}
\label{sec:Mminus1ptShifts}

\begin{figure}[!t]
\centering
\resizebox{12cm}{!}{%
\begin{tikzpicture}[thick]
\begin{scope}
\node at (0,0) {$\mathcal{O}^{\prime}_{2}(x_2)$};
\draw[-] (0.8,0)--(1.8,0);
\draw[-] (1.8,0)--+(60:1) node[circle,draw,above right,minimum size=24pt] {$3^{\prime}$};
\draw[-] (1.8,0)--+(-60:1) node[circle,draw,below right,minimum size=24pt]  {$M^{\prime}$};
\end{scope}
\begin{scope}[xshift=6cm]
\node at (0,0) {$\mathcal{O}_{2}(x_2)$};
\draw[-] (1,0)--(4.5,0);
\draw[-] (2,0)--(2,1) node at (2,1.5) {$\mathcal{O}_{1}(x_1)$};
\draw[-] (3.5,0)--(3.5,1) node[circle,draw,above,minimum size=24pt] {$3$};
\draw[-] (3.5,0)--(4.5,0) node[circle,draw,right,minimum size=24pt]  {\footnotesize{$M$}};
\node at (3,-0.5) {$\mathcal{O}_{k_{1}}$};
\end{scope}
\end{tikzpicture}
}
\caption{The desired $M$-point conformal block (right), in which external (internal) operators $\mathcal{O}_a(x_a)$ ($\mathcal{O}_{k_a}$) have conformal dimensions $\Delta_a$ ($\Delta_{k_a}$) and the $(M-1)$-point conformal block (left) related to \eqref{InductiveBMins1st}, in which external (internal) operators $\mathcal{O}^{\prime}_a(x_a)$ ($\mathcal{O}^{\prime}_{k_a}$) have conformal dimensions $\Delta^{\prime}_a=\Delta_{a}-\lambda_{a}$ for $2\leq a\leq M$ ($\Delta^{\prime}_{k_a}=\Delta_{k_a}-\lambda_{k_a}$ for $2\leq a\leq M-3$). }
\label{FigM-1&M}
\end{figure}
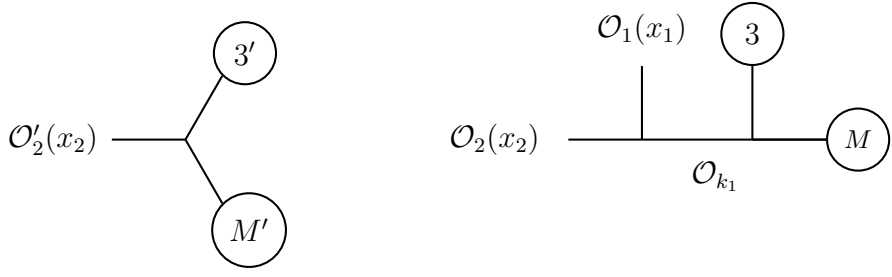

Our first non-trivial step is to relate \eqref{BMins1st} with the corresponding $(M-1)$-point quantity. Since the Mellin parameters $\gamma_{1a}$ should not appear in the $(M-1)$-point AdS diagram, they must be properly absorbed into other quantities. Specifically, to absorb $\gamma_{1a}$ appearing in $\delta(\sum_{\substack{b=1\\b\neq a}}^M \gamma_{ab} - \Delta_a)$, we define the primed conformal dimensions $\Delta^{\prime}_a$ for $2\leq a\leq M$ by 
\eqn{DeltaPrime}{
\Delta_a^{\prime}=\Delta_a-\gamma_{1a}=:\Delta_a-\lambda_a.
}
Thus $s_a$ can be re-expressed as\footnote{Since $J_{k_a}^c$ contains $\mathcal{O}_1(x_1)$ and the latter does not appear in the $(M-1)$-point AdS diagram, we compute $s_a$ from $J_{k_a}$.}
\begin{align}
   \nonumber s_a=&\sum_{b\in J_{k_a}}\Delta_{b}-2\sum_{\substack{b,c\in J_{k_a}\\b<c}}\gamma_{bc}=\sum_{b\in J_{k_a}}\Delta^{\prime}_{b}-2\sum_{\substack{b,c\in J_{k_a}\\b<c}}\gamma_{bc}+\sum_{b\in J_{k_a}}\gamma_{1b}\\\nonumber
       =&s^{\prime}_a+\sum_{b\in J_{k_a}}\gamma_{1b},
\end{align}
where we define 
\begin{align}
    s_a^{\prime}=\sum_{b\in J_{k_a}}\Delta^{\prime}_{b}-2\sum_{\substack{b,c\in J_{k_a}\\b<c}}\gamma_{bc}.
\end{align}
Moreover, we also define 
\eqn{DeltaKPrime}{
\Delta^{\prime}_{k_a}=\Delta_{k_a}-\sum_{b\in J_{k_a}}\gamma_{1b}=:\Delta_{k_a}-\lambda_{k_a}
}
such that $\Delta_{k_a}-s_a=\Delta^{\prime}_{k_a}-s^{\prime}_a$. As a result, all $\gamma_{1a}$ in \eqref{BMins1} can be absorbed into $\Delta^{\prime}_a$ and $\Delta^{\prime}_{k_a}$, leading to
\eqn{BMins1stPrime}{
B^{\text{s.t.}}_{M-1} &= \left[\left(\prod_{(rs) \in {\cal U}^\prime \bigcup {\cal V}^\prime \bigcup {\cal D}^\prime} \int{ d \gamma_{rs} \over 2\pi i} {\Gamma(\gamma_{rs}) \over (x_{rs}^2)^{\gamma_{rs}} } \right)
\left(\prod_{a=2}^M 2\pi i\delta(\sum_{\substack{b=2\\b\neq a}}^M \gamma_{ab} - \Delta^{\prime}_a) \right) 
\left(\prod_{a=2}^{M-3}  {1 \over {\Delta^{\prime}_{k_a}-s^{\prime}_a \over 2} +m_a} \right)\right]_{\text{s.t.}} \!\!.
}
For the inductive step, we assume that $B^{\text{s.t.}}_{M-1}$ satisfies the Feynman-like rules~\eno{RulesBMst}, \textit{i.e.} 
\eqn{InductiveBMins1st}{
B^{\text{s.t.}}_{M-1} &= 
L^{\prime}(\Delta^{\prime}_2,\ldots,\Delta^{\prime}_M)\left(\prod_{a=2}^{M-3}(u_a^{\prime})^{\frac{\Delta^{\prime}_{k_a}}{2}+m_a}\right)\left(\prod_{(rs)\in\mathcal{V}^{\prime}}\sum_{j_{rs}=0}^{\infty}\frac{(1-v^{\prime}_{rs})^{j_{rs}}}{j_{rs}!}\right)\left(\prod_{a=2}^{M-2}\hat{V}^{\prime}_a\right)\left(\prod_{a=2}^{M-3}\hat{E}^{\prime}_a\right)\!,
}
where the $\hat{V}^{\prime}_a$ and $\hat{E}^{\prime}_a$ are “Gamma-vertex” and “Gamma-edge” factors for an $(M-1)$-point conformal blocks of the same topology and labelling as the original $M$-point diagram, except with the $\Delta_1$ external leg removed and with the external (internal) legs having shifted-dimensions $\Delta^{\prime}_a$ for $2\leq a\leq M$ ($\Delta^{\prime}_{k_a}$ for $2\leq a\leq M-3$), as shown in Figure~\ref{fig:AdS} [see also Figure~\ref{FigM-1&M} for the associated $(M-1)$- and $M$-point conformal blocks]. 

\subsubsection{Edge- and Vertex-factors of the $(M-1)$-point Block}
\label{sec:Mminus1ptGamma}

As a consequence, the leg factor $L^{\prime}(\Delta^{\prime}_2,\ldots,\Delta^{\prime}_M)$ of the $(M-1)$-point diagram, built from our rules, has exponents which have linear dependence on $\lambda_a=\gamma_{1a}$ variables. This will be relevant for constructing the additional cross ratios of the desired $M$-point topology. Schematically, the Gamma-vertex and Gamma-edge factors take the form
\eqn{VE}{
\hat{V}^{\prime}_a &=\Gamma(\Delta^{\prime}_{\sigma_1\sigma_2,\sigma_3}+m_{\sigma_1\sigma_2,\sigma_3}+\frac{1}{2}\ell^{\prime}_{\sigma_1\sigma_2,\sigma_3})\Gamma(\Delta^{\prime}_{\sigma_2\sigma_3,\sigma_1}+m_{\sigma_2\sigma_3,\sigma_1}+\frac{1}{2}\ell^{\prime}_{\sigma_2\sigma_3,\sigma_1})\cr
&\qquad\times\Gamma(\Delta^{\prime}_{\sigma_3\sigma_1,\sigma_2}+m_{\sigma_3\sigma_1,\sigma_2}+\frac{1}{2}\ell^{\prime}_{\sigma_3\sigma_1,\sigma_2})\cr
\hat{E}^{\prime}_a&=\frac{1}{\Gamma(\Delta^{\prime}_{k_a}+2m_{a}+\ell^{\prime}_{k_a})},
}
where $\ell^{\prime}_a$ are the post-Mellin parameters of the $(M-1)$-point topology, and $\sigma_1$, $\sigma_2$, $\sigma_3$ label the operators incident on the internal vertex. It is worth noting that $\ell^{\prime}_a$ do not contain any $j_{1b}$ which appear in the post-Mellin parameters $\ell_a$ for the $M$-point block. Moreover, it can be proved that the $\hat{V}^{\prime}_a$ ($\hat{E}^{\prime}_a$), which are not inside the initial comb structure,\footnote{The definition of the initial comb structure is discussed in more detail in Section~\ref{sec:FirstBarnes}.} can be obtained from the corresponding $M$-point quantities $\hat{V}_a$ ($\hat{E}_a$) through replacing all $j_{1b}$ in $\hat{V}_a$ ($\hat{E}_a$) by $-\lambda_b$. To prove this, we first note that the following relations hold 

\eqn{llPrime}{
&\ell_1=\sum_{a=4}^{M}j_{1a}\cr
&\ell_a=\ell^{\prime}_a\quad {\rm for}\quad a=2,3\cr
&\ell_a=\ell^{\prime}_a+j_{1a}=:\ell^{\prime}_a+\hat{j}_a\quad {\rm for} \quad 4\leq a\leq M\,,
}
where the fact that $\mathcal{V}=\mathcal{V}^{\prime}\cup\{(1a):4\leq a\leq M\}$ has been used. Substituting \eqref{llPrime} into the definition 
\eqn{}{
\ell_{k_a}\overset{2\mathcal{J}}{=}\sum_{b\in J_{k_a}}\ell_b
}
where the $2{\cal J}$-equality is defined above~\eno{PostMellinInternal1}, 
leads to
\eqn{}{
\ell_{k_a}\overset{2\mathcal{J}}{=}\sum_{b\in J_{k_a}}j_{1b}+\sum_{b\in J_{k_a}}\ell^{\prime}_b\,,
}
when $3$ is not in $J_{k_a}$. Since all $j_{1b}$ on the right-hand side of the above equality only appear once, we conclude that when $3$ is not in $J_{k_a}$, then $\ell_{k_a}$ can be written as
\eqn{Lk}{
\ell_{k_a}=\sum_{b\in J_{k_a}}j_{1b}+\ell^{\prime}_{k_a}=:\hat{j}_{k_a}+\ell^{\prime}_{k_a}\,.
}
It is worth noting that when $3$ is not in $J_{s_a}$, then $\hat{j}_{s_a}$ can be built from $\lambda_{s_a}$ through replacing all $\lambda_b$ in $\lambda_{s_a}$ by $j_{1b}$.  With the help of \eqref{llPrime} and \eqref{Lk}, the $\hat{V}_i$ and $\hat{E}_i$, which are not inside the initial comb structure, are given by
\eqn{ViEi}{
\hat{V}_a&=\Gamma(\Delta_{\sigma_1\sigma_2,\sigma_3}+m_{\sigma_1\sigma_2,\sigma_3}+\frac{1}{2}\ell_{\sigma_1\sigma_2,\sigma_3})\Gamma(\Delta_{\sigma_2\sigma_3,\sigma_1}+m_{\sigma_2\sigma_3,\sigma_1}+\frac{1}{2}\ell_{\sigma_2\sigma_3,\sigma_1})\cr
&\qquad\times\Gamma(\Delta_{\sigma_3\sigma_1,\sigma_2}+m_{\sigma_3\sigma_1,\sigma_2}+\frac{1}{2}\ell_{\sigma_3\sigma_1,\sigma_2})\cr
&=\Gamma(\Delta_{\sigma_1\sigma_2,\sigma_3}+m_{\sigma_1\sigma_2,\sigma_3}+\frac{1}{2}\ell^{\prime}_{\sigma_1\sigma_2,\sigma_3}+\frac{1}{2}\hat{j}_{\sigma_1\sigma_2,\sigma_3})\Gamma(\Delta_{\sigma_2\sigma_3,\sigma_1}+m_{\sigma_2\sigma_3,\sigma_1}+\frac{1}{2}\ell^{\prime}_{\sigma_2\sigma_3,\sigma_1}+\frac{1}{2}\hat{j}_{\sigma_2\sigma_3,\sigma_1})\cr
&\qquad\times\Gamma(\Delta_{\sigma_3\sigma_1,\sigma_2}+m_{\sigma_3\sigma_1,\sigma_2}+\frac{1}{2}\ell^{\prime}_{\sigma_3\sigma_1,\sigma_2}+\frac{1}{2}\hat{j}_{\sigma_3\sigma_1,\sigma_2})\cr
\hat{E}_a&=\frac{1}{\Gamma(\Delta_{k_a}+2m_{a}+\ell_{k_a})}=\frac{1}{\Gamma(\Delta_{k_a}+2m_{a}+\ell^{\prime}_{k_a}+\hat{j}_{k_a})}\,,
}
where $\sigma_1$, $\sigma_2$, $\sigma_3$ label the operators incident on the internal vertex and we defined $\hat{j}_{\sigma_a\sigma_b,\sigma_c}$ as
\eqn{}{
\hat{j}_{\sigma_a\sigma_b,\sigma_c}=:\hat{j}_{\sigma_a}+\hat{j}_{\sigma_b}-\hat{j}_{\sigma_c}.
}
On the other hand, substituting \eqref{DeltaPrime} and \eqref{DeltaKPrime} into \eqref{VE} leads to
\eqn{}{
\hat{V}^{\prime}_a&=\Gamma(\Delta_{\sigma_1\sigma_2,\sigma_3}+m_{\sigma_1\sigma_2,\sigma_3}+\frac{1}{2}\ell^{\prime}_{\sigma_1\sigma_2,\sigma_3}-\frac{1}{2}\lambda_{\sigma_1\sigma_2,\sigma_3})\Gamma(\Delta_{\sigma_2\sigma_3,\sigma_1}+m_{\sigma_2\sigma_3,\sigma_1}+\frac{1}{2}\ell^{\prime}_{\sigma_2\sigma_3,\sigma_1}-\frac{1}{2}\lambda_{\sigma_2\sigma_3,\sigma_1})\cr
&\qquad\times\Gamma(\Delta_{\sigma_3\sigma_1,\sigma_2}+m_{\sigma_3\sigma_1,\sigma_2}+\frac{1}{2}\ell^{\prime}_{\sigma_3\sigma_1,\sigma_2}-\frac{1}{2}\lambda_{\sigma_3\sigma_1,\sigma_2}),\cr
\hat{E}_a^{\prime}&=\frac{1}{\Gamma(\Delta_{k_a}+2m_{a}+\ell^{\prime}_{k_a}-\lambda_{k_a})}\,.
}
Since $\hat{j}_{\sigma_a}$ are the same as $\lambda_{\sigma_a}$, except with all $\lambda_b$ in $\lambda_{\sigma_a}$ replaced by $j_{1b}$, we can conclude that the $\hat{V}^{\prime}_a$ ($\hat{E}^{\prime}_a$) which are not inside the initial comb structure can be obtained from the corresponding $M$-point quantities $\hat{V}_a$ ($\hat{E}_a$) through replacing all $j_{1b}$ in $\hat{V}_a$ ($\hat{E}_a$) by $-\lambda_b$.

\subsubsection{Turning to the $M$-point Block: Projecting Out the Single-trace Contribution}
\label{sec:SingelTraceProj}

Substituting \eqref{InductiveBMins1st} into \eqref{BMst}, we find that $B^{\text{s.t.}}_M$ can be written as
\eqn{InductiveBMst}{
B^{\text{s.t.}}_M &= \left[\left(\prod_{a=2}^M \int {d\gamma_{1a} \over 2\pi i} {\Gamma(\gamma_{1a}) \over (x_{1a}^2)^{\gamma_{1a}} } \right) 
2\pi i\delta(\sum_{a=2}^M \gamma_{1a} - \Delta_1) 
 {1 \over {\Delta_{k_1}-s_1 \over 2} + m_1}L^{\prime}(\Delta^{\prime}_2,\ldots,\Delta^{\prime}_M)\right. \cr 
&\times
\left.\left(\prod_{a=2}^{M-3}(u_a^{\prime})^{\frac{\Delta^{\prime}_{k_a}}{2}+m_a}\right)\left(\prod_{(rs)\in\mathcal{V}^{\prime}}\sum_{j_{rs}=0}^{\infty}\frac{(1-v^{\prime}_{rs})^{j_{rs}}}{j_{rs}!}\right)\left(\prod_{a=2}^{M-2}\hat{V}^{\prime}_a\right)\left(\prod_{a=2}^{M-3}\hat{E}^{\prime}_a\right)\right]_{\text{s.t.}}.
}
Evaluating the residue at the final single-trace pole $\Delta_{k_1}+2m_1=s_1=\Delta_1+\Delta_2-2\gamma_{12}$ gets essentially all instances of $\gamma_{12}$ replaced with
\eqn{gamma12}{
\gamma_{12}=\Delta_{12,k_1}-m_1\,.
}
Let us also perform the integral over $\gamma_{13}$, the delta function forces all instances of $\gamma_{13}$ to become
\eqn{gamma13}{
\gamma_{13}=\Delta_1-\gamma_{12}-\sum_{a=4}^{M}\gamma_{1a}=\Delta_{1k_1,2}+m_1-\sum_{a=4}^{M}\gamma_{1a}\,,
}
where \eqref{gamma12} has been used to replace $\gamma_{12}$. We define the substitution $\mathcal{S}$ by 
\eqn{Sub}{
\mathcal{S}: \gamma_{12}\rightarrow \Delta_{12,k_1}-m_1\,, \gamma_{13}\rightarrow \Delta_{1k_1,2}+m_1-\sum_{a=4}^{M}\gamma_{1a}\,.
}
As a consequence, $B^{\text{s.t.}}_M$ can be written as
\eqn{BMHM}{
B^{\text{s.t.}}_M &= \left(\prod_{(rs)\in\mathcal{V}^{\prime}}\sum_{j_{rs}=0}^{\infty}\frac{(1-v^{\prime}_{rs})^{j_{rs}}}{j_{rs}!}\right)\left(\prod_{a=4}^M \int {d\gamma_{1a} \over 2\pi i}\right)L^{\prime}(\Delta^{\prime}_2,\ldots,\Delta^{\prime}_M)\cr
&\times\left[\left(\prod_{a=2}^M {\Gamma(\gamma_{1a}) \over (x_{1a}^2)^{\gamma_{1a}} } \right)\left(\prod_{a=2}^{M-3}(u_a^{\prime})^{\frac{\Delta^{\prime}_{k_a}}{2}+m_a}\right)\left(\prod_{a=2}^{M-2}\hat{V}^{\prime}_a\right)\left(\prod_{a=2}^{M-3}\hat{E}^{\prime}_a\right)\right]_{\mathcal{S}}\cr
&=\left(\prod_{(rs)\in\mathcal{V}^{\prime}}\sum_{j_{rs}=0}^{\infty}\frac{(1-v^{\prime}_{rs})^{j_{rs}}}{j_{rs}!}\right)\left(\prod_{a=4}^M \int {d\gamma_{1a} \over 2\pi i}\right)\cr
&\times H_M^{\text{s.t.}}\left[\left(\prod_{a=2}^M \Gamma(\gamma_{1a}) \right)\left(\prod_{a=2}^{M-2}\hat{V}^{\prime}_a\right)\left(\prod_{a=2}^{M-3}\hat{E}^{\prime}_a\right)\right]_{\mathcal{S}},
}
where from the first equality to the second we separated the terms in square bracket into a coordinate dependent part $H_{M}^{\text{s.t.}}$  
\eqn{HMst}{
H_{M}^{\text{s.t.}} :=\left[L^{\prime}(\Delta^{\prime}_2,\ldots,\Delta^{\prime}_M)\left(\prod_{a=2}^M {1 \over (x_{1a}^2)^{\gamma_{1a}} } \right)\left(\prod_{a=2}^{M-3}(u_a^{\prime})^{\frac{\Delta^{\prime}_{k_a}}{2}+m_a}\right)\right]_{\mathcal{S}},
}
and a \text{Gamma} function dependent part.

\subsubsection{Recovering the $M$-point Cross Ratios}
\label{sec:RecovMptCrossRatios}

\begin{figure}[t]
\centering
\resizebox{8cm}{!}{%
\begin{tikzpicture}[thick]
\begin{scope}
\draw[-] (0,0)--+(90:1) node[above]{$\mathcal{O}_{1}(x_{1})$};
\draw[-] (0,0)--+(180:1) node[left]{$\mathcal{O}_{2}(x_{2})$};
\draw[-] (0,0)--+(0:1) node[right]{$\mathcal{O}_{k_{1}}(x_{3})$};
\end{scope}
\end{tikzpicture}
}
\caption{The new $1I$ OPE vertex.}
\label{FigNewOPE}
\end{figure}
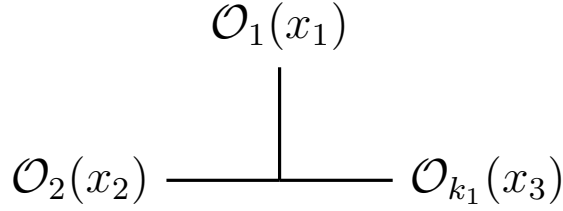

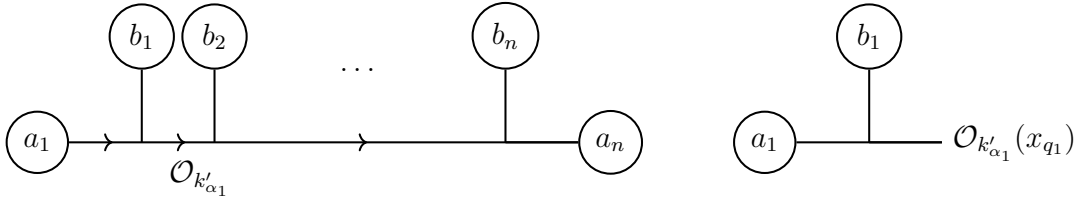
\begin{figure}[!t]
\centering
\resizebox{14.5cm}{!}{%
\begin{tikzpicture}[thick]
\begin{scope}
\node[circle,draw,left,minimum size=24pt] at (1,0) {$a_1$};
\draw[-,postaction={decorate,decoration={markings,mark=at position 0.6cm with {\arrow{>}},mark=at position 1.6cm with {\arrow{>}},mark=at position 4.1cm with {\arrow{>}}}}] (1,0)--(8,0);
\draw[-] (2,0)--(2,1) node[circle,draw,above,minimum size=24pt] {$b_1$};
\node at (2.8,-0.5) {$\mathcal{O}_{k^{\prime}_{\alpha_1}}$};
\draw[-] (3,0)--(3,1) node[circle,draw,above,minimum size=24pt] {$b_2$};
\node at (5,1) {$\dots$};
\draw[-] (7,0)--(7,1) node[circle,draw,above,minimum size=24pt] {$b_n$};
\draw[-] (7,0)--(8,0) node[circle,draw,right,minimum size=24pt] {$a_n$};
\end{scope}
\begin{scope}[xshift=10cm]
\node[circle,draw,left,minimum size=24pt] at (1,0) {$a_1$};
\draw[-] (1,0)--(3,0);
\draw[-] (2,0)--(2,1) node[circle,draw,above,minimum size=24pt] {$b_1$};
\draw[-] (2,0)--(3,0) node at (4,0) {$\mathcal{O}_{k^{\prime}_{\alpha_1}}(x_{q_1})$};
\end{scope}
\end{tikzpicture}
}
\caption{The diagram that defines the cross ratios in \eqref{uProduct} (left) and the OPE vertex (right). Here the arrows represent the fact that $x_{a_1}$ flows all the way to the last vertex on the rightmost side.}
\label{FiguProduct}
\end{figure}

To proceed further, we first prove that $H_{M}^{\text{s.t.}}$ can be rewritten as
\eqn{HMst1}{
H_{M}^{\text{s.t.}}=L(\Delta_1,\ldots,\Delta_M)\left(\prod_{a=1}^{M-3}u_a^{\frac{\Delta_{k_a}}{2}+m_a}\right)\left(\prod_{a=4}^{M}v_{1a}^{-\gamma_{1a}}\right),
}
where $L(\Delta_1,\ldots,\Delta_M)$, $u_a$, and $v_{1a}$ are built from our rules for the leg factor, $u$-type and $v$-type conformal cross ratios of the $M$-point conformal block (see Figure \ref{FigM-1&M}). To prove \eqref{HMst1}, the new OPE vertex which appears when going from the $(M-1)$- to the $M$-point conformal block must be specified. Without loss of generality, we assume that the new $1I$ OPE vertex is as shown in Figure \ref{FigNewOPE}.
From Figures \ref{FigM-1&M} and \ref{FigNewOPE}, one can directly write down the conformal cross ratio $u_1$ per our rules, given by
\eqn{u1}{
u_{1}=\frac{x_{12}^2x_{3M}^2}{x_{13}^2x_{2M}^2}\,.
}
Since the OPE vertices of the $(M-1)$-point conformal block are also OPE vertices of the $M$-point conformal block except with the conformal dimensions being shifted, other $u$-type conformal cross ratios $u_a$ for $2\leq a\leq M-3$ as well as the part of $v$-type conformal cross ratios $v_{rs}$ for $(rs)\in\mathcal{V}^{\prime}$ of the $M$-point conformal block coincide with the corresponding conformal cross ratios $u^{\prime}_a$ and $v^{\prime}_{rs}$ of the $(M-1)$-point conformal block, \textit{i.e.}
\eqn{uPrime=u}{
&u_{a}=u^{\prime}_a\,,\hspace{0.5cm}2\leq a\leq M-3\cr
&v_{rs}=v^{\prime}_{rs}\,,\hspace{0.5cm}(rs)\in\mathcal{V}^{\prime}\,.
} 
Moreover, with the knowledge of the $1I$ OPE vertex Figure \ref{FigNewOPE}, according to our rules, it is easy to check that the following relation between $L^{\prime}(\Delta^{\prime}_2,\ldots,\Delta^{\prime}_M)$ and $L(\Delta_1,\ldots,\Delta_M)$ holds,
\eqn{LPrime=L}{
L^{\prime}(\Delta^{\prime}_2,\ldots,\Delta^{\prime}_M)=X^{-\frac{\Delta_1}{2}}_{23,1}u_1^{\frac{\Delta_2}{2}}L^{\prime}(-\gamma_{12},\ldots,-\gamma_{1M})L(\Delta_1,\ldots,\Delta_M)\,,
}
where \eqref{DeltaPrime} has been used and $X_{ab,c}$ is defined in~\eno{X}. Substituting \eqref{DeltaKPrime}, \eqref{uPrime=u}, and \eqref{LPrime=L} into \eqref{HMst}, we find that proving \eqref{HMst1} is equivalent to proving the following identity 
\eqn{Identity}{
\left[X^{-\frac{\Delta_1}{2}}_{23,1}u_1^{\Delta_2}K^{\prime}\right]_{\mathcal{S}}=u_1^{\frac{\Delta_{k_1}}{2}+m_1}\left(\prod_{a=4}^{M}v_{1a}^{-\gamma_{1a}}\right),
}
where we defined $K^{\prime}$ as
\eqn{KPrime}{
K^{\prime}=L^{\prime}(-\gamma_{12},\ldots,-\gamma_{1M})\left(\prod_{a=2}^M {1 \over (x_{1a}^2)^{\gamma_{1a}} } \right)\left(\prod_{a=2}^{M-3}u_a^{\prime-\frac{\lambda_{k_a}}{2}}\right).
}
To compute $K^{\prime}$, we note that terms in $K^{\prime}$ can be classified according to their powers. Specifically, in $K^{\prime}$ there are three types of terms with powers $\gamma_{12}$, $\gamma_{13}$, and $\gamma_{1a}$ for $4\leq a\leq M$, respectively. To prove \eqref{Identity}, we need to compute each type of terms. Before computing these terms, we introduce the following identity
\eqn{uProduct}{
X_{b_1q_1,a_1}\prod_{i=1}^{n-1}u^{\prime}_{k_{\alpha_i}}=X_{a_nb_n,a_1}=\frac{x^2_{a_nb_n}}{x^2_{a_1a_n}x^2_{a_1b_n}}\,,
}
which will be repeatedly used in our computations. Here, the cross ratios $u^{\prime}_{k_{\alpha_i}}$, which are constructed from the Feynman-like rules, are related to the exchanged operators illustrated in Figure \ref{FiguProduct} and $x_{q_1}$ is the coordinate that appears in the vertex containing $x_{a_1}$ and $x_{b_1}$ with exchange operator $\mathcal{O}^{\prime}_{k_{\alpha_1}}$. 
The identity~\eqref{uProduct} can be easily proved by induction which we do not show here. 

\begin{figure}[t]
\centering
\resizebox{6cm}{!}{%
\begin{tikzpicture}[thick]
\begin{scope}
\draw[-] (0,0)--+(90:1) node[above]{$\mathcal{O}^{\prime}_{\sigma_1}(x_{a_1})$};
\draw[-] (0,0)--+(180:1) node[left]{$\mathcal{O}^{\prime}_{3}(x_{3})$};
\draw[-] (0,0)--+(0:1) node[right]{$\mathcal{O}^{\prime}_{k_{\alpha_1}}(x_{q_{1}})$};
\end{scope}
\end{tikzpicture}
}
\caption{The OPE vertex containing $\mathcal{O}^{\prime}_{3}(x_3)$.}
\label{FigO3OPE}
\end{figure}
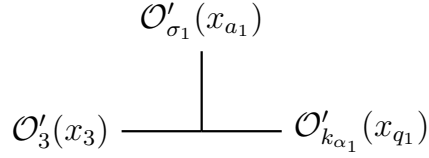

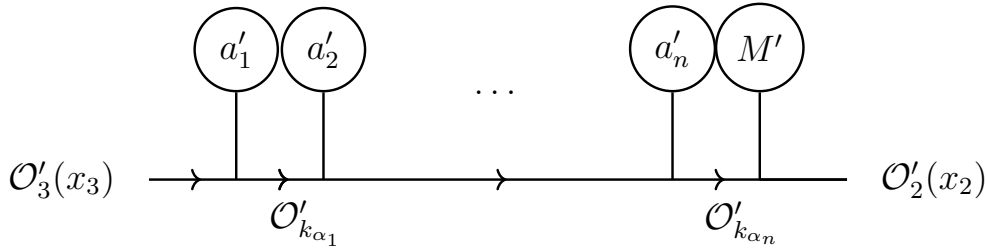
\begin{figure}[!t]
\centering
\resizebox{13.5cm}{!}{%
\begin{tikzpicture}[thick]
\begin{scope}
\node at (0.0,0) {$\mathcal{O}^{\prime}_{3}(x_3)$};
\draw[-,postaction={decorate,decoration={markings,mark=at position 0.6cm with {\arrow{>}},mark=at position 1.6cm with {\arrow{>}},mark=at position 4.1cm with {\arrow{>}},mark=at position 1.6cm with {\arrow{>}},mark=at position 6.6cm with {\arrow{>}}}}] (1,0)--(9,0);
\draw[-] (2,0)--(2,1) node[circle,draw,above,minimum size=24pt] {$a^{\prime}_1$};
\node at (2.8,-0.5) {$\mathcal{O}^{\prime}_{k_{\alpha_1}}$};
\draw[-] (3,0)--(3,1) node[circle,draw,above,minimum size=24pt] {$a^{\prime}_2$};
\node at (5,1) {$\dots$};
\draw[-] (7,0)--(7,1) node[circle,draw,above,minimum size=24pt] {$a^{\prime}_n$};
\draw[-] (8,0)--(8,1) node[circle,draw,above,minimum size=24pt] {$M^{\prime}$};
\node at (7.8,-0.5) {$\mathcal{O}^{\prime}_{k_{\alpha_n}}$};
\draw[-] (8,0)--(9,0) node at (10,0) {$\mathcal{O}^{\prime}_{2}(x_2)$};
\end{scope}
\end{tikzpicture}
}
\caption{The $(M-1)$-point conformal blocks where we expand the circle filled with $3^{\prime}$ in Figure \ref{FigM-1&M}. Here the arrows represent the fact that $x_3$ flows all the way to the last vertex on the rightmost side.}
\label{FigM-1ExpansedU}
\end{figure}

\begin{figure}[!t]
\centering
\resizebox{13.5cm}{!}{%
\begin{tikzpicture}[thick]
\begin{scope}
\node at (0.0,0) {$\mathcal{O}^{\prime}_{a}(x_a)$};
\draw[-] (1,0)--(6,0);
\draw[-,postaction={decorate,decoration={markings,mark=at position 0.6cm with {\arrow{>}},mark=at position 2.1cm with {\arrow{>}},mark=at position 3.6cm with {\arrow{>}}}}] (6,0)--(10,0);
\draw[-] (2,0)--(2,1) node[circle,draw,above,minimum size=24pt] {$a^{\prime}_1$};
\node at (2.8,-0.5) {$\mathcal{O}^{\prime}_{k_{\alpha_1}}$};
\draw[-] (3,0)--(3,1) node[circle,draw,above,minimum size=24pt] {$a^{\prime}_2$};
\node at (4,1) {$\dots$};
\draw[-] (5,0)--(5,1) node[circle,draw,above,minimum size=24pt] {$a^{\prime}_p$};
\node at (5.8,-0.5) {$\mathcal{O}^{\prime}_{k_{\alpha_p}}$};
\draw[-,postaction={decorate,decoration={markings,mark=at position 0.6cm with {\arrow{<}},mark=at position 1.6cm with {\arrow{>>}},mark=at position 2.6cm with {\arrow{>>}}}}] (6,0)--(6,1) node[circle,draw,above,minimum size=24pt] {$3^{\prime}$};
\node at (6.8,-0.5) {$\mathcal{O}^{\prime}_{k_{\alpha_{p+1}}}$};
\draw[-] (7,0)--(7,1) node[circle,draw,above,minimum size=24pt] {$b^{\prime}_1$};
\node at (8,1) {$\dots$};
\draw[-] (9,0)--(9,1) node[circle,draw,above,minimum size=24pt] {$b^{\prime}_q$};
\draw[-] (10,0)--(10,1) node[circle,draw,above,minimum size=24pt] {$M^{\prime}$};
\node at (10,-0.5) {$\mathcal{O}^{\prime}_{k_{\alpha_{p+q+1}}}$};
\draw[-] (10,0)--(11,0) node at (12,0) {$\mathcal{O}^{\prime}_{2}(x_2)$};
\end{scope}
\end{tikzpicture}
}
\caption{The $(M-1)$-point conformal blocks where we expand the circle filled with $3^{\prime}$ in Figure \ref{FigM-1&M}. Here the arrow means that $x_{3}$ flows.}
\label{FigM-1ExpansedV}
\end{figure}
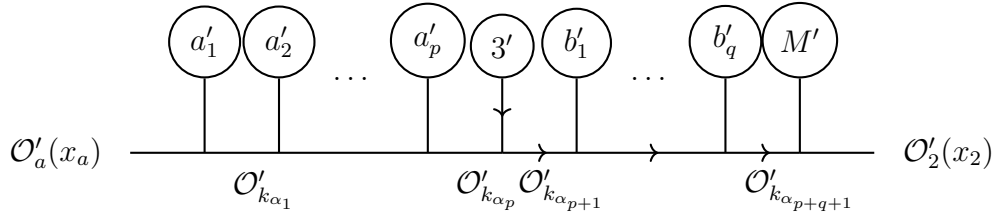

\begin{figure}[!t]
\centering
\resizebox{13.5cm}{!}{%
\begin{tikzpicture}[thick]
\begin{scope}
\node at (0.0,0) {$\mathcal{O}_{k_{\alpha_{r_1-1}}}(x_a)$};
\draw[-] (1.5,0)--(10.5,0);
\draw[-] (2.5,0)--(2.5,1) node[circle,draw,above,minimum size=24pt] {$a_{r_1}$};
\node at (3.5,1) {$\dots$};
\draw[-] (4.5,0)--(4.5,1) node[circle,draw,above,minimum size=24pt] {$3$};
\draw[-] (5.5,0)--(5.5,1) node[circle,draw,above,minimum size=24pt] {$b_1$};
\node at (6.5,1) {$\dots$};
\draw[-] (7.5,0)--(7.5,1) node[circle,draw,above,minimum size=24pt] {$b_q$};
\draw[-] (8.5,0)--(8.5,1) node[circle,draw,above,minimum size=24pt] {$M$};
\draw[-] (9.5,0)--(9.5,1) node at (9.5,1.5) {$\mathcal{O}_2$};
\draw[-] (9.5,0)--(10.5,0) node at (11,0) {$\mathcal{O}_{1}$};
\end{scope}
\end{tikzpicture}
}
\caption{The comb structure which can be used to construct the cross ratio $v_{1j}$ following our rules.}
\label{FigVCombProof}
\end{figure}
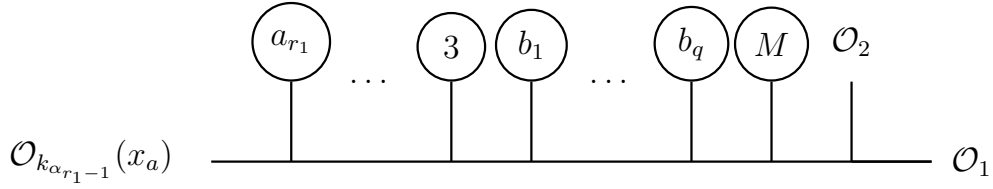

Now, we are ready to compute $K^{\prime}$. Since $K^{\prime}$ only contains quantities from the $(M-1)$-point block, in the following computations we focus on the $(M-1)$-point diagram only. We first extract the terms with powers $\gamma_{12}$ and $\gamma_{13}$. From Figure \ref{FigM-1&M}, we find that the leg factor $L^{\prime}(-\gamma_{12},\ldots,-\gamma_{1M})$ has contribution
\eqn{uProofLegs}{
X_{3M,2}^{-\frac{\gamma_{12}}{2}}X_{a_1q_1,3}^{-\frac{\gamma_{13}}{2}}=\bigg(\frac{x_{3M}^2}{x_{23}^2x_{2M}^2}\bigg)^{-\frac{\gamma_{12}}{2}}\bigg(\frac{x^2_{a_1q_1}}{x^2_{3a_1}x^2_{3q_1}}\bigg)^{-\frac{\gamma_{13}}{2}},
}
where we assume that the vertex involving $\mathcal{O}^{\prime}_{3}(x_3)$ is as depicted in Figure \ref{FigO3OPE}.   
Moreover, due to the fact that $\lambda_{k_a}=\sum_{b\in J_{k_a}}\gamma_{1b}$ with $J_{k_a}\subseteq\{3,\ldots,M\}$, those $u^{\prime}$ cross ratios related to internal lines which directly connect $\mathcal{O}^{\prime}_{3}(x_3)$ and $\mathcal{O}^{\prime}_2(x_2)$ have powers $\gamma_{13}$ (see Figure \ref{FigM-1ExpansedU}). 
In other words, the following terms with powers $\gamma_{13}$ can be extracted
\eqn{ugamma13}{
\prod_{1\leq i\leq n}(u^{\prime}_{\alpha_i})^{-\frac{\gamma_{13}}{2}}=X^{-\frac{\gamma_{13}}{2}}_{2M,3}X^{\frac{\gamma_{13}}{2}}_{a_1q_1,3},
}
where \eqref{uProduct} has been used. Together with $(x^2_{12})^{-\gamma_{12}}(x^2_{13})^{-\gamma_{13}}$, we find that terms with powers $\gamma_{12}$ are given by
\eqn{PowerGamma12}{
X_{3M,2}^{-\frac{\gamma_{12}}{2}}(x^2_{12})^{-\gamma_{12}},
}
while terms with powers $\gamma_{13}$ are given by
\eqn{PowerGamma13}{
X_{2M,3}^{-\frac{\gamma_{13}}{2}}(x^2_{13})^{-\gamma_{13}}.
}
The last group we need to compute includes terms with powers $\gamma_{1a}$ for $4\leq a\leq M$. To compute those terms, we assume that the external operator $\mathcal{O}^{\prime}_{a}(x_{a})$ is inside the circle labeled by $3^{\prime}$ and the vertex containing $\mathcal{O}^{\prime}_{a}(x_a)$ is the left-most OPE vertex of Figure~\ref{FigM-1ExpansedV} with $\mathcal{O}^{\prime}_{k_{\alpha_1}}$ at $x_{q_1}$.\footnote{The following computations can be easily generalized to the case when $\mathcal{O}^{\prime}_a(x_a)$ is inside the circle labeled by $M^{\prime}$.} After expanding the circle labeled by $3^{\prime}$ as depicted in Figure \ref{FigM-1ExpansedV},
we find that terms with powers $\gamma_{1a}$ include:
\begin{enumerate}
\item The leg factor $L^{\prime}_{a}(-\gamma_{1a})$ given by
\eqn{LaGamma1j}{
    L^{\prime}_{a}(-\gamma_{1a})=X_{a_1q_1,a}^{-\frac{\gamma_{1a}}{2}} \,,
}
\item The cross ratios $(u^{\prime}_{\alpha_i})^{-\frac{\lambda_{k_{\alpha_i}}}{2}}$ that relate to the exchanged operators depicted in Figure \ref{FigM-1ExpansedV}, given by
\eqn{uGamma1a}{
    \prod_{i=1}^{p+q+1}(u^{\prime}_{\alpha_i})^{-\frac{\lambda_{k_{\alpha_i}}}{2}}=\prod_{i=1}^{p+q+1}(u^{\prime}_{\alpha_i})^{-\frac{\gamma_{1a}}{2}-\dots},
}
\item And the spacetime coordinates $(x_{1a}^{2})^{-\gamma_{1a}}$. \end{enumerate}
Combining the three contributions, we find that the terms with powers $\gamma_{1a}$ are given by
\begin{align*}
(x^2_{1a})^{-\gamma_{1a}}\bigg(X_{a_1q_1,a}\prod_{i=1}^{p+q+1}u^{\prime}_{\alpha_i}\bigg)^{-\frac{\gamma_{1a}}{2}}. 
\end{align*}
We note that $X_{a_1q_1,a}\prod_{i=1}^{p+q+1}u^{\prime}_{\alpha_i}$ can be computed through repeated use of \eqref{uProduct}. Specifically, we assume that the boundary vertices in Figure \ref{FigM-1ExpansedV} are given by $\bar{T}^{\prime a2}_{a_{r_i}a_{r_{i+1}}p_{i+1}}$ [as described around \eqref{wDef}] for $0\leq i\leq \sigma_a-1$ with the understanding that $a_{r_0}\equiv a$ and $a_{r_{\sigma_a}}\equiv3$. In other words, $x_{a}$ flows until the vertex involving circle $a^{\prime}_{r_1}$ at which point $x_{a_{r_1}}$ starts flowing. Then $x_{a_{r_1}}$ flows until the vertex involving the circle $a^{\prime}_{r_2}$ at which point $x_{a_{r_2}}$ starts flowing.  By continuing this procedure until the last boundary vertex which contains the circle $3^{\prime}$ is reached, we find that
\begin{align*}
    X_{a_1q_1,a}\prod_{i=1}^{p+q+1}u^{\prime}_{\alpha_i}=&X_{a_1q_1,a}\prod_{i=1}^{r_{1}-1}u^{\prime}_{\alpha_{i}}\prod_{b=1}^{\sigma_a-2}\bigg(\prod_{c=r_b}^{r_{b+1}-1}u^{\prime}_{\alpha_{c}}\bigg)\prod_{t=r_{\sigma_a-1}}^{p}u^{\prime}_{\alpha_{t}}\prod_{\ell=p+1}^{p+q+1}u^{\prime}_{\alpha_{\ell}}\\
    =&X_{a_{r_1}p_1,a}\prod_{b=1}^{\sigma_a-2}\bigg(\prod_{c=r_b}^{r_{b+1}-1}u^{\prime}_{\alpha_{c}}\bigg)\prod_{t=r_{\sigma_a-1}}^{p}u^{\prime}_{\alpha_{t}}\prod_{\ell=p+1}^{p+q+1}u^{\prime}_{\alpha_{\ell}}\\
    =&X_{2M,3}\prod_{i=0}^{\sigma_a}\bigg(X_{a_{r_{i}}p_i,a_{r_{i-1}}}X^{-1}_{a_{r_{i-1}}p_{i},a_{r_{i}}}\bigg)\\
    =&X_{2M,3}\bigg(\frac{x^2_{a_{r_1}p_1}}{x^2_{ap_1}}\prod_{i=1}^{\sigma_a-1}w^{\prime a2}_{a_{r_{i}}a_{r_{i+1}}p_{i+1}}\bigg)^2,
\end{align*}
where $w^{ab}_{abc}$ was defined in~\eno{wDef} and \eqref{uProduct} has been used.  Thus the terms with powers $\gamma_{1a}$ for $4\leq a\leq M$ are given by
\eqn{PowerGamma1a}{
(x^2_{1a})^{-\gamma_{1a}}X^{-\frac{\gamma_{1a}}{2}}_{2M,3}\bigg(\frac{x^2_{a_{r_1}p_1}}{x^2_{ap_1}}\prod_{i=1}^{\sigma_a-1}w^{\prime a2}_{a_{r_{i}}a_{r_{i+1}}p_{i+1}}\bigg)^{-\gamma_{1a}}, \hspace{0.5cm}4\leq a\leq M\,.
}
Multiplying \eqref{PowerGamma12}, \eqref{PowerGamma13}, and \eqref{PowerGamma1a} together leads to the final expression for $K^{\prime}$, given by
\eqn{KPrime1}{
K^{\prime}=X_{3M,2}^{-\frac{\gamma_{12}}{2}}(x^2_{12})^{-\gamma_{12}}X_{2M,3}^{-\frac{\gamma_{13}}{2}}(x^2_{13})^{-\gamma_{13}}\left(\prod_{a=4}^{M-4}(x^2_{1a})^{-\gamma_{1a}}X^{-\frac{\gamma_{1a}}{2}}_{2M,3}\bigg(\frac{x^2_{a_{r_1}p_1}}{x^2_{ap_1}}\prod_{i=1}^{\sigma_a-1}w^{\prime a2}_{a_{r_{i}}a_{r_{i+1}}p_{i+1}}\bigg)^{-\gamma_{1a}}\right).
}
After performing the substitution $\mathcal{S}$ \eqref{Sub}, we find that
\eqn{}{
\left[X^{-\frac{\Delta_1}{2}}_{23,1}u_1^{\frac{\Delta_2}{2}}K^{\prime}\right]_{\mathcal{S}}=u_1^{\frac{\Delta_{k_1}}{2}+m_1}\prod_{a=4}^{M-4}\bigg(\frac{x^2_{1a}x^2_{a_{r_1}p_1}}{x^2_{13}x^2_{ap_1}}\prod_{i=1}^{\sigma_a-1}w^{\prime a2}_{a_{r_{i}}a_{r_{i+1}}p_{i+1}}\bigg)^{-\gamma_{1a}}.
}
We note that $\frac{x^2_{1a}x^2_{a_{r_1}p_1}}{x^2_{13}x^2_{ap_1}}\prod_{i=1}^{\sigma_a-1}w^{\prime a2}_{a_{r_{i}}a_{r_{i+1}}p_{i+1}}$ is exactly $v_{1a}$ which can be built from our rules by looking at the comb structure in Figure \ref{FigVCombProof} inside the $M$-point conformal block.
As a consequence, we find that
\eqn{}{
\left[X^{-\frac{\Delta_1}{2}}_{23,1}u_1^{\frac{\Delta_2}{2}}K^{\prime}\right]_{\mathcal{S}}=u_1^{\frac{\Delta_{k_1}}{2}+m_1}\prod_{a=4}^{M-4}v_{1a}^{-\gamma_{1a}},
}
which completes the proof of \eqref{Identity} and thus leads to a proof of \eqref{HMst1}.

\subsubsection{Setting Up to Recover the $M$-point Edge- and Vertex-factors}
\label{sec:SetUpMptGamma}

Substituting \eqref{HMst1} and \eqref{uPrime=u} into \eqref{BMHM}, $B^{\text{s.t.}}_M$ then becomes
\eqn{}{
B^{\text{s.t.}}_M&=L(\Delta_1,\ldots,\Delta_M)\left(\prod_{a=1}^{M-3}u_a^{\frac{\Delta_{k_a}}{2}+m_a}\right)\left(\prod_{(rs)\in\mathcal{V}^{\prime}}\sum_{j_{rs}=0}^{\infty}\frac{(1-v_{rs})^{j_{rs}}}{j_{rs}!}\right)\left(\prod_{a=4}^M \int {d\lambda_{a} \over 2\pi i}\right)\cr
&\times\left(\prod_{a=4}^{M}v_{1a}^{-\lambda_{a}}\right)\left[\left(\prod_{a=2}^M \Gamma(\lambda_{a}) \right)\left(\prod_{a=2}^{M-2}\hat{V}^{\prime}_a\right)\left(\prod_{a=2}^{M-3}\hat{E}^{\prime}_i\right)\right]_{\mathcal{S}},
}
where the fact that $\lambda_{a}=\gamma_{1a}$ for $2\leq a\leq M$ has been used [see \eqref{DeltaPrime}]. We rewrite each factor of $v^{-\lambda_{a}}_{1a}$ above by introducing an additional contour integral, as follows
\eqn{}{
v^{-\lambda_{a}}_{1a}=\frac{1}{\Gamma(\lambda_{a})}\int\frac{d\tilde{\lambda}_{a}}{2\pi i}\Gamma(-\tilde{\lambda}_{a})\Gamma(\lambda_{a}+\tilde{\lambda}_{a})(1-v_{1a})^{\tilde{\lambda}_{a}}\,,
}
where the $\tilde{\lambda}_{a}$ contour runs vertically such that it separates the semi-infinite sequence of poles running to the left and to right of the origin. Then $B^{\text{s.t.}}_M$ can be rewritten as
\eqn{}{
B^{\text{s.t.}}_M&=L(\Delta_1,\ldots,\Delta_M)\left(\prod_{a=1}^{M-3}u_a^{\frac{\Delta_{k_a}}{2}+m_a}\right)\left(\prod_{(rs)\in\mathcal{V}^{\prime}}\sum_{j_{rs}=0}^{\infty}\frac{(1-v_{rs})^{j_{rs}}}{j_{rs}!}\right)\cr
&\times\Gamma(\Delta_{12,k_1}-m_1)\left(\prod_{a=4}^M \int {d\lambda_{a} \over 2\pi i}\int\frac{d\tilde{\lambda}_{a}}{2\pi i}\Gamma(-\tilde{\lambda}_{a})\Gamma(\lambda_{a}+\tilde{\lambda}_{a})(1-v_{1a})^{\tilde{\lambda}_{a}}\right)\cr
&\times\Gamma(\Delta_{1k_1,2}+m_1-\sum_{a=4}^{M}\lambda_{a})\left[\left(\prod_{a=2}^{M-2}\hat{V}^{\prime}_a\right)\left(\prod_{a=2}^{M-3}\hat{E}^{\prime}_a\right)\right]_{\mathcal{S}}.
}
We now focus on the integral over $\lambda_{a}$ for $4\leq a\leq M$ and define $I$ as
\eqn{I}{
I:=\left(\prod_{a=4}^M \int {d\lambda_{a} \over 2\pi i}\Gamma(\lambda_{a}+\tilde{\lambda}_{a})\right)\Gamma(\Delta_{1k_1,2}+m_1-\sum_{a=4}^{M}\lambda_{a})\left[\left(\prod_{a=2}^{M-2}\hat{V}^{\prime}_a\right)\left(\prod_{a=2}^{M-3}\hat{E}^{\prime}_a\right)\right]_{\mathcal{S}},
}
such that
\eqn{BMst=TildeLambdaI}{
B^{\text{s.t.}}_M&=L(\Delta_1,\ldots,\Delta_M)\left(\prod_{a=1}^{M-3}u_a^{\frac{\Delta_{k_a}}{2}+m_a}\right)\left(\prod_{(rs)\in\mathcal{V}^{\prime}}\sum_{j_{rs}=0}^{\infty}\frac{(1-v_{rs})^{j_{rs}}}{j_{rs}!}\right)\cr
&\times\Gamma(\Delta_{12,k_1}-m_1)\left(\prod_{a=4}^M \int\frac{d\tilde{\lambda}_{a}}{2\pi i}\Gamma(-\tilde{\lambda}_{a})(1-v_{1a})^{\tilde{\lambda}_{a}}\right)I.
}
\subsubsection{Applying the First Barnes Lemma}
\label{sec:FirstBarnes}
To prove the Feynman-like rules, our next step is to compute $I$. As we will see later, all $\lambda_{a}$ integrals can be evaluated through repeated use of the first Barnes lemma leaving us with trivial-to-evaluate $\tilde{\lambda}_{a}$ contour integrals. 

Before doing real computations, we note that any $(M-1)$-point topology can be reached by gluing a set of comb structures.\footnote{Since $I$ only involves Gamma-vertices and Gamma-edges of the $(M-1)$-point block, the $(M-1)$-point diagram will be exclusively considered in the computations for $I$.} Specifically, starting with the OPE vertex containing $\mathcal{O}^{\prime}_3(x_3)$, we can glue $2I$ and $3I$ OPE vertices in the proper order until we reach another OPE vertex containing $\mathcal{O}^{\prime}_2(x_2)$, where this procedure stops.  This procedure produces a comb topology, dubbed the ``initial comb structure'', which connects $\mathcal{O}^{\prime}_3(x_3)$ and $\mathcal{O}^{\prime}_2(x_2)$, but some of the teeth of this comb correspond to internal lines that need to be glued further.

For $2I$ OPE vertices included in this initial comb structure, there is nothing further to do since the corresponding tooth represents an external operator. However, this is not the case for $3I$ OPE vertices.  From this initial comb topology, we select one of the $3I$ OPE vertices and repeat the procedure above by gluing $2I$ and $3I$ OPE vertices in the correct order until we reach another $1I$ OPE vertex. To obtain the desired $(M-1)$-point conformal block, we continue this procedure with each additional comb structure until all the $3I$ OPE vertices have been completely glued, \textit{i.e.}\ until the number of $3I$ OPE vertices added in the corresponding comb structure is zero. We dub the comb structure, which is not the initial comb structure and does not contain $3I$ OPE vertices, the ``final comb structures''. We stress that even when the initial comb structure does not contain any $3I$ OPE vertex, \textit{i.e.} the conformal block is in the comb topology, it is not a final comb structure by our definition. 

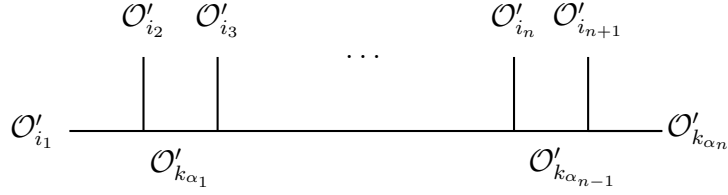
\begin{figure}[!t]
\centering
\resizebox{10cm}{!}{%
\begin{tikzpicture}[thick]
\begin{scope}
\node at (0.0,0) {$\mathcal{O}^{\prime}_{i_1}$};
\draw[-] (0.5,0)--(8.5,0);
\draw[-] (1.5,0)--(1.5,1) node at (1.5,1.5) {$\mathcal{O}^{\prime}_{i_2}$};
\node at (2,-0.5) {$\mathcal{O}^{\prime}_{k_{\alpha_1}}$};
\draw[-] (2.5,0)--(2.5,1) node at (2.5,1.5) {$\mathcal{O}^{\prime}_{i_3}$};
\node at (4.5,1) {$\dots$};
\draw[-] (6.5,0)--(6.5,1) node at (6.5,1.5) {$\mathcal{O}^{\prime}_{i_{n}}$};
\node at (7.3,-0.5) {$\mathcal{O}^{\prime}_{k_{\alpha_{n-1}}}$};
\draw[-] (7.5,0)--(7.5,1) node at (7.5,1.5) {$\mathcal{O}^{\prime}_{i_{n+1}}$};
\draw[-] (7.5,0)--(8.5,0) node at (9,0) {$\mathcal{O}^{\prime}_{k_{\alpha_{n}}}$};
\end{scope}
\end{tikzpicture}
}
\caption{The final comb structure. Here $\mathcal{O}^{\prime}_{i_a}\equiv\mathcal{O}^{\prime}_{i_a}(x_{i_a})$ with $i_a\neq 2, 3$ are internal operators, while $\mathcal{O}^{\prime}_{k_{\alpha_a}}$ are exchange operators.}
\label{FigFinalComb}
\end{figure}

\begin{figure}[!t]
\centering
\resizebox{10cm}{!}{%
\begin{tikzpicture}[thick]
\begin{scope}
\node at (0.0,0) {$\lambda_{i_1}$};
\draw[-] (0.5,0)--(8.5,0);
\draw[-] (1.5,0)--(1.5,1) node at (1.5,1.5) {$\lambda_{i_2}$};
\node at (2,-0.5) {$\lambda_{k_{\alpha_1}}$};
\draw[-] (2.5,0)--(2.5,1) node at (2.5,1.5) {$\lambda_{i_3}$};
\node at (4.5,1) {$\dots$};
\draw[-] (6.5,0)--(6.5,1) node at (6.5,1.5) {$\lambda_{i_{n}}$};
\node at (7.3,-0.5) {$\lambda_{k_{\alpha_{n-1}}}$};
\draw[-] (7.5,0)--(7.5,1) node at (7.5,1.5) {$\lambda_{i_{n+1}}$};
\draw[-] (7.5,0)--(8.5,0) node at (9,0) {$\lambda_{k_{\alpha_{n}}}$};
\end{scope}
\end{tikzpicture}
}
\caption{The $\lambda_{j}$ dependence that comes from $\Delta^{\prime}_{i_a}=\Delta_{i_a}-\lambda_{j}$ as well as $\Delta^{\prime}_{k_{\alpha_a}}=\Delta_{k_{\alpha_a}}-\lambda_{k_{\alpha_a}}$ in the final comb structure Figure \ref{FigFinalComb}.}
\label{FigGamma1a}
\end{figure}
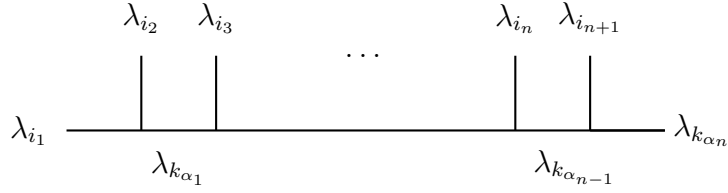

Now, we are ready to compute $I$. Let us start with one final comb structure, depicted in Figure \ref{FigFinalComb}. Since we want to evaluate the $\lambda_{a}$ integral, we also depict the $\lambda_{a}$ appearing in this final comb structure in Figure \ref{FigGamma1a}. 
There are $n$ Gamma-vertices $\hat{V}^{\prime}_a$ and $n-1$ Gamma-edges $\hat{E}^{\prime}_a$ associated with this final comb structure. Without loss of generality, we label the Gamma-vertices from the leftmost side to the rightmost side in Figure \ref{FigFinalComb} by $\hat{V}^{\prime}_{i_1}$, $\hat{V}^{\prime}_{i_2}$, to $\hat{V}^{\prime}_{i_{n}}$. With these conventions, using \eqref{VE}, we find that
\eqn{VPrime2}{
\hat{V}^{\prime}_{i_1}&=\Gamma(\Delta^{\prime}_{i_1i_2,k_{\alpha_1}}-m_{\alpha_1}+\frac{1}{2}\ell^{\prime}_{i_1i_2,k_{\alpha_1}})\Gamma(\Delta^{\prime}_{i_2k_{\alpha_1},i_1}+m_{\alpha_1}+\frac{1}{2}\ell^{\prime}_{i_2k_{\alpha_1},i_1})\Gamma(\Delta^{\prime}_{k_{\alpha_1}i_1,i_2}+m_{\alpha_1}+\frac{1}{2}\ell^{\prime}_{k_{\alpha_1}i_1,i_2})\cr
&=\Gamma(\Delta_{i_1i_2,k_{\alpha_1}}-m_{\alpha_1}+\frac{1}{2}\ell^{\prime}_{i_1i_2,k_{\alpha_1}})\Gamma(\Delta_{i_2k_{\alpha_1},i_1}-\lambda_{i_2}+m_{\alpha_1}+\frac{1}{2}\ell^{\prime}_{i_2k_{\alpha_1},i_1})\cr
&\qquad\qquad\times\Gamma(\Delta_{k_{\alpha_1}i_1,i_2}-\lambda_{i_1}+m_{\alpha_1}+\frac{1}{2}\ell^{\prime}_{k_{\alpha_1}i_1,i_2})\cr
\hat{E}^{\prime}_{\alpha_1}&=\frac{1}{\Gamma(\Delta^{\prime}_{k_{\alpha_1}}+2m_{\alpha_1}+\ell^{\prime}_{k_{\alpha_1}})}=\frac{1}{\Gamma(\Delta_{k_{\alpha_1}}-\lambda_{k_{\alpha_1}}+2m_{\alpha_1}+\ell^{\prime}_{k_{\alpha_1}})}\cr
&=\frac{1}{\Gamma(\Delta_{k_{\alpha_1}}-\lambda_{i_1}-\lambda_{i_2}+2m_{\alpha_1}+\ell^{\prime}_{k_{\alpha_1}})},
}
where we used the relations
\eqn{}{
&\Delta^{\prime}_{i_1}=\Delta_{i_1}-\lambda_{i_1},\cr
&\Delta^{\prime}_{i_2}=\Delta_{i_2}-\lambda_{i_2},\cr
&\Delta^{\prime}_{k_{\alpha_1}}=\Delta_{k_{\alpha_1}}-\lambda_{k_{\alpha_1}}=\Delta_{k_{\alpha_1}}-\lambda_{i_1}-\lambda_{1i_2}.
}
We note that since $\mathcal{O}^{\prime}_{i_1}$ and $\mathcal{O}^{\prime}_{i_2}$ dwell at the same $1I$ OPE vertex, $\lambda_{k_a}$ for $1\leq a\leq M-3$ either do not contain $\lambda_{i_1}$ and $\lambda_{i_2}$, or contain $\lambda_{k_{\alpha_1}}=\lambda_{i_1}+\lambda_{i_2}$. Thus, after changing the $\lambda_{i_2}$ integral to the $\lambda_{k_{\alpha_1}}$ integral by using 
\eqn{NewVariable}{
\lambda_{i_2}=\lambda_{k_{\alpha_1}}-\lambda_{i_1},
}
the terms in $I$ which contain $\lambda_{i_1}$ are given by
\eqn{}{
&\Gamma(\tilde{\lambda}_{i_1}+\lambda_{i_1})\Gamma(\Delta_{i_2k_{\alpha_1},i_1}-\lambda_{k_{\alpha_1}}+\lambda_{i_1}+m_{\alpha_1}+\frac{1}{2}\ell^{\prime}_{i_2k_{\alpha_1},i_1})\cr
&\qquad\times\Gamma(\tilde{\lambda}_{i_2}+\lambda_{k_{\alpha_1}}-\lambda_{i_1})\Gamma(\Delta_{k_{\alpha_1}i_1,i_2}-\lambda_{i_1}+m_{\alpha_1}+\frac{1}{2}\ell^{\prime}_{k_{\alpha_1}i_1,i_2}).
}
With the help of the first Barnes lemma
\eqn{FirstBarnes}{
\int_{-i\infty}^{+i\infty}\frac{d\lambda}{2\pi i}\Gamma(a_1+\lambda)\Gamma(a_2+\lambda)\Gamma(b_1-\lambda)\Gamma(b_2-\lambda)=\frac{\Gamma(a_1+b_1)\Gamma(a_1+b_2)\Gamma(a_2+b_1)\Gamma(a_2+b_2)}{\Gamma(a_1+a_2+b_1+b_2)},
}
we can evaluate the integral over $\lambda_{i_1}$, leading to
\eqn{Lambdai1}{
&\Gamma(\tilde{\lambda}_{k_{\alpha_1}}+\lambda_{k_{\alpha_1}})\Gamma(\Delta_{k_{\alpha_1}i_1,i_2}+\tilde{\lambda}_{i_1}+m_{\alpha_1}+\frac{1}{2}\ell^{\prime}_{k_{\alpha_1}i_1,i_2})\cr
&\times\frac{\Gamma(\Delta_{i_2k_{\alpha_1},i_1}+\tilde{\lambda}_{i_2}+m_{\alpha_1}+\frac{1}{2}\ell^{\prime}_{i_2k_{\alpha_1},i_1})\Gamma(\Delta_{k_{\alpha_1}}-\lambda_{k_{\alpha_1}}+2m_{\alpha_1}+\ell^{\prime}_{k_{\alpha_1}})}{\Gamma(\Delta_{k_{\alpha_1}}+\tilde{\lambda}_{k_{\alpha_1}}+2m_{\alpha_1}+\ell^{\prime}_{k_{\alpha_1}})},
}
with $\tilde{\lambda}_{k_{\alpha_1}}=\tilde{\lambda}_{i_1}+\tilde{\lambda}_{i_2}$. 

\begin{figure}[!t]
\centering
\resizebox{10cm}{!}{%
\begin{tikzpicture}[thick]
\begin{scope}
\node at (0.0,0) {$\mathcal{O}^{\prime}_{k_{\alpha_1}}$};
\draw[-] (0.5,0)--(8.5,0);
\draw[-] (1.5,0)--(1.5,1) node at (1.5,1.5) {$\mathcal{O}^{\prime}_{i_3}$};
\node at (2,-0.5) {$\mathcal{O}^{\prime}_{k_{\alpha_2}}$};
\draw[-] (2.5,0)--(2.5,1) node at (2.5,1.5) {$\mathcal{O}^{\prime}_{i_4}$};
\node at (4.5,1) {$\dots$};
\draw[-] (6.5,0)--(6.5,1) node at (6.5,1.5) {$\mathcal{O}^{\prime}_{i_{n}}$};
\node at (7.3,-0.5) {$\mathcal{O}^{\prime}_{k_{\alpha_{n-1}}}$};
\draw[-] (7.5,0)--(7.5,1) node at (7.5,1.5) {$\mathcal{O}^{\prime}_{i_{n+1}}$};
\draw[-] (7.5,0)--(8.5,0) node at (9,0) {$\mathcal{O}^{\prime}_{k_{\alpha_{n}}}$};
\end{scope}
\begin{scope}[yshift=-3cm]
\node at (0.0,0) {$\lambda_{k_{\alpha_1}}$};
\draw[-] (0.5,0)--(8.5,0);
\draw[-] (1.5,0)--(1.5,1) node at (1.5,1.5) {$\lambda_{i_3}$};
\node at (2,-0.5) {$\lambda_{k_{\alpha_2}}$};
\draw[-] (2.5,0)--(2.5,1) node at (2.5,1.5) {$\lambda_{i_4}$};
\node at (4.5,1) {$\dots$};
\draw[-] (6.5,0)--(6.5,1) node at (6.5,1.5) {$\lambda_{i_{n}}$};
\node at (7.3,-0.5) {$\lambda_{k_{\alpha_{n-1}}}$};
\draw[-] (7.5,0)--(7.5,1) node at (7.5,1.5) {$\lambda_{i_{n+1}}$};
\draw[-] (7.5,0)--(8.5,0) node at (9,0) {$\lambda_{k_{\alpha_{n}}}$};
\end{scope}
\end{tikzpicture}
}
\caption{The final comb structure after evaluating the $\lambda_{i_1}$ integral.}
\label{FigIntGamma1i1}
\end{figure}
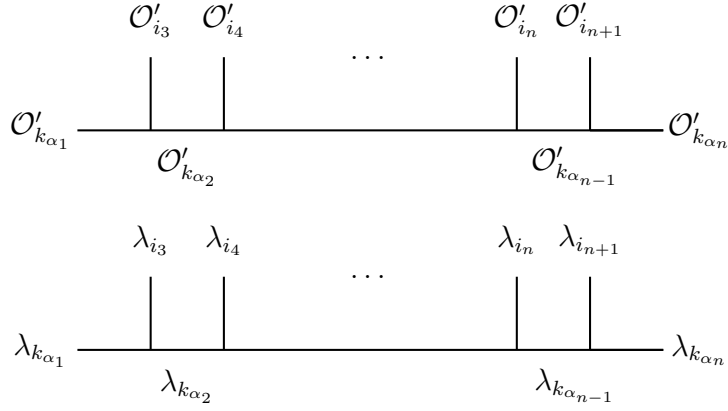

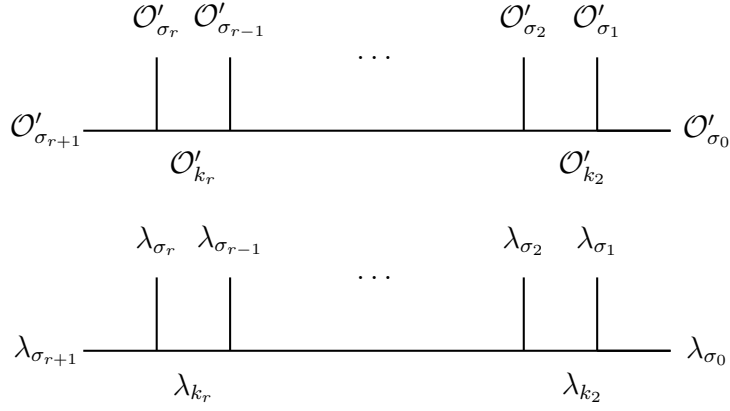
\begin{figure}[!t]
\centering
\resizebox{10cm}{!}{%
\begin{tikzpicture}[thick]
\begin{scope}
\node at (0.0,0) {$\mathcal{O}^{\prime}_{\sigma_{r+1}}$};
\draw[-] (0.5,0)--(8.5,0);
\draw[-] (1.5,0)--(1.5,1) node at (1.5,1.5) {$\mathcal{O}^{\prime}_{\sigma_r}$};
\node at (2,-0.5) {$\mathcal{O}^{\prime}_{k_{r}}$};
\draw[-] (2.5,0)--(2.5,1) node at (2.5,1.5) {$\mathcal{O}^{\prime}_{\sigma_{r-1}}$};
\node at (4.5,1) {$\dots$};
\draw[-] (6.5,0)--(6.5,1) node at (6.5,1.5) {$\mathcal{O}^{\prime}_{\sigma_{2}}$};
\node at (7.3,-0.5) {$\mathcal{O}^{\prime}_{k_{2}}$};
\draw[-] (7.5,0)--(7.5,1) node at (7.5,1.5) {$\mathcal{O}^{\prime}_{\sigma_1}$};
\draw[-] (7.5,0)--(8.5,0) node at (9,0) {$\mathcal{O}^{\prime}_{\sigma_0}$};
\end{scope}
\begin{scope}[yshift=-3cm]
\node at (0.0,0) {$\lambda_{\sigma_{r+1}}$};
\draw[-] (0.5,0)--(8.5,0);
\draw[-] (1.5,0)--(1.5,1) node at (1.5,1.5) {$\lambda_{\sigma_r}$};
\node at (2,-0.5) {$\lambda_{k_{r}}$};
\draw[-] (2.5,0)--(2.5,1) node at (2.5,1.5) {$\lambda_{\sigma_{r-1}}$};
\node at (4.5,1) {$\dots$};
\draw[-] (6.5,0)--(6.5,1) node at (6.5,1.5) {$\lambda_{\sigma_2}$};
\node at (7.3,-0.5) {$\lambda_{k_2}$};
\draw[-] (7.5,0)--(7.5,1) node at (7.5,1.5) {$\lambda_{\sigma_1}$};
\draw[-] (7.5,0)--(8.5,0) node at (9,0) {$\lambda_{\sigma_0}$};
\end{scope}
\end{tikzpicture}
}
\caption{The initial comb structure after removing all other comb structures in the $(M-1)$-point diagram. Here, it should be understood that $\sigma_0\equiv2$, $\sigma_1\equiv M$, and $\sigma_{r+1}\equiv 3$. Other $\mathcal{O}^{\prime}_{\sigma_a}$ can be either external or internal operators.}
\label{FigInitialComb}
\end{figure}

Substituting the above result into $I$ and noting that $\Gamma(\Delta_{k_{\alpha_1}}-\lambda_{k_{\alpha_1}}+2m_{\alpha_1}+\ell^{\prime}_{k_{\alpha_1}})$ in the numerator of \eqref{Lambdai1} cancel $\hat{E}^{\prime}_{\alpha_1}$, we find that $I$ becomes
\eqn{}{
I&=\left(\prod_{\substack{a=4\\a\neq i_1, i_2}}^M \int {d\lambda_{a} \over 2\pi i}\Gamma(\lambda_{a}+\tilde{\lambda}_{a})\right)\int\frac{d\lambda_{k_{\alpha_1}}}{2\pi i}\Gamma(\tilde{\lambda}_{k_{\alpha_1}}+\lambda_{k_{\alpha_1}})\Gamma(\Delta_{1k_1,2}+m_1-\lambda_{k_{\alpha_1}}-\sum_{\substack{a=4\\a\neq i_1,i_2}}^{M}\lambda_{a})\cr
&\times\frac{\Gamma(\Delta_{i_2k_{\alpha_1},i_1}+\tilde{\lambda}_{i_2}+m_{\alpha_1}+\frac{1}{2}\ell^{\prime}_{i_2k_{\alpha_1},i_1})\Gamma(\Delta_{k_{\alpha_1}i_1,i_2}+\tilde{\lambda}_{i_1}+m_{\alpha_1}+\frac{1}{2}\ell^{\prime}_{k_{\alpha_1}i_1,i_2})}{\Gamma(\Delta_{k_{\alpha_1}}+\tilde{\lambda}_{k_{\alpha_1}}+2m_{\alpha_1}+\ell^{\prime}_{k_{\alpha_1}})}\cr
&\times\Gamma(\Delta_{i_1i_2,k_{\alpha_1}}-m_{\alpha_1}+\frac{1}{2}\ell^{\prime}_{i_1i_2,k_{\alpha_1}})\left[\left(\prod_{\substack{a=2\\a\neq i_1}}^{M-2}\hat{V}^{\prime}_a\right)\left(\prod_{\substack{a=2\\a\neq \alpha_1}}^{M-3}\hat{E}^{\prime}_a\right)\right]_{\mathcal{S}}.
}
We define $\tilde{V}_{i_1}$ and $\tilde{E}_{\alpha_1}$ as
\eqn{TildeV2E2}{
\tilde{V}_{i_1}&=\Gamma(\Delta_{i_1i_2,k_{\alpha_1}}-m_{\alpha_1}+\frac{1}{2}\ell^{\prime}_{i_1i_2,k_{\alpha_1}})\Gamma(\Delta_{i_2k_{\alpha_1},i_1}+\tilde{\lambda}_{i_2}+m_{\alpha_1}+\frac{1}{2}\ell^{\prime}_{i_2k_{\alpha_1},i_1})\cr
&\qquad\qquad\times\Gamma(\Delta_{k_{\alpha_1}i_1,i_2}+\tilde{\lambda}_{i_1}+m_{\alpha_1}+\frac{1}{2}\ell^{\prime}_{k_{\alpha_1}i_1,i_2}),\cr
\tilde{E}_{\alpha_1}&=\frac{1}{\Gamma(\Delta_{k_{\alpha_1}}+\tilde{\lambda}_{k_{\alpha_1}}+2m_{\alpha_1}+\ell^{\prime}_{k_{\alpha_1}})}.
}
We note that $\tilde{V}_{i_1}$ ($\tilde{E}_{\alpha_1}$) is the same as $\hat{V}^{\prime}_{i_1}$ ($\hat{E}^{\prime}_{\alpha_1}$), except with all $-\lambda_{i_1}$ and $-\lambda_{i_2}$ in $\hat{V}^{\prime}_{i_1}$ ($\hat{E}^{\prime}_{\alpha_1}$) replaced by $\tilde{\lambda}_{i_1}$ and $\tilde{\lambda}_{i_2}$, respectively. Since the $\hat{V}^{\prime}_{i_1}$ ($\hat{E}^{\prime}_{\alpha_1}$) can be obtained from the corresponding $M$-point quantities $\hat{V}_{i_1}$ ($\hat{E}_{\alpha_1}$) through replacing all $j_{1b}$ in $\hat{V}_{i_1}$ ($\hat{E}_{\alpha_1}$) by $-\lambda_b$, we can also define $\tilde{V}_{i_1}$ and $\tilde{E}_{\alpha_1}$ by
\eqn{}{
&\tilde{V}_{i_1}=:\left(\hat{V}_{i_1}\right)|_{j_{1b}\rightarrow \tilde{\lambda}_b},\cr
&\tilde{E}_{\alpha_1}=\left(\hat{E}_{\alpha_1}\right)|_{j_{1b}\rightarrow \tilde{\lambda}_b}.
}
For future use, we  generalize the above definition and define $\tilde{V}_a$ for $1\leq a\leq M-2$ and $\tilde{E}_a$ for $1\leq a\leq M_3$ by
\eqn{TildeViEj}{
&\tilde{V}_a=\left(\hat{V}_a\right)|_{j_{1b}\rightarrow \tilde{\lambda}_b},\hspace{0.5cm}1\leq a\leq M-2\cr
&\tilde{E}_a=\left(\hat{E}_a\right)|_{j_{1b}\rightarrow \tilde{\lambda}_b},\hspace{0.5cm}1\leq a\leq M-3.
}
As a consequence, $I$ can be written as
\eqn{IGamma1i1}{
I&=\tilde{V}_{i_1}\tilde{E}_{\alpha_1}\left(\prod_{\substack{a=4\\a\neq i_1, i_2}}^M \int {d\lambda_{a} \over 2\pi i}\Gamma(\lambda_{a}+\tilde{\lambda}_{a})\right)\int\frac{d\lambda_{k_{\alpha_1}}}{2\pi i}\Gamma(\tilde{\lambda}_{k_{\alpha_1}}+\lambda_{k_{\alpha_1}})\cr
&\times\Gamma(\Delta_{1k_1,2}+m_1-\lambda_{k_{\alpha_1}}-\sum_{\substack{a=4\\a\neq i_1,i_2}}^{M}\lambda_{a})\left[\left(\prod_{\substack{a=2\\a\neq i_1}}^{M-2}\hat{V}^{\prime}_a\right)\left(\prod_{\substack{a=2\\a\neq \alpha_1}}^{M-3}\hat{E}^{\prime}_a\right)\right]_{\mathcal{S}}.
} 
Comparing \eqref{IGamma1i1} with \eqref{I}, we find that the integrand in \eqref{IGamma1i1} corresponds to the initial integrand but with the leftmost vertex in Figure \ref{FigGamma1a} removed, as shown in Figure \ref{FigIntGamma1i1}.
Now, we can change the $\lambda_{i_3}$ integral to $\lambda_{k_{\alpha_2}}$ integral by noting that
\eqn{}{
\lambda_{i_3}=\lambda_{k_{\alpha_2}}-\lambda_{k_{\alpha_1}},
}
and then evaluate the $\lambda_{k_{\alpha_1}}$ integral. Again, the $\lambda_{k_{\alpha_1}}$ integral just replace all $-\lambda_{k_{\alpha_1}}$ and $-\lambda_{i_3}$ in $\hat{V}^{\prime}_{i_2}$ and $\hat{E}^{\prime}_{\alpha_2}$ by $\tilde{\lambda}_{k_{\alpha_1}}$ and $\tilde{\lambda}_{i_3}$, respectively, leading to
\eqn{}{
I&=\tilde{V}_{i_1}\tilde{V}_{i_2}\tilde{E}_{\alpha_1}\tilde{E}_{\alpha_2}\left(\prod_{\substack{a=4\\a\neq i_1,i_2,i_3}}^M \int {d\lambda_{a} \over 2\pi i}\Gamma(\lambda_{a}+\tilde{\lambda}_{a})\right)\int\frac{d\lambda_{k_{\alpha_2}}}{2\pi i}\Gamma(\tilde{\lambda}_{k_{\alpha_2}}+\lambda_{k_{\alpha_2}})\cr
&\times\Gamma(\Delta_{1k_1,2}+m_1-\lambda_{k_{\alpha_2}}-\sum_{\substack{a=4\\j\neq i_1,i_2,i_3}}^{M}\lambda_{a})\left[\left(\prod_{\substack{a=2\\a\neq i_1,i_2}}^{M-2}\hat{V}^{\prime}_a\right)\left(\prod_{\substack{a=2\\a\neq \alpha_1,\alpha_2}}^{M-3}\hat{E}^{\prime}_a\right)\right]_{\mathcal{S}},
} 
where $\tilde{V}_{i_2}$ and $\tilde{E}_{\alpha_2}$ are defined in \eqref{TildeViEj} and $\tilde{\lambda}_{k_{\alpha_2}}=\sum_{a\in J_{k_{\alpha_2}}}\tilde{\lambda}_{a}=\tilde{\lambda}_{i_1}+\tilde{\lambda}_{i_2}+\tilde{\lambda}_{i_3}$. After performing the $\lambda_{k_{\alpha_2}}$ integral, the leftmost vertex in Figure \ref{FigIntGamma1i1} gets removed. We can continue this procedure until the whole final comb structure in Figure \ref{FigFinalComb} is removed, \textit{i.e.} we evaluate all of the integrals over $\lambda_{i_a}$ for $1\leq a\leq n+1$. Similarly, we can eliminate the remaining final comb structures in the $(M-1)$-point topology. Indeed, after removing all final comb structures in the $(M-1)$-point diagram, a set of new final comb structures emerge in the remaining diagram. Then we repeat the above procedure to remove these new final comb structures. We can keep going on until we reach the initial comb structure, depicted in Figure \ref{FigInitialComb}. 

In Figure \ref{FigInitialComb}, it should be understood that $\mathcal{O}^{\prime}_{\sigma_0}\equiv\mathcal{O}^{\prime}_{2}$, $\mathcal{O}^{\prime}_{\sigma_1}\equiv\mathcal{O}^{\prime}_{M}$, and $\mathcal{O}^{\prime}_{\sigma_{r+1}}\equiv\mathcal{O}^{\prime}_{3}$. Other $\mathcal{O}^{\prime}_{\sigma_i}$ for $1\leq i\leq r$ can be either external or internal operators, depending on the topology of the $(M-1)$ block. We also note that $\lambda_{\sigma_i}=\lambda_{a}$ if $\mathcal{O}^{\prime}_{\sigma_i}=\mathcal{O}^{\prime}_a$ is an external operator, while $\lambda_{\sigma_i}=\sum_{b\in J_{k_a}}\lambda_b$ if $\mathcal{O}^{\prime}_{\sigma_i}=\mathcal{O}^{\prime}_{k_a}$ is an internal operator. Moreover, it is easy to check that the following identities hold for $\lambda_{\sigma_i}$ and $\tilde{\lambda}_{\sigma_i}$
\eqn{SumLambda}{
\sum_{a=1}^r\lambda_{\sigma_a}=\sum_{a=3}^{M}\lambda_a,\cr
\sum_{a=1}^r\tilde{\lambda}_{\sigma_a}=\sum_{a=3}^{M}\tilde{\lambda}_a,
}
where we defined $\tilde{\lambda}_{k_a}$ as
\eqn{}{
\tilde{\lambda}_{k_a}=\sum_{b\in J_{k_a}}\tilde{\lambda}_b.
}

Now, if we label the vertices from the rightmost side to the leftmost side in Figure \ref{FigInitialComb} by $\hat{V}^{\prime}_2$, $\hat{V}^{\prime}_3$, \dots, $\hat{V}^{\prime}_{r}$, then $I$ related to Figure \ref{FigInitialComb} can be written as
\eqn{IInitialComb}{
I&=\left(\prod_{a=r+2}^{M-2}\tilde{V}_a\right)\left(\prod_{a=r+1}^{M-3}\tilde{E}_{a}\right)\left(\prod_{\substack{a=1}}^r \int {d\lambda_{\sigma_a} \over 2\pi i}\Gamma(\lambda_{\sigma_a}+\tilde{\lambda}_{\sigma_a})\right)\cr
&\times\Gamma(\Delta_{1k_1,2}+m_1-\sum_{\substack{a=1}}^{r}\lambda_{\sigma_a})\left[\left(\prod_{a=2}^{r+1}\hat{V}^{\prime}_a\right)\left(\prod_{\substack{a=2}}^{r}\hat{E}^{\prime}_a\right)\right]_{\mathcal{S}},
} 
with $\sigma_0\equiv2$, $\sigma_1\equiv M$, and $\sigma_{r+1}\equiv 3$. We stress that $\tilde{V}_a$ ($\tilde{E}_a$) are obtained by replacing all $-\lambda_b$ and $-\lambda_{k_b}$ in $\hat{V}_a^{\prime}$ ($\hat{E}^{\prime}_a$) by $\tilde{\lambda}_a$ and $\tilde{\lambda}_{k_a}$. From Figure \ref{FigInitialComb}, we can deduce the following relations
\eqn{}{
&\lambda_{k_{a}}=\sum_{b=a}^{r+1}\lambda_{\sigma_{b}}=\Delta_{1k_1,2}+m_1-\sum_{b=1}^{a-1}\lambda_{\sigma_b}, \hspace{0.5cm}2\leq a\leq r,
}
where we used the fact that $\lambda_{\sigma_{r+1}}=\lambda_3=\Delta_{1k_1,2}+m_1-\sum_{a=1}^{r}\lambda_{\sigma_a}$. As a consequence, $\lambda_{\sigma_r}$ does not appear in any $\lambda_{k_a}$ for $2\leq a\leq r$ and the terms in \eqref{IInitialComb} which contain $\lambda_{\sigma_r}$ are given by
\eqn{}{
&\Gamma(\lambda_{\sigma_r}+\tilde{\lambda}_{\sigma_r})\Gamma(\Delta_{1k_1,2}+m_1-\sum_{\substack{a=1}}^{r}\lambda_{\sigma_a})\Gamma(\Delta_{\sigma_rk_r,3}+m_{\sigma_rr,}-\lambda_{\sigma_r}+\frac{1}{2}\ell^{\prime}_{\sigma_rk_r,3})\cr
&\times\Gamma(\Delta_{3k_r,\sigma_r}-\Delta_{1k_1,2}-m_1+\sum_{a=1}^r\lambda_{\sigma_a}+m_{r,\sigma_r}+\frac{1}{2}\ell^{\prime}_{3k_r,\sigma_r})
}
Using the first Barnes lemma \eqref{FirstBarnes}, we can evaluate the $\lambda_{\sigma_r}$ integral, leading to
\eqn{}{
I&=\left(\prod_{a=r+2}^{M-2}\tilde{V}_a\right)\left(\prod_{a=r+1}^{M-3}\tilde{E}_{a}\right)\left(\prod_{\substack{a=1}}^r \int {d\lambda_{\sigma_a} \over 2\pi i}\Gamma(\lambda_{\sigma_a}+\tilde{\lambda}_{\sigma_a})\right)\cr
&\times\Gamma(\Delta_{1k_1,2}+m_1+\tilde{\lambda}_{\sigma_r}-\sum_{a=1}^{r-1}\lambda_{\sigma_a})\left[\left(\prod_{a=2}^{r}\hat{V}^{\prime}_a\right)\left(\prod_{\substack{a=2}}^{r-1}\hat{E}^{\prime}_a\right)\right]_{\mathcal{S}}\cr
&\times\Gamma(\Delta_{3\sigma_r,k_r}+m_{\sigma_r,r}+\frac{1}{2}\ell^{\prime}_{3\sigma_r,k_r})\Gamma(\Delta_{\sigma_rk_r,3}+m_{\sigma_rr,}+\tilde{\lambda}_{\sigma_r}+\frac{1}{2}\ell^{\prime}_{\sigma_rk_r,3})\cr
&\times\frac{\Gamma(\Delta_{3k_r,\sigma_r}+m_{r,\sigma_r}+\frac{1}{2}\ell^{\prime}_{3k_r,\sigma_r})}{\Gamma(\Delta_{k_r}+2m_r+\ell^{\prime}_{k_r}+\tilde{\lambda}_{\sigma_r})}.
}
We note that with the help of \eqref{llPrime} and \eqref{Lk}, $\ell_{k_a}$ for $2\leq a\leq r$ can be evaluated as
\eqn{}{
\ell_{k_a}\overset{2J}{=}\ell^{\prime}_3+\sum_{b=a}^{r}\ell^{\prime}_{\sigma_b}+\sum_{b=a}^r\hat{j}_{\sigma_b},
}
leading to
\eqn{llprimek}{
\ell_{k_a}=\ell^{\prime}_{k_a}+\sum_{b=a}^r\hat{j}_{\sigma_b},\hspace{0.5cm}2\leq a\leq r.
}
As a consequence, we find that
\eqn{}{
&\ell_{3\sigma_r,k_r}=\ell^{\prime}_3+\ell^{\prime}_{\sigma_r}+\hat{j}_{\sigma_r}-\ell^{\prime}_{k_r}-\hat{j}_{\sigma_r}=\ell^{\prime}_{3\sigma_r,k_r},\cr
&\ell_{\sigma_rk_r,3}=\ell^{\prime}_{\sigma_r}+\hat{j}_{\sigma_r}+\ell^{\prime}_{k_r}+\hat{j}_{\sigma_r}+\ell^{\prime}_3=2\hat{j}_{\sigma_r}+\ell^{\prime}_{\sigma_rk_r,3},\cr
&\ell_{3k_r,\sigma_r}=\ell^{\prime}_3+\ell^{\prime}_{k_r}+\hat{j}_{\sigma_r}-\ell^{\prime}_{\sigma_r}-\hat{j}_{\sigma_r}=\ell^{\prime}_{3k_r,\sigma_r},\cr
&\ell_{k_r}=\ell^{\prime}_{k_r}+\hat{j}_{\sigma_r},
}
which implies that
\eqn{}{
&\frac{\Gamma(\Delta_{3\sigma_r,k_r}+m_{\sigma_r,r}+\frac{1}{2}\ell^{\prime}_{3\sigma_r,k_r})\Gamma(\Delta_{\sigma_rk_r,3}+m_{\sigma_rr,}+\tilde{\lambda}_{\sigma_r}+\frac{1}{2}\ell^{\prime}_{\sigma_rk_r,3})\Gamma(\Delta_{3k_r,\sigma_r}+m_{r,\sigma_r}+\frac{1}{2}\ell^{\prime}_{3k_r,\sigma_r})}{\Gamma(\Delta_{k_r}+2m_r+\ell^{\prime}_{k_r}+\tilde{\lambda}_{\sigma_r})}\cr
&=\tilde{V}_{r+1}\tilde{E}_r.
}
Thus, after evaluating the $\lambda_{\sigma_r}$ integral, $I$ becomes
\eqn{}{
I&=\left(\prod_{a=r+1}^{M-2}\tilde{V}_a\right)\left(\prod_{a=r}^{M-3}\tilde{E}_{a}\right)\left(\prod_{\substack{a=1}}^{r-1} \int {d\lambda_{\sigma_a} \over 2\pi i}\Gamma(\lambda_{\sigma_a}+\tilde{\lambda}_{\sigma_a})\right)\cr
&\times\Gamma(\Delta_{1k_1,2}+m_1+\tilde{\lambda}_{\sigma_r}-\sum_{\substack{a=1}}^{r-1}\lambda_{\sigma_a})\left[\left(\prod_{a=2}^{r}\hat{V}^{\prime}_a\right)\left(\prod_{\substack{a=2}}^{r-1}\hat{E}^{\prime}_a\right)\right]_{\mathcal{S}}.
} 
Repeating the above procedure, we can evaluate all other integrals over $\lambda_{\sigma_{a}}$ for $2\leq a\leq r-1$, leading to\footnote{We note that the integrals should be computed in the following order: we first evaluate the $\lambda_{\sigma_{r-1}}$ integral, and then the $\lambda_{\sigma_{r-2}}$ integral, up to the final $\lambda_{\sigma_2}$ integral.}
\eqn{IV2}{
I&=\left(\prod_{a=3}^{M-2}\tilde{V}_a\right)\left(\prod_{a=2}^{M-3}\tilde{E}_{a}\right) \int {d\lambda_{M} \over 2\pi i}\Gamma(\lambda_{M}+\tilde{\lambda}_{M})\cr
&\times\Gamma(\Delta_{1k_1,2}+m_1+\sum_{a=2}^{r}\tilde{\lambda}_{\sigma_a}-\lambda_{M})\left[\hat{V}^{\prime}_2\right]_{\mathcal{S}},
} 
where $\sigma_1\equiv M$ and $\sigma_0\equiv 2$ have been used. To compute the integral over $\lambda_{\sigma_1}\equiv\lambda_M$, we note that
\eqn{}{
\hat{V}^{\prime}_2&=\Gamma(\Delta_{Mk_2,2}-\frac{1}{2}\lambda_{Mk_2,2}+m_{2}+\frac{1}{2}\ell^{\prime}_{Mk_2,2})\Gamma(\Delta_{2M,k_2}-\frac{1}{2}\lambda_{2M,k_2}-m_2+\frac{1}{2}\ell^{\prime}_{2M,k_2})\cr
&\qquad\times\Gamma(\Delta_{2k_2,M}-\frac{1}{2}\lambda_{2k_2,M}+m_2+\frac{1}{2}\ell^{\prime}_{2k_2,M}),
}
where we defined $\lambda_{ab,c}=\lambda_a+\lambda_b-\lambda_c$. Since $\lambda_{k_2}=\sum_{a=3}^{M-1}\lambda_a$, after performing the substitution $\mathcal{S}$, $\hat{V}^{\prime}_2$  becomes
\eqn{}{
\left[\hat{V}^{\prime}_2\right]_{\mathcal{S}}&=\Gamma(\Delta_{Mk_1,k_2}-\lambda_{M}+m_{1,2}+\frac{1}{2}\ell^{\prime}_{2M,k_2})\Gamma(\Delta_{2k_2,1M}+\lambda_{M}+m_2+\frac{1}{2}\ell^{\prime}_{2k_2,M})\cr
&\qquad\qquad\times\Gamma(\Delta_{Mk_2,k_1}+m_{2,1}+\frac{1}{2}\ell^{\prime}_{Mk_2,2}).
}
Thus evaluating the integral over $\lambda_{\sigma_1}\equiv\lambda_{M}$ leads to
\eqn{FinalI}{
I&=\frac{1}{\Gamma(\Delta_{12,k_1}-m_1)}\left(\prod_{a=2}^{M-2}\tilde{V}_a\right)\left(\prod_{a=2}^{M-3}\tilde{E}_{a}\right)\cr
&\times\frac{\Gamma(\Delta_{12,k_1}-m_1)\Gamma(\Delta_{1k_1,2}+m_1+\sum_{a=4}^{M}\tilde{\lambda}_{a})\Gamma(\Delta_{2k_1,1}+m_{1}+\ell^{\prime}_{2})}{\Gamma(\Delta_{k_1}+2m_1+\sum_{a=4}^{M}\tilde{\lambda}_{a}+\ell_2^{\prime})}, 
} 
where \eqref{SumLambda} has been used to get $\sum_{a=4}^M\tilde{\lambda}_{a}$. Using \eqref{llPrime} and noting that
\eqn{}{
\ell_{k_1}\overset{2J}{=}\ell_1+\ell_2=\sum_{a=4}^Mj_{1a}+\ell^{\prime}_2,
}
we find that
\eqn{}{
&\ell_{12,k_1}=0,\cr
&\ell_{1k_1,2}=2\sum_{a=4}^Mj_{1a},\cr
&\ell_{2k_1,1}=2\ell_2=2\ell^{\prime}_2.
}
As a consequence, according to \eqref{TildeViEj}, the second line in \eqref{FinalI} can be rewritten as $\tilde{V}_1\tilde{E}_1$, leading to
\eqn{FinalI1}{
I&=\frac{1}{\Gamma(\Delta_{12,k_1}-m_1)}\left(\prod_{a=1}^{M-2}\tilde{V}_a\right)\left(\prod_{a=1}^{M-3}\tilde{E}_{a}\right).
} 
Substituting \eqref{FinalI1} into \eqref{BMst=TildeLambdaI} leads to
\eqn{BMst=TildeLambda}{
B^{\text{s.t.}}_M&=L(\Delta_1,\ldots,\Delta_M)\left(\prod_{a=1}^{M-3}u_a^{\frac{\Delta_{k_a}}{2}+m_a}\right)\left(\prod_{(rs)\in\mathcal{V}^{\prime}}\sum_{j_{rs}=0}^{\infty}\frac{(1-v_{rs})^{j_{rs}}}{j_{rs}!}\right)\cr
&\times\left(\prod_{a=4}^M \int\frac{d\tilde{\lambda}_{a}}{2\pi i}\Gamma(-\tilde{\lambda}_{a})(1-v_{1a})^{\tilde{\lambda}_{a}}\right)\left(\prod_{a=1}^{M-2}\tilde{V}_a\right)\left(\prod_{a=1}^{M-3}\tilde{E}_{a}\right).
}
\subsubsection{Recovering the $M$-point Edge- and Vertex-factors}
\label{sec:RecovEdgeVert}
Finally, we evaluate the remaining $\tilde{\lambda}_a$ integrals via the Cauchy residue theorem. We close all $\tilde{\lambda}_a$ contours to the right to be able to drop the contribution from the arc at infinity, picking the lone semi-infinite sequence of poles starting at the origin, at $\tilde{\lambda}_a=j_{1a}$ for $j_{1a} \in \mathbb{Z}^{\geq0}$ for each $4\leq a\leq M$. These poles come from the poles of $\Gamma(-\tilde{\lambda}_a)$ in \eqref{BMst=TildeLambda}, and the residues, which are elementary to compute, introduce $M-3$ additional infinite sums such that
\eqn{FinalBMst}{
B^{\text{s.t.}}_M&=L(\Delta_1,\ldots,\Delta_M)\left(\prod_{a=1}^{M-3}u_a^{\frac{\Delta_{k_a}}{2}+m_a}\right)\left(\prod_{(rs)\in\mathcal{V}}\sum_{j_{rs}=0}^{\infty}\frac{(1-v_{rs})^{j_{rs}}}{j_{rs}!}\right)\cr
&\times\left(\prod_{a=1}^{M-2}\left(\tilde{V}_a\right)|_{\tilde{\lambda}_a\rightarrow j_{1a}}\right)\left(\prod_{a=1}^{M-3}\left(\tilde{E}_{a}\right)|_{\tilde{\lambda}_a\rightarrow j_{1a}}\right)\cr
&=L(\Delta_1,\ldots,\Delta_M)\left(\prod_{a=1}^{M-3}u_a^{\frac{\Delta_{k_a}}{2}+m_a}\right)\left(\prod_{(rs)\in\mathcal{V}}\sum_{j_{rs}=0}^{\infty}\frac{(1-v_{rs})^{j_{rs}}}{j_{rs}!}\right)\left(\prod_{a=1}^{M-2}\hat{V}_a\right)\left(\prod_{a=1}^{M-3}\hat{E}_{a}\right),
}
thus completing the proof by induction of the Feynman rules.


\section{Nine-point Scalar Block and its Symmetries}
\label{9-point}

\begin{figure}[h!]
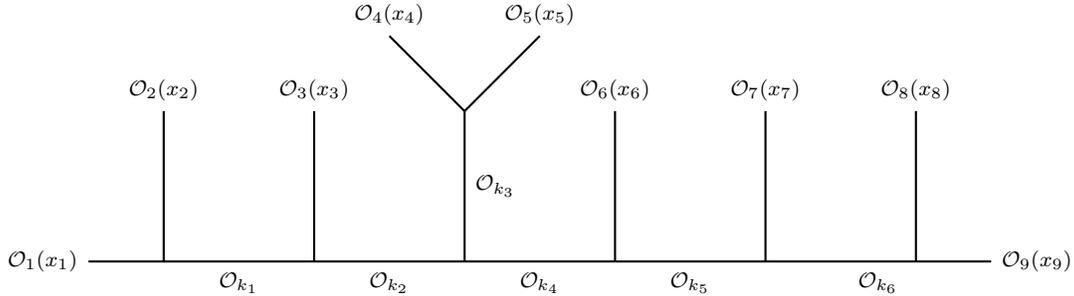

    \centering
     \[  \musepic{\figNineMixedCB} \]
    \caption{9-point asymmetrical mixed conformal block.}
    \label{fig:NineMixedCB}
\end{figure}

In this section we put to use our Feynman-like rules in a concrete, previously unknown higher-point example: the nine-point conformal block for arbitrary external and internal scalars in an asymmetric topology shown in Figure \ref{fig:NineMixedCB}.

\subsection{\texorpdfstring{$u$}{u}-type Cross Ratios}

To illustrate how to use the rules for generating the cross ratios as described in Section~\ref{INDEXCROSSRATIOS} consider the 9-point asymmetrical conformal block with the topology and labeling shown in Figure \ref{fig:NineMixedCB}.  In the case of solving for both the $u$-type cross ratios and the $v$-type cross ratios, our method of generating the cross ratios requires making a flow diagram as laid out in Section \ref{INDEXCROSSRATIOS}. Flow diagrams are not unique, and any diagram that meets the conditions above is valid. 

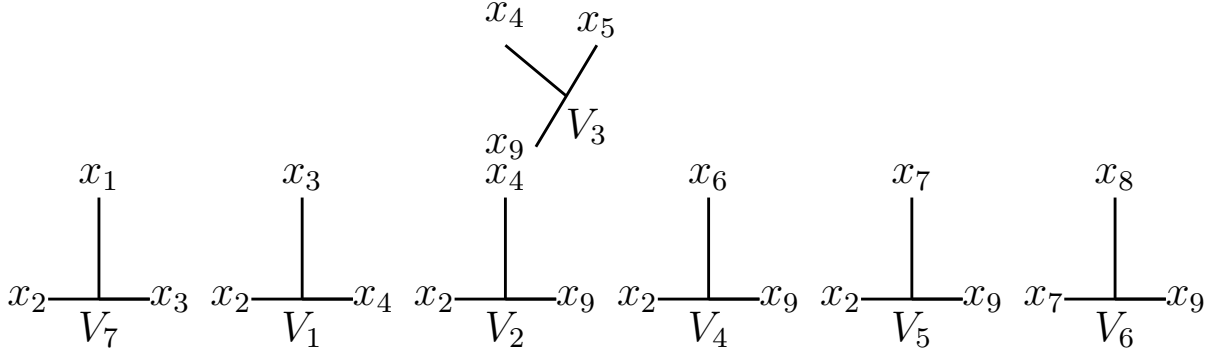
\begin{figure}[h!]
\centering
\resizebox{16.5cm}{!}{%
\begin{tikzpicture}[thick]
\begin{scope}
\node at (-.2,0) {$x_{2}$};
\node at (.5,-.3) {$V_{7}$};
\draw[-] (0.0,0)--(1,0);
\draw[-] (.5,0)--(.5,1) node at (.5,1.2) {$x_{1}$};
\draw[-] (.5,0)--(1,0) node at (1.2,0) {$x_{3}$};
\end{scope}
\begin{scope}[xshift=2cm]
\node at (-.2,0) {$x_{2}$};
\node at (.5,-.3) {$V_{1}$};
\draw[-] (0.0,0)--(1,0);
\draw[-] (.5,0)--(.5,1) node at (.5,1.2) {$x_{3}$};
\draw[-] (.5,0)--(1,0) node at (1.2,0) {$x_{4}$};
\end{scope}
\begin{scope}[xshift=4cm]
\node at (-.2,0) {$x_{2}$};
\node at (.5,-.3) {$V_{2}$};
\draw[-] (0.0,0)--(1,0);
\draw[-] (.5,0)--(.5,1) node at (.5,1.2) {$x_{4}$};
\draw[-] (.5,0)--(1,0) node at (1.2,0) {$x_{9}$};
\end{scope}
\begin{scope}[xshift=4cm]
\node at (.5,1.5) {$x_{9}$};
\node at (1.3,1.7) {$V_{3}$};
\draw[-] (.8,1.5)--(1.4,2.5);
\draw[-] (1.1,2)--(.5,2.5) node at (.5,2.8) {$x_{4}$};
\node at (1.4,2.7) {$x_{5}$};
\end{scope}
\begin{scope}[xshift=6cm]
\node at (-.2,0) {$x_{2}$};
\node at (.5,-.3) {$V_{4}$};
\draw[-] (0.0,0)--(1,0);
\draw[-] (.5,0)--(.5,1) node at (.5,1.2) {$x_{6}$};
\draw[-] (.5,0)--(1,0) node at (1.2,0) {$x_{9}$};
\end{scope}
\begin{scope}[xshift=8cm]
\node at (-.2,0) {$x_{2}$};
\node at (.5,-.3) {$V_{5}$};
\draw[-] (0.0,0)--(1,0);
\draw[-] (.5,0)--(.5,1) node at (.5,1.2) {$x_{7}$};
\draw[-] (.5,0)--(1,0) node at (1.2,0) {$x_{9}$};
\end{scope}
\begin{scope}[xshift=10cm]
\node at (-.2,0) {$x_{7}$};
\node at (.5,-.3) {$V_{6}$};
\draw[-] (0.0,0)--(1,0);
\draw[-] (.5,0)--(.5,1) node at (.5,1.2) {$x_{8}$};
\draw[-] (.5,0)--(1,0) node at (1.2,0) {$x_{9}$};
\end{scope}
\end{tikzpicture}
}
\caption{Flow diagram using rules in Section \ref{CROSSFEYN} for the 9-point asymmetrical block.}
\label{flow diagram 9}
\end{figure}

The first step is choosing an OPE flow.
For this conformal block we use the flow diagram as shown in Figure \ref{flow diagram 9} where we labeled our vertices $V_{i}$ and show the flow by depicting which coordinates continue moving throughout the block. We say a coordinate does not flow when it does not have another comb structure to connect to that contains the same coordinate (for example $x_{1}$ never flows in this diagram). According to Figure \ref{FigU4pt} in order to write down the $u$-type cross ratios, we first need to identify all of the four-point structures within our diagram. The figures below outline the corresponding four-point structures for Figure \ref{fig:NineMixedCB} and the respective $u$-type cross ratios that arise from each structure [see Section~\ref{INDEXCROSSRATIOS} for the general discussion and Sections~\ref{FOURCROSSRATIOS} and \ref{LOWPOINT} for a detailed explanation of the diagrammatic notations that follow]:\footnote{We know that there will be $(M-3)$ $u$-type cross ratios, thus in this case we expect to see $6$ four-point structures that are each used to generate a $u$-type cross ratio. }
\eqn{ucross9}{ 
&
\begin{tikzpicture}[thick]
\begin{scope}
\node at (-.2,0) {$x_{2}$};
\node at (.5,-.3) {$V_{7}$};
\node at (1.6,-0.5) {$\mathcal{O}_{k_{1}}$};
\draw[-] (0.0,0)--(1,0);
\draw[-] (.5,0)--(.5,1) node at (.5,1.2) {$x_{1}$};
\draw[-] (.5,0)--(1,0) node at (1.2,0) {$x_3$};
\draw[black] (.45,1.15) circle (.3 cm);
\draw[dashed,  red]  (.55,1) to[bend right=5] (.8, .1);
\draw[dashed,  teal]  (0,.08) to[bend right=5] (.45,1);
\end{scope}
\begin{scope}[xshift=2cm]
\node at (-.2,0) {$x_2$};
\node at (.5,-.3) {$V_{1}$};
\draw[-] (0.0,0)--(1,0);
\draw[-] (.5,0)--(.5,1) node at (.5,1.2) {$x_{3}$};
\draw[-] (.5,0)--(1,0) node at (1.2,0) {$x_4$};
\draw[black] (1.25,-.05) circle (.3 cm);
\draw[dashed,  red]  (0,-.08) to[bend right=5] (.9,-.08);
\draw[dashed,  teal]  (.55,1) to[bend right=5] (.8, .1);
\end{scope}
\end{tikzpicture}
\quad : u_{1}=\frac{x_{12}^2x_{43}^2}{x_{13}^2 x_{42}^2}
\qquad 
\begin{tikzpicture}[thick]
\begin{scope}
\node at (-.2,0) {$x_{2}$};
\node at (.5,-.3) {$V_{1}$};
\node at (1.6,-0.5) {$\mathcal{O}_{k_{2}}$};
\draw[-] (0.0,0)--(1,0);
\draw[-] (.5,0)--(.5,1) node at (.5,1.2) {$x_{3}$};
\draw[-] (.5,0)--(1,0) node at (1.2,0) {$x_4$};
\draw[black] (.45,1.15) circle (.3 cm);
\draw[dashed,  red]  (.55,1) to[bend right=5] (.8, .1);
\draw[dashed,  teal]  (0,.08) to[bend right=5] (.45,1);
\end{scope}
\begin{scope}[xshift=2cm]
\node at (-.2,0) {$x_2$};
\node at (.5,-.3) {$V_{2}$};
\draw[-] (0.0,0)--(1,0);
\draw[-] (.5,0)--(.5,1) node at (.5,1.2) {$x_{4}$};
\draw[-] (.5,0)--(1,0) node at (1.2,0) {$x_9$};
\draw[black] (1.25,-.05) circle (.3 cm);
\draw[dashed,  red]  (0,-.08) to[bend right=5] (.9,-.08);
\draw[dashed,  teal]  (.55,1) to[bend right=5] (.8, .1);
\end{scope}
\end{tikzpicture}
\quad : u_{2}= \frac{ x_{32}^2x_{94}^2}{x_{34}^2x_{92}^2}
}

\eqn{}{ 
&
\begin{tikzpicture}[thick]
\begin{scope}
\node at (-.2,0) {$x_{2}$};
\node at (.5,-.3) {$V_{2}$};
\node at (1.6,-0.5) {$\mathcal{O}_{k_{3}}$};
\draw[-] (0.0,0)--(1,0);
\draw[-] (.5,0)--(.5,1) node at (.5,1.2) {$x_{9}$};
\draw[-] (.5,0)--(1,0) node at (1.2,0) {$x_4$};
\draw[black] (-.25,-.05) circle (.3 cm);
\draw[dashed,  red]  (0,-.08) to[bend right=5] (.9,-.08);
\draw[dashed,  teal]  (0,.08) to[bend right=5] (.45,1);
\end{scope}
\begin{scope}[xshift=2cm]
\node at (-.2,0) {$x_9$};
\node at (.5,-.3) {$V_{3}$};
\draw[-] (0.0,0)--(1,0);
\draw[-] (.5,0)--(.5,1) node at (.5,1.2) {$x_{4}$};
\draw[-] (.5,0)--(1,0) node at (1.2,0) {$x_5$};
\draw[black] (1.25,-.05) circle (.3 cm);
\draw[dashed,  red]  (0,-.08) to[bend right=5] (.9,-.08);
\draw[dashed,  teal]  (.55,1) to[bend right=5] (.8, .1);
\end{scope}
\end{tikzpicture}
\quad : u_{3}=\frac{x^2_{29 }x^2_{54}} {x_{24}^2x_{59}^2}
\qquad
\begin{tikzpicture}[thick]
\begin{scope}
\node at (-.2,0) {$x_{2}$};
\node at (.5,-.3) {$V_{2}$};
\node at (1.6,-0.5) {$\mathcal{O}_{k_{4}}$};
\draw[-] (0.0,0)--(1,0);
\draw[-] (.5,0)--(.5,1) node at (.5,1.2) {$x_{4}$};
\draw[-] (.5,0)--(1,0) node at (1.2,0) {$x_9$};
\draw[black] (.45,1.15) circle (.3 cm);
\draw[dashed,  red]  (.55,1) to[bend right=5] (.8, .1);
\draw[dashed,  teal]  (0,.08) to[bend right=5] (.45,1);
\end{scope}
\begin{scope}[xshift=2cm]
\node at (-.2,0) {$x_2$};
\node at (.5,-.3) {$V_{4}$};
\draw[-] (0.0,0)--(1,0);
\draw[-] (.5,0)--(.5,1) node at (.5,1.2) {$x_{6}$};
\draw[-] (.5,0)--(1,0) node at (1.2,0) {$x_9$};
\draw[black] (.45,1.15) circle (.3 cm);
\draw[dashed,  red]  (0,.08) to[bend right=5] (.45,1);
\draw[dashed,  teal]  (.55,1) to[bend right=5] (.8, .1);
\end{scope}
\end{tikzpicture}
\quad : u_{4}=\frac{x^2_{42}x^2_{69}}{x_{49}^2x_{62}^2}
}

\eqn{}{ 
&
\begin{tikzpicture}[thick]
\begin{scope}
\node at (-.2,0) {$x_{2}$};
\node at (.5,-.3) {$V_{4}$};
\node at (1.6,-0.5) {$\mathcal{O}_{k_{5}}$};
\draw[-] (0.0,0)--(1,0);
\draw[-] (.5,0)--(.5,1) node at (.5,1.2) {$x_{6}$};
\draw[-] (.5,0)--(1,0) node at (1.2,0) {$x_9$};
\draw[black] (.45,1.15) circle (.3 cm);
\draw[dashed,  red]  (.55,1) to[bend right=5] (.8, .1);
\draw[dashed,  teal]  (0,.08) to[bend right=5] (.45,1);
\end{scope}
\begin{scope}[xshift=2cm]
\node at (-.2,0) {$x_2$};
\node at (.5,-.3) {$V_{5}$};
\draw[-] (0.0,0)--(1,0);
\draw[-] (.5,0)--(.5,1) node at (.5,1.2) {$x_{7}$};
\draw[-] (.5,0)--(1,0) node at (1.2,0) {$x_9$};
\draw[black] (.45,1.15) circle (.3 cm);
\draw[dashed,  red]  (0,.08) to[bend right=5] (.45,1);
\draw[dashed,  teal]  (.55,1) to[bend right=5] (.8, .1);
\end{scope}
\end{tikzpicture}
\quad : u_{5}=\frac{x^2_{62 }x^2_{79}}{x_{69}^2x_{72}^2}
\qquad
\begin{tikzpicture}[thick]
\begin{scope}
\node at (-.2,0) {$x_{2}$};
\node at (.5,-.3) {$V_{5}$};
\node at (1.6,-0.5) {$\mathcal{O}_{k_{6}}$};
\draw[-] (0.0,0)--(1,0);
\draw[-] (.5,0)--(.5,1) node at (.5,1.2) {$x_{7}$};
\draw[-] (.5,0)--(1,0) node at (1.2,0) {$x_9$};
\draw[black] (-.25,-.05) circle (.3 cm);
\draw[dashed,  red]  (0,-.08) to[bend right=5] (.9,-.08);
\draw[dashed,  teal]  (0,.08) to[bend right=5] (.45,1);
\end{scope}
\begin{scope}[xshift=2cm]
\node at (-.2,0) {$x_7$};
\node at (.5,-.3) {$V_{6}$};
\draw[-] (0.0,0)--(1,0);
\draw[-] (.5,0)--(.5,1) node at (.5,1.2) {$x_{8}$};
\draw[-] (.5,0)--(1,0) node at (1.2,0) {$x_9$};
\draw[black] (.45,1.15) circle (.3 cm);
\draw[dashed,  red]  (0,.08) to[bend right=5] (.45,1);
\draw[dashed,  teal]  (.55,1) to[bend right=5] (.8, .1);
\end{scope}
\end{tikzpicture}
\quad : u_{6}=\frac{x^2_{2 7}x^2_{89}}{x_{29}^2x_{87}^2}
}

\subsection{\texorpdfstring{$v$}{v}-type Cross Ratios}

Using our flow diagram Figure \ref{flow diagram 9} we can also write down the $v$-type cross ratios. To do that we first find the comb structures for any $V_{i}V_{j}$ pair. To illustrate the rules explained around Equation~\eno{vCR} consider the resulting $v_{rs}$ cross ratios that correspond to all possible vertex pairings with the vertex labeled $V_7$ in the 9-point block of Figure~\ref{flow diagram 9}. 

\begin{enumerate}
    \item $V_7V_1$: In the following flow diagram $x_1$ and $x_{4}$ do not flow. Thus we will label the cross ratio associated with the $V_{7}V_{1}$ pairing, $v_{14}$. Consequently, a factor of $x_{14}^2$ appears in the numerator of $v_{14}$. To determine the rest of the numerator we circle the $x_{i}$ and $x_{j}$'s connecting the flow diagram. In this case it is only $x_{2}$ and $x_{3}$. Finally, we draw geodesics between the non-flowing legs and assign these pairwise to the denominator, in this case $x_{13}^3$ and $x_{24}^2$,

\eqn{}{ 
\begin{tikzpicture}[thick][r]
\begin{scope}
\node at (-.2,0) {$x_{2}$};
\node at (.5,-.3) {$V_{7}$};
\draw[-] (0.0,0)--(1,0);
\draw[-] (.5,0)--(.5,1) node at (.5,1.2) {$x_{1}$};
\draw[-] (.5,0)--(1,0) node at (1.2,0) {$x_{3}$};
\draw[cyan] (1.5,.1) circle (.6 cm);
\draw[dashed, orange]  (.5,1.1) to[bend right=30] (1,.15);
\end{scope}
\begin{scope}[xshift=2cm]
\node at (-.2,0) {$x_{2}$};
\node at (.5,-.3) {$V_{1}$};
\draw[-] (0.0,0)--(1,0);
\draw[-] (.5,0)--(.5,1) node at (.5,1.2) {$x_{3}$};
\draw[-] (.5,0)--(1,0) node at (1.2,0) {$x_{4}$};
\draw[dashed,  orange]  (0,-.08) to[bend right=5] (.9,-.08);
\end{scope}
\end{tikzpicture}
: V_{7}V_{1}: v_{14}= {x_{14}^2x_{32}^2\over x_{13}^2x_{24}^2}
}

\item $V_7V_2$: As is clear from the following comb structure, $x_{1}$ and $x_{9}$ do not flow. Thus the numerator contains $x_{19}^2$ and the $x_{i}x_{j}$ pairs circled in blue, $x_{23}^2$ and $x_{24}^2$, while the denominator contains the pairs of the geodesic between $x_{1}$ and $x_{9}$, which are $x_{13}^2$, $x_{24}^2$ and $x_{29}^2$. Note that for this block we see a cancellation of the $x_{24}^2$ factor in the final cross ratio. 
\eqn{}{ 
&\begin{tikzpicture}[thick][r]
\begin{scope}
\node at (-.2,0) {$x_{2}$};
\node at (.5,-.3) {$V_{7}$};
\draw[-] (0.0,0)--(1,0);
\draw[-] (.5,0)--(.5,1) node at (.5,1.2) {$x_{1}$};
\draw[-] (.5,0)--(1,0) node at (1.2,0) {$x_{3}$};
\draw[cyan] (1.5,.1) circle (.6 cm);
\draw[dashed,  orange]  (.5,1.1) to[bend right=30] (1,.15);
\end{scope}
\begin{scope}[xshift=2cm]
\node at (-.2,0) {$x_{2}$};
\node at (.5,-.3) {$V_{1}$};
\draw[-] (0.0,0)--(1,0);
\draw[-] (.5,0)--(.5,1) node at (.5,1.2) {$x_{3}$};
\draw[-] (.5,0)--(1,0) node at (1.2,0) {$x_{4}$};
\draw[cyan] (1.5,.1) circle (.6 cm);
\draw[dashed,  orange]  (0,-.08) to[bend right=5] (.9,-.08);
\end{scope}
\begin{scope}[xshift=4cm]
\node at (-.2,0) {$x_{2}$};
\node at (.5,-.3) {$V_{2}$};
\draw[-] (0.0,0)--(1,0);
\draw[-] (.5,0)--(.5,1) node at (.5,1.2) {$x_{4}$};
\draw[-] (.5,0)--(1,0) node at (1.2,0) {$x_{9}$};
\draw[dashed,  orange]  (0,-.08) to[bend right=5] (.9,-.08);
\end{scope}
\end{tikzpicture}
\qquad
: 
V_{7}V_{2}: v_{19}= {x_{19}^2 x_{32}^2 x_{24}^2 \over x_{13}^2 x_{24}^2 x_{29}^2 }= {x_{19}^2 x_{32}^2 \over x_{13}^2 x_{29}^2}
}

\item $V_7V_3$: $x_{1}$ and $x_{5}$ do not flow. Thus the numerator contains $x_{15}^2$ and the $x_{i}x_{j}$ pairs circled in blue, $x_{23}^2$, $x_{24}^2$ and $x_{49}^2$, while the denominator contains the pairs of the geodesic between $x_{1}$ and $x_{5}$, which are $x_{13}^2$, $x_{24}^2$, $x_{24}^2$ and  $x_{59}^2$. We also see a cancellation of the $x_{24}^2$ factor, though the numerator and denominator of the final simplified cross ratio still contains three $x_{ij}^2$ factors each. 
\eqn{}{ 
\begin{tikzpicture}[thick][r]
\begin{scope}
\node at (-.2,0) {$x_{2}$};
\node at (.5,-.3) {$V_{7}$};
\draw[-] (0.0,0)--(1,0);
\draw[-] (.5,0)--(.5,1) node at (.5,1.2) {$x_{1}$};
\draw[-] (.5,0)--(1,0) node at (1.2,0) {$x_{3}$};
\draw[cyan] (1.5,.1) circle (.6 cm);
\draw[dashed,  orange]  (.5,1.1) to[bend right=30] (1,.15);
\end{scope}
\begin{scope}[xshift=2cm]
\node at (-.2,0) {$x_{2}$};
\node at (.5,-.3) {$V_{1}$};
\draw[-] (0.0,0)--(1,0);
\draw[-] (.5,0)--(.5,1) node at (.5,1.2) {$x_{3}$};
\draw[-] (.5,0)--(1,0) node at (1.2,0) {$x_{4}$};
\draw[cyan] (1.5,.1) circle (.6 cm);
\draw[dashed,  orange]  (0,-.08) to[bend right=5] (.9,-.08);
\end{scope}
\begin{scope}[xshift=4cm]
\node at (-.2,0) {$x_{2}$};
\node at (.5,-.3) {$V_{2}$};
\draw[-] (0.0,0)--(1,0);
\draw[-] (.5,0)--(.5,1) node at (.5,1.2) {$x_{4}$};
\draw[-] (.5,0)--(1,0) node at (1.2,0) {$x_{9}$};
\draw[cyan] (.5,1.3) circle (.5 cm);
\draw[dashed,  orange]  (.1,.15) to[bend right=30] (.5,1.1);
\end{scope}
\begin{scope}[xshift=4cm]
\node at (.5,1.5) {$x_{9}$};
\node at (1.3,1.7) {$V_{3}$};
\draw[-] (.8,1.5)--(1.4,2.5);
\draw[-] (1.1,2)--(.5,2.5) node at (.5,2.8) {$x_{4}$};
\node at (1.4,2.7) {$x_{5}$};
\draw[dashed,  orange]  (.9,1.5) to[bend right=5] (1.2,2.4);
\end{scope}
\end{tikzpicture}
: 
V_{7}V_{3}: v_{15}= {x_{15}^2 x_{32}^2 x_{24}^2 x_{49}^2 \over x_{13}^2 x_{24}^2 x_{24}^2 x_{95}^2 }= {x_{15}^2 x_{32}^2  x_{49}^2 \over x_{13}^2 x_{24}^2 x_{59}^2 }
}

\item $V_7V_4$, $V_7V_5$ and $V_7V_6$: Finally, we finish the rest of the $V_7$ $V_{i}$ possible combinations as follows
\eqn{}{ 
\begin{tikzpicture}[thick][r]
\begin{scope}
\node at (-.2,0) {$x_{2}$};
\node at (.5,-.3) {$V_{7}$};
\draw[-] (0.0,0)--(1,0);
\draw[-] (.5,0)--(.5,1) node at (.5,1.2) {$x_{1}$};
\draw[-] (.5,0)--(1,0) node at (1.2,0) {$x_{3}$};
\draw[cyan] (1.5,.1) circle (.6 cm);
\draw[dashed,  orange]  (.5,1.1) to[bend right=30] (1,.15);
\end{scope}
\begin{scope}[xshift=2cm]
\node at (-.2,0) {$x_{2}$};
\node at (.5,-.3) {$V_{1}$};
\draw[-] (0.0,0)--(1,0);
\draw[-] (.5,0)--(.5,1) node at (.5,1.2) {$x_{3}$};
\draw[-] (.5,0)--(1,0) node at (1.2,0) {$x_{4}$};
\draw[cyan] (1.5,.1) circle (.6 cm);
\draw[dashed,  orange]  (0,-.08) to[bend right=5] (.9,-.08);
\end{scope}
\begin{scope}[xshift=4cm]
\node at (-.2,0) {$x_{2}$};
\node at (.5,-.3) {$V_{2}$};
\draw[-] (0.0,0)--(1,0);
\draw[-] (.5,0)--(.5,1) node at (.5,1.2) {$x_{4}$};
\draw[-] (.5,0)--(1,0) node at (1.2,0) {$x_{9}$};
\draw[cyan] (1.5,.1) circle (.6 cm);
\draw[dashed,  orange]  (0,-.08) to[bend right=5] (.9,-.08);
\end{scope}
\begin{scope}[xshift=6cm]
\node at (-.2,0) {$x_{2}$};
\node at (.5,-.3) {$V_{4}$};
\draw[-] (0.0,0)--(1,0);
\draw[-] (.5,0)--(.5,1) node at (.5,1.2) {$x_{6}$};
\draw[-] (.5,0)--(1,0) node at (1.2,0) {$x_{9}$};
\draw[dashed,  orange]  (.1,.15) to[bend right=30] (.5,1.1);
\end{scope}
\end{tikzpicture}
: V_{7}V_{4}: v_{16}= {x_{16}^2 x_{23}^2 x_{24}^2 x_{29}^2 \over x_{13}^2 x_{24}^2 x_{29}^2 x_{26}^2 }= {x_{16}^2 x_{32}^2 \over x_{13}^2 x_{26}^2 }
}

\eqn{}{ 
\begin{tikzpicture}[thick][r]
\begin{scope}
\node at (-.2,0) {$x_{2}$};
\node at (.5,-.3) {$V_{7}$};
\draw[-] (0.0,0)--(1,0);
\draw[-] (.5,0)--(.5,1) node at (.5,1.2) {$x_{1}$};
\draw[-] (.5,0)--(1,0) node at (1.2,0) {$x_{3}$};
\draw[cyan] (1.5,.1) circle (.6 cm);
\draw[dashed,  orange]  (.5,1.1) to[bend right=30] (1,.15);
\end{scope}
\begin{scope}[xshift=2cm]
\node at (-.2,0) {$x_{2}$};
\node at (.5,-.3) {$V_{1}$};
\draw[-] (0.0,0)--(1,0);
\draw[-] (.5,0)--(.5,1) node at (.5,1.2) {$x_{3}$};
\draw[-] (.5,0)--(1,0) node at (1.2,0) {$x_{4}$};
\draw[cyan] (1.5,.1) circle (.6 cm);
\draw[dashed,  orange]  (0,-.08) to[bend right=5] (.9,-.08);
\end{scope}
\begin{scope}[xshift=4cm]
\node at (-.2,0) {$x_{2}$};
\node at (.5,-.3) {$V_{2}$};
\draw[-] (0.0,0)--(1,0);
\draw[-] (.5,0)--(.5,1) node at (.5,1.2) {$x_{4}$};
\draw[-] (.5,0)--(1,0) node at (1.2,0) {$x_{9}$};
\draw[cyan] (1.5,.1) circle (.6 cm);
\draw[dashed,  orange]  (0,-.08) to[bend right=5] (.9,-.08);
\end{scope}
\begin{scope}[xshift=6cm]
\node at (-.2,0) {$x_{2}$};
\node at (.5,-.3) {$V_{4}$};
\draw[-] (0.0,0)--(1,0);
\draw[-] (.5,0)--(.5,1) node at (.5,1.2) {$x_{6}$};
\draw[-] (.5,0)--(1,0) node at (1.2,0) {$x_{9}$};
\draw[cyan] (1.5,.1) circle (.6 cm);
\draw[dashed,  orange]  (0,-.08) to[bend right=5] (.9,-.08);
\end{scope}
\begin{scope}[xshift=8cm]
\node at (-.2,0) {$x_{2}$};
\node at (.5,-.3) {$V_{5}$};
\draw[-] (0.0,0)--(1,0);
\draw[-] (.5,0)--(.5,1) node at (.5,1.2) {$x_{7}$};
\draw[-] (.5,0)--(1,0) node at (1.2,0) {$x_{9}$};
\draw[dashed,  orange]  (.1,.15) to[bend right=30] (.5,1.1);
\end{scope}
\end{tikzpicture}
\cr 
: V_{7}V_{5}: v_{17}= {x_{17}^2 x_{32}^2 x_{24}^2 x_{92}^2 x_{92}^2 \over x_{13}^2 x_{24}^2 x_{29}^2 x_{29}^2 x_{27}^2 }= {x_{17}^2 x_{32}^2 \over x_{13}^2 x_{27}^2 }
}

\eqn{}{ 
\begin{tikzpicture}[thick][r]
\begin{scope}
\node at (-.2,0) {$x_{2}$};
\node at (.5,-.3) {$V_{7}$};
\draw[-] (0.0,0)--(1,0);
\draw[-] (.5,0)--(.5,1) node at (.5,1.2) {$x_{1}$};
\draw[-] (.5,0)--(1,0) node at (1.2,0) {$x_{3}$};
\draw[cyan] (1.5,.1) circle (.6 cm);
\draw[dashed,  orange]  (.5,1.1) to[bend right=30] (1,.15);
\end{scope}
\begin{scope}[xshift=2cm]
\node at (-.2,0) {$x_{2}$};
\node at (.5,-.3) {$V_{1}$};
\draw[-] (0.0,0)--(1,0);
\draw[-] (.5,0)--(.5,1) node at (.5,1.2) {$x_{3}$};
\draw[-] (.5,0)--(1,0) node at (1.2,0) {$x_{4}$};
\draw[cyan] (1.5,.1) circle (.6 cm);
\draw[dashed,  orange]  (0,-.08) to[bend right=5] (.9,-.08);
\end{scope}
\begin{scope}[xshift=4cm]
\node at (-.2,0) {$x_{2}$};
\node at (.5,-.3) {$V_{2}$};
\draw[-] (0.0,0)--(1,0);
\draw[-] (.5,0)--(.5,1) node at (.5,1.2) {$x_{4}$};
\draw[-] (.5,0)--(1,0) node at (1.2,0) {$x_{9}$};
\draw[cyan] (1.5,.1) circle (.6 cm);
\draw[dashed,  orange]  (0,-.08) to[bend right=5] (.9,-.08);
\end{scope}
\begin{scope}[xshift=6cm]
\node at (-.2,0) {$x_{2}$};
\node at (.5,-.3) {$V_{4}$};
\draw[-] (0.0,0)--(1,0);
\draw[-] (.5,0)--(.5,1) node at (.5,1.2) {$x_{6}$};
\draw[-] (.5,0)--(1,0) node at (1.2,0) {$x_{9}$};
\draw[cyan] (1.5,.1) circle (.6 cm);
\draw[dashed,  orange]  (0,-.08) to[bend right=5] (.9,-.08);
\end{scope}
\begin{scope}[xshift=8cm]
\node at (-.2,0) {$x_{2}$};
\node at (.5,-.3) {$V_{5}$};
\draw[-] (0.0,0)--(1,0);
\draw[-] (.5,0)--(.5,1) node at (.5,1.2) {$x_{7}$};
\draw[-] (.5,0)--(1,0) node at (1.2,0) {$x_{9}$};
\draw[cyan] (1.5,.1) circle (.6 cm);
\draw[dashed,  orange]  (0,-.08) to[bend right=5] (.9,-.08);
\end{scope}
\begin{scope}[xshift=10cm]
\node at (-.2,0) {$x_{7}$};
\node at (.5,-.3) {$V_{6}$};
\draw[-] (0.0,0)--(1,0);
\draw[-] (.5,0)--(.5,1) node at (.5,1.2) {$x_{8}$};
\draw[-] (.5,0)--(1,0) node at (1.2,0) {$x_{9}$};
\draw[dashed,  orange]  (.1,.15) to[bend right=30] (.5,1.1);
\end{scope}
\end{tikzpicture}
\cr 
: V_{7}V_{6}: v_{18}= {x_{18}^2 x_{32}^2 x_{24}^2 x_{29}^2 x_{29}^2 x_{97}^2 \over x_{13}^2 x_{24}^2 x_{29}^2  x_{29}^2 x_{29}^2 x_{78}^2 }= {x_{18}^2 x_{32}^2 x_{79}^2 \over x_{13}^2  x_{29}^2  x_{78}^2 }
}
\end{enumerate}

Having completed all possible $V_7$ pairings in the 9-point block of Figure~\ref{flow diagram 9}, it remains to find the cross ratios that result from the subset of the flow diagram of the 9-point block without the $V_7$ vertex. But these are precisely the cross ratios of an 8-point block with the choice of OPE flow as in Figure~\ref{flow diagram 9} sans the $V_7$ vertex. 
Indeed, in Section~\ref{PROOF} we proved that with a convenient choice of OPE flows, one can always arrange for the cross ratios of an $M$-point block to form a superset of the cross ratios of the associated $(M-1)$-point block obtained by removing one external leg but preserving the labeling. The remaining 15 cross ratios are given by:

\begingroup\makeatletter\def\f@size{10}\check@mathfonts
\allowdisplaybreaks
\begin{align*} 
\begin{tikzpicture}[thick]
\begin{scope}[xshift=0cm]
\node at (-.2,0) {$x_{2}$};
\node at (.5,-.3) {$V_{1}$};
\draw[-] (0.0,0)--(1,0);
\draw[-] (.5,0)--(.5,1) node at (.5,1.2) {$x_{3}$};
\draw[-] (.5,0)--(1,0) node at (1.2,0) {$x_{4}$};
\draw[cyan] (1.5,.1) circle (.6 cm);
\draw[dashed,  orange]  (.5,1.1) to[bend right=30] (1,.15);
\end{scope}
\begin{scope}[xshift=2cm]
\node at (-.2,0) {$x_{2}$};
\node at (.5,-.3) {$V_{2}$};
\draw[-] (0.0,0)--(1,0);
\draw[-] (.5,0)--(.5,1) node at (.5,1.2) {$x_{4}$};
\draw[-] (.5,0)--(1,0) node at (1.2,0) {$x_{9}$};
\draw[dashed,  orange]  (0,-.08) to[bend right=5] (.9,-.08);
\end{scope}
\end{tikzpicture}
: V_{1}V_{2}=v_{39}= {x_{39}^2x_{24}^2\over x_{34}^2x_{29}^2} 
\\
\begin{tikzpicture}[thick]
\begin{scope}[xshift=0cm]
\node at (-.2,0) {$x_{2}$};
\node at (.5,-.3) {$V_{1}$};
\draw[-] (0.0,0)--(1,0);
\draw[-] (.5,0)--(.5,1) node at (.5,1.2) {$x_{3}$};
\draw[-] (.5,0)--(1,0) node at (1.2,0) {$x_{4}$};
\draw[cyan] (1.5,.1) circle (.6 cm);
\draw[dashed,  orange]  (.5,1.1) to[bend right=30] (1,.15);
\end{scope}
\begin{scope}[xshift=2cm]
\node at (-.2,0) {$x_{2}$};
\node at (.5,-.3) {$V_{2}$};
\draw[-] (0.0,0)--(1,0);
\draw[-] (.5,0)--(.5,1) node at (.5,1.2) {$x_{4}$};
\draw[-] (.5,0)--(1,0) node at (1.2,0) {$x_{9}$};
\draw[cyan] (.5,1.3) circle (.5 cm);
\draw[dashed,  orange]  (.1,.15) to[bend right=30] (.5,1.1);
\end{scope}
\begin{scope}[xshift=2cm]
\node at (.5,1.5) {$x_{9}$};
\node at (1.3,1.7) {$V_{3}$};
\draw[-] (.8,1.5)--(1.4,2.5);
\draw[-] (1.1,2)--(.5,2.5) node at (.5,2.8) {$x_{4}$};
\node at (1.4,2.7) {$x_{5}$};
\draw[dashed,  orange]  (.9,1.5) to[bend right=5] (1.2,2.4);
\end{scope}
\end{tikzpicture}
: V_{1}V_{3}:v_{35}= {x_{35}^2 x_{49}^2 \over x_{34}^2 x_{59}^2 }
\\
\begin{tikzpicture}[thick]
\begin{scope}[xshift=2cm]
\node at (-.2,0) {$x_{2}$};
\node at (.5,-.3) {$V_{1}$};
\draw[-] (0.0,0)--(1,0);
\draw[-] (.5,0)--(.5,1) node at (.5,1.2) {$x_{3}$};
\draw[-] (.5,0)--(1,0) node at (1.2,0) {$x_{4}$};
\draw[cyan] (1.5,.1) circle (.6 cm);
\draw[dashed,  orange]  (.5,1.1) to[bend right=30] (1,.15);
\end{scope}
\begin{scope}[xshift=4cm]
\node at (-.2,0) {$x_{2}$};
\node at (.5,-.3) {$V_{2}$};
\draw[-] (0.0,0)--(1,0);
\draw[-] (.5,0)--(.5,1) node at (.5,1.2) {$x_{4}$};
\draw[-] (.5,0)--(1,0) node at (1.2,0) {$x_{9}$};
\draw[cyan] (1.5,.1) circle (.6 cm);
\draw[dashed,  orange]  (0,-.08) to[bend right=5] (.9,-.08);
\end{scope}
\begin{scope}[xshift=6cm]
\node at (-.2,0) {$x_{2}$};
\node at (.5,-.3) {$V_{4}$};
\draw[-] (0.0,0)--(1,0);
\draw[-] (.5,0)--(.5,1) node at (.5,1.2) {$x_{6}$};
\draw[-] (.5,0)--(1,0) node at (1.2,0) {$x_{9}$};
\draw[dashed,  orange]  (.1,.15) to[bend right=30] (.5,1.1);
\end{scope}
\end{tikzpicture}
: V_{1}V_{4}: v_{36}= {x_{36}^2 x_{24}^2 \over x_{34}^2 x_{26}^2 }
\\
\begin{tikzpicture}[thick]
\begin{scope}[xshift=2cm]
\node at (-.2,0) {$x_{2}$};
\node at (.5,-.3) {$V_{1}$};
\draw[-] (0.0,0)--(1,0);
\draw[-] (.5,0)--(.5,1) node at (.5,1.2) {$x_{3}$};
\draw[-] (.5,0)--(1,0) node at (1.2,0) {$x_{4}$};
\draw[cyan] (1.5,.1) circle (.6 cm);
\draw[dashed,  orange]  (.5,1.1) to[bend right=30] (1,.15);
\end{scope}
\begin{scope}[xshift=4cm]
\node at (-.2,0) {$x_{2}$};
\node at (.5,-.3) {$V_{2}$};
\draw[-] (0.0,0)--(1,0);
\draw[-] (.5,0)--(.5,1) node at (.5,1.2) {$x_{4}$};
\draw[-] (.5,0)--(1,0) node at (1.2,0) {$x_{9}$};
\draw[cyan] (1.5,.1) circle (.6 cm);
\draw[dashed,  orange]  (0,-.08) to[bend right=5] (.9,-.08);
\end{scope}
\begin{scope}[xshift=6cm]
\node at (-.2,0) {$x_{2}$};
\node at (.5,-.3) {$V_{4}$};
\draw[-] (0.0,0)--(1,0);
\draw[-] (.5,0)--(.5,1) node at (.5,1.2) {$x_{6}$};
\draw[-] (.5,0)--(1,0) node at (1.2,0) {$x_{9}$};
\draw[cyan] (1.5,.1) circle (.6 cm);
\draw[dashed,  orange]  (0,-.08) to[bend right=5] (.9,-.08);
\end{scope}
\begin{scope}[xshift=8cm]
\node at (-.2,0) {$x_{2}$};
\node at (.5,-.3) {$V_{5}$};
\draw[-] (0.0,0)--(1,0);
\draw[-] (.5,0)--(.5,1) node at (.5,1.2) {$x_{7}$};
\draw[-] (.5,0)--(1,0) node at (1.2,0) {$x_{9}$};
\draw[dashed,  orange]  (.1,.15) to[bend right=30] (.5,1.1);
\end{scope}
\end{tikzpicture}
: V_{1}V_{5}: v_{37}= {x_{37}^2 x_{24}^2 \over x_{34}^2 x_{27}^2 }
\\
\begin{tikzpicture}[thick]
\begin{scope}[xshift=2cm]
\node at (-.2,0) {$x_{2}$};
\node at (.5,-.3) {$V_{1}$};
\draw[-] (0.0,0)--(1,0);
\draw[-] (.5,0)--(.5,1) node at (.5,1.2) {$x_{3}$};
\draw[-] (.5,0)--(1,0) node at (1.2,0) {$x_{4}$};
\draw[cyan] (1.5,.1) circle (.6 cm);
\draw[dashed,  orange]  (.5,1.1) to[bend right=30] (1,.15);
\end{scope}
\begin{scope}[xshift=4cm]
\node at (-.2,0) {$x_{2}$};
\node at (.5,-.3) {$V_{2}$};
\draw[-] (0.0,0)--(1,0);
\draw[-] (.5,0)--(.5,1) node at (.5,1.2) {$x_{4}$};
\draw[-] (.5,0)--(1,0) node at (1.2,0) {$x_{9}$};
\draw[cyan] (1.5,.1) circle (.6 cm);
\draw[dashed,  orange]  (0,-.08) to[bend right=5] (.9,-.08);
\end{scope}
\begin{scope}[xshift=6cm]
\node at (-.2,0) {$x_{2}$};
\node at (.5,-.3) {$V_{4}$};
\draw[-] (0.0,0)--(1,0);
\draw[-] (.5,0)--(.5,1) node at (.5,1.2) {$x_{6}$};
\draw[-] (.5,0)--(1,0) node at (1.2,0) {$x_{9}$};
\draw[cyan] (1.5,.1) circle (.6 cm);
\draw[dashed,  orange]  (0,-.08) to[bend right=5] (.9,-.08);
\end{scope}
\begin{scope}[xshift=8cm]
\node at (-.2,0) {$x_{2}$};
\node at (.5,-.3) {$V_{5}$};
\draw[-] (0.0,0)--(1,0);
\draw[-] (.5,0)--(.5,1) node at (.5,1.2) {$x_{7}$};
\draw[-] (.5,0)--(1,0) node at (1.2,0) {$x_{9}$};
\draw[cyan] (1.5,.1) circle (.6 cm);
\draw[dashed,  orange]  (0,-.08) to[bend right=5] (.9,-.08);
\end{scope}
\begin{scope}[xshift=10cm]
\node at (-.2,0) {$x_{7}$};
\node at (.5,-.3) {$V_{6}$};
\draw[-] (0.0,0)--(1,0);
\draw[-] (.5,0)--(.5,1) node at (.5,1.2) {$x_{8}$};
\draw[-] (.5,0)--(1,0) node at (1.2,0) {$x_{9}$};
\draw[dashed,  orange]  (.1,.15) to[bend right=30] (.5,1.1);
\end{scope}
\end{tikzpicture}
: V_{1}V_{6}: v_{38}= {x_{38}^2 x_{24}^2 x_{79}^2 \over x_{34}^2 x_{29}^2 x_{78}^2  }
\\
\begin{tikzpicture}[thick]
\begin{scope}[xshift=4cm]
\node at (-.2,0) {$x_{2}$};
\node at (.5,-.3) {$V_{2}$};
\draw[-] (0.0,0)--(1,0);
\draw[-] (.5,0)--(.5,1) node at (.5,1.2) {$x_{4}$};
\draw[-] (.5,0)--(1,0) node at (1.2,0) {$x_{9}$};
\draw[cyan] (.5,1.3) circle (.5 cm);
\draw[dashed,  orange]  (.1,.15) to[bend right=30] (.5,1.1);
\end{scope}
\begin{scope}[xshift=4cm]
\node at (.5,1.5) {$x_{9}$};
\node at (1.3,1.7) {$V_{3}$};
\draw[-] (.8,1.5)--(1.4,2.5);
\draw[-] (1.1,2)--(.5,2.5) node at (.5,2.8) {$x_{4}$};
\node at (1.4,2.7) {$x_{5}$};
\draw[dashed,  orange]  (.9,1.5) to[bend right=5] (1.2,2.4);
\end{scope}
\end{tikzpicture}
: V_{2}V_{3}: v_{25}= {x_{25}^2 x_{49}^2 \over x_{24}^2 x_{59}^2 }
\\ 
\begin{tikzpicture}[thick]
\begin{scope}[xshift=4cm]
\node at (-.2,0) {$x_{2}$};
\node at (.5,-.3) {$V_{2}$};
\draw[-] (0.0,0)--(1,0);
\draw[-] (.5,0)--(.5,1) node at (.5,1.2) {$x_{4}$};
\draw[-] (.5,0)--(1,0) node at (1.2,0) {$x_{9}$};
\draw[cyan] (1.5,.1) circle (.6 cm);
\draw[dashed,  orange]  (.5,1.1) to[bend right=30] (1,.15);
\end{scope}
\begin{scope}[xshift=6cm]
\node at (-.2,0) {$x_{2}$};
\node at (.5,-.3) {$V_{4}$};
\draw[-] (0.0,0)--(1,0);
\draw[-] (.5,0)--(.5,1) node at (.5,1.2) {$x_{6}$};
\draw[-] (.5,0)--(1,0) node at (1.2,0) {$x_{9}$};
\draw[dashed,  orange]  (.1,.15) to[bend right=30] (.5,1.1);
\end{scope}
\end{tikzpicture}
: V_{2}V_{4}: v_{46}= {x_{46}^2 x_{29}^2 \over x_{49}^2 x_{26}^2  }
\\
\begin{tikzpicture}[thick]
\begin{scope}[xshift=4cm]
\node at (-.2,0) {$x_{2}$};
\node at (.5,-.3) {$V_{2}$};
\draw[-] (0.0,0)--(1,0);
\draw[-] (.5,0)--(.5,1) node at (.5,1.2) {$x_{4}$};
\draw[-] (.5,0)--(1,0) node at (1.2,0) {$x_{9}$};
\draw[cyan] (1.5,.1) circle (.6 cm);
\draw[dashed,  orange]  (.5,1.1) to[bend right=30] (1,.15);
\end{scope}
\begin{scope}[xshift=6cm]
\node at (-.2,0) {$x_{2}$};
\node at (.5,-.3) {$V_{4}$};
\draw[-] (0.0,0)--(1,0);
\draw[-] (.5,0)--(.5,1) node at (.5,1.2) {$x_{6}$};
\draw[-] (.5,0)--(1,0) node at (1.2,0) {$x_{9}$};
\draw[cyan] (1.5,.1) circle (.6 cm);
\draw[dashed,  orange]  (0,-.08) to[bend right=5] (.9,-.08);
\end{scope}
\begin{scope}[xshift=8cm]
\node at (-.2,0) {$x_{2}$};
\node at (.5,-.3) {$V_{5}$};
\draw[-] (0.0,0)--(1,0);
\draw[-] (.5,0)--(.5,1) node at (.5,1.2) {$x_{7}$};
\draw[-] (.5,0)--(1,0) node at (1.2,0) {$x_{9}$};
\draw[dashed,  orange]  (.1,.15) to[bend right=30] (.5,1.1);
\end{scope}
\end{tikzpicture}
: V_{2}V_{5}: v_{47}= {x_{47}^2 x_{29}^2 \over x_{49}^2 x_{27}^2 }
\\ 
\begin{tikzpicture}[thick]
\begin{scope}[xshift=4cm]
\node at (-.2,0) {$x_{2}$};
\node at (.5,-.3) {$V_{2}$};
\draw[-] (0.0,0)--(1,0);
\draw[-] (.5,0)--(.5,1) node at (.5,1.2) {$x_{4}$};
\draw[-] (.5,0)--(1,0) node at (1.2,0) {$x_{9}$};
\draw[cyan] (1.5,.1) circle (.6 cm);
\draw[dashed,  orange]  (.5,1.1) to[bend right=30] (1,.15);
\end{scope}
\begin{scope}[xshift=6cm]
\node at (-.2,0) {$x_{2}$};
\node at (.5,-.3) {$V_{4}$};
\draw[-] (0.0,0)--(1,0);
\draw[-] (.5,0)--(.5,1) node at (.5,1.2) {$x_{6}$};
\draw[-] (.5,0)--(1,0) node at (1.2,0) {$x_{9}$};
\draw[cyan] (1.5,.1) circle (.6 cm);
\draw[dashed,  orange]  (0,-.08) to[bend right=5] (.9,-.08);
\end{scope}
\begin{scope}[xshift=8cm]
\node at (-.2,0) {$x_{2}$};
\node at (.5,-.3) {$V_{5}$};
\draw[-] (0.0,0)--(1,0);
\draw[-] (.5,0)--(.5,1) node at (.5,1.2) {$x_{7}$};
\draw[-] (.5,0)--(1,0) node at (1.2,0) {$x_{9}$};
\draw[cyan] (1.5,.1) circle (.6 cm);
\draw[dashed,  orange]  (0,-.08) to[bend right=5] (.9,-.08);
\end{scope}
\begin{scope}[xshift=10cm]
\node at (-.2,0) {$x_{7}$};
\node at (.5,-.3) {$V_{6}$};
\draw[-] (0.0,0)--(1,0);
\draw[-] (.5,0)--(.5,1) node at (.5,1.2) {$x_{8}$};
\draw[-] (.5,0)--(1,0) node at (1.2,0) {$x_{9}$};
\draw[dashed,  orange]  (.1,.15) to[bend right=30] (.5,1.1);
\end{scope}
\end{tikzpicture}
: V_{2}V_{6}: v_{48}= {x_{48}^2 x_{79}^2 \over x_{49}^2 x_{78}^2  }
\\
\begin{tikzpicture}[thick]
\begin{scope}[xshift=0cm]
\node at (-.2,0) {$x_{2}$};
\node at (.5,-.3) {$V_{2}$};
\draw[-] (0.0,0)--(1,0);
\draw[-] (.5,0)--(.5,1) node at (.5,1.2) {$x_{4}$};
\draw[-] (.5,0)--(1,0) node at (1.2,0) {$x_{9}$};
\draw[cyan] (.5,1.3) circle (.5 cm);
\draw[dashed,  orange]  (.5,1.1) to[bend right=30] (1,.15);
\end{scope}
\begin{scope}[xshift=0cm]
\node at (.5,1.5) {$x_{9}$};
\node at (1.3,1.7) {$V_{3}$};
\draw[-] (.8,1.5)--(1.4,2.5);
\draw[-] (1.1,2)--(.5,2.5) node at (.5,2.8) {$x_{4}$};
\node at (1.4,2.7) {$x_{5}$};
\draw[cyan] (1.5,.1) circle (.6 cm);
\draw[dashed,  orange]  (1.3,2.5) to[bend right=10] (.65,1.6);
\end{scope}
\begin{scope}[xshift=2cm]
\node at (-.2,0) {$x_{2}$};
\node at (.5,-.3) {$V_{4}$};
\draw[-] (0.0,0)--(1,0);
\draw[-] (.5,0)--(.5,1) node at (.5,1.2) {$x_{6}$};
\draw[-] (.5,0)--(1,0) node at (1.2,0) {$x_{9}$};
\draw[dashed,  orange]  (.1,.15) to[bend right=30] (.5,1.1);
\end{scope}
\end{tikzpicture}
: V_{3}V_{4}: v_{56}= {x_{56}^2 x_{29}^2 \over x_{59}^2 x_{26}^2  }
\\ 
\begin{tikzpicture}[thick]
\begin{scope}[xshift=0cm]
\node at (-.2,0) {$x_{2}$};
\node at (.5,-.3) {$V_{2}$};
\draw[-] (0.0,0)--(1,0);
\draw[-] (.5,0)--(.5,1) node at (.5,1.2) {$x_{4}$};
\draw[-] (.5,0)--(1,0) node at (1.2,0) {$x_{9}$};
\draw[cyan] (.5,1.3) circle (.5 cm);
\draw[dashed,  orange]  (.5,1.1) to[bend right=30] (1,.15);
\end{scope}
\begin{scope}[xshift=0cm]
\node at (.5,1.5) {$x_{9}$};
\node at (1.3,1.7) {$V_{3}$};
\draw[-] (.8,1.5)--(1.4,2.5);
\draw[-] (1.1,2)--(.5,2.5) node at (.5,2.8) {$x_{4}$};
\node at (1.4,2.7) {$x_{5}$};
\draw[cyan] (1.5,.1) circle (.6 cm);
\draw[dashed,  orange]  (1.3,2.5) to[bend right=10] (.65,1.6);
\end{scope}
\begin{scope}[xshift=2cm]
\node at (-.2,0) {$x_{2}$};
\node at (.5,-.3) {$V_{4}$};
\draw[-] (0.0,0)--(1,0);
\draw[-] (.5,0)--(.5,1) node at (.5,1.2) {$x_{6}$};
\draw[-] (.5,0)--(1,0) node at (1.2,0) {$x_{9}$};
\draw[cyan] (1.5,.1) circle (.6 cm);
\draw[dashed,  orange]  (0,-.08) to[bend right=5] (.9,-.08);
\end{scope}
\begin{scope}[xshift=4cm]
\node at (-.2,0) {$x_{2}$};
\node at (.5,-.3) {$V_{5}$};
\draw[-] (0.0,0)--(1,0);
\draw[-] (.5,0)--(.5,1) node at (.5,1.2) {$x_{7}$};
\draw[-] (.5,0)--(1,0) node at (1.2,0) {$x_{9}$};
\draw[dashed,  orange]  (.1,.15) to[bend right=30] (.5,1.1);
\end{scope}
\end{tikzpicture}
:V_{3}V_{5} : v_{57}= {x_{57}^2 x_{29}^2 \over x_{59}^2 x_{27}^2  }
\\
\begin{tikzpicture}[thick]
\begin{scope}[xshift=4cm]
\node at (-.2,0) {$x_{2}$};
\node at (.5,-.3) {$V_{2}$};
\draw[-] (0.0,0)--(1,0);
\draw[-] (.5,0)--(.5,1) node at (.5,1.2) {$x_{4}$};
\draw[-] (.5,0)--(1,0) node at (1.2,0) {$x_{9}$};
\draw[cyan] (.5,1.3) circle (.5 cm);
\draw[dashed,  orange]  (.5,1.1) to[bend right=30] (1,.15);
\end{scope}
\begin{scope}[xshift=4cm]
\node at (.5,1.5) {$x_{9}$};
\node at (1.3,1.7) {$V_{3}$};
\draw[-] (.8,1.5)--(1.4,2.5);
\draw[-] (1.1,2)--(.5,2.5) node at (.5,2.8) {$x_{4}$};
\node at (1.4,2.7) {$x_{5}$};
\draw[cyan] (1.5,.1) circle (.6 cm);
\draw[dashed,  orange]  (1.3,2.5) to[bend right=10] (.65,1.6);
\end{scope}
\begin{scope}[xshift=6cm]
\node at (-.2,0) {$x_{2}$};
\node at (.5,-.3) {$V_{4}$};
\draw[-] (0.0,0)--(1,0);
\draw[-] (.5,0)--(.5,1) node at (.5,1.2) {$x_{6}$};
\draw[-] (.5,0)--(1,0) node at (1.2,0) {$x_{9}$};
\draw[cyan] (1.5,.1) circle (.6 cm);
\draw[dashed,  orange]  (0,-.08) to[bend right=5] (.9,-.08);
\end{scope}
\begin{scope}[xshift=8cm]
\node at (-.2,0) {$x_{2}$};
\node at (.5,-.3) {$V_{5}$};
\draw[-] (0.0,0)--(1,0);
\draw[-] (.5,0)--(.5,1) node at (.5,1.2) {$x_{7}$};
\draw[-] (.5,0)--(1,0) node at (1.2,0) {$x_{9}$};
\draw[cyan] (1.5,.1) circle (.6 cm);
\draw[dashed,  orange]  (0,-.08) to[bend right=5] (.9,-.08);
\end{scope}
\begin{scope}[xshift=10cm]
\node at (-.2,0) {$x_{7}$};
\node at (.5,-.3) {$V_{6}$};
\draw[-] (0.0,0)--(1,0);
\draw[-] (.5,0)--(.5,1) node at (.5,1.2) {$x_{8}$};
\draw[-] (.5,0)--(1,0) node at (1.2,0) {$x_{9}$};
\draw[dashed,  orange]  (.1,.15) to[bend right=30] (.5,1.1);
\end{scope}
\end{tikzpicture}
: V_{3}V_{6}: v_{58}= {x_{58}^2 x_{97}^2 \over x_{78}^2 x_{59}^2  }
\\ 
\begin{tikzpicture}[thick]
\begin{scope}[xshift=6cm]
\node at (-.2,0) {$x_{2}$};
\node at (.5,-.3) {$V_{4}$};
\draw[-] (0.0,0)--(1,0);
\draw[-] (.5,0)--(.5,1) node at (.5,1.2) {$x_{6}$};
\draw[-] (.5,0)--(1,0) node at (1.2,0) {$x_{9}$};
\draw[cyan] (1.5,.1) circle (.6 cm);
\draw[dashed,  orange]  (.5,1.1) to[bend right=30] (1,.15);
\end{scope}
\begin{scope}[xshift=8cm]
\node at (-.2,0) {$x_{2}$};
\node at (.5,-.3) {$V_{5}$};
\draw[-] (0.0,0)--(1,0);
\draw[-] (.5,0)--(.5,1) node at (.5,1.2) {$x_{7}$};
\draw[-] (.5,0)--(1,0) node at (1.2,0) {$x_{9}$};
\draw[dashed,  orange]  (.1,.15) to[bend right=30] (.5,1.1);
\end{scope}
\end{tikzpicture} 
: V_{4}V_{5}: v_{67}= {x_{67}^2 x_{29}^2 \over x_{69}^2 x_{27}^2 }
\\
\begin{tikzpicture}[thick]
\begin{scope}[xshift=6cm]
\node at (-.2,0) {$x_{2}$};
\node at (.5,-.3) {$V_{4}$};
\draw[-] (0.0,0)--(1,0);
\draw[-] (.5,0)--(.5,1) node at (.5,1.2) {$x_{6}$};
\draw[-] (.5,0)--(1,0) node at (1.2,0) {$x_{9}$};
\draw[cyan] (1.5,.1) circle (.6 cm);
\draw[dashed,  orange]  (.5,1.1) to[bend right=30] (1,.15);
\end{scope}
\begin{scope}[xshift=8cm]
\node at (-.2,0) {$x_{2}$};
\node at (.5,-.3) {$V_{5}$};
\draw[-] (0.0,0)--(1,0);
\draw[-] (.5,0)--(.5,1) node at (.5,1.2) {$x_{7}$};
\draw[-] (.5,0)--(1,0) node at (1.2,0) {$x_{9}$};
\draw[cyan] (1.5,.1) circle (.6 cm);
\draw[dashed,  orange]  (0,-.08) to[bend right=5] (.9,-.08);
\end{scope}
\begin{scope}[xshift=10cm]
\node at (-.2,0) {$x_{7}$};
\node at (.5,-.3) {$V_{6}$};
\draw[-] (0.0,0)--(1,0);
\draw[-] (.5,0)--(.5,1) node at (.5,1.2) {$x_{8}$};
\draw[-] (.5,0)--(1,0) node at (1.2,0) {$x_{9}$};
\draw[dashed,  orange]  (.1,.15) to[bend right=30] (.5,1.1);
\end{scope}
\end{tikzpicture}
: V_{4}V_{6}: v_{68}= {x_{68}^2 x_{79}^2 \over x_{69}^2 x_{78}^2  }
\\ 
\begin{tikzpicture}[thick]
\begin{scope}[xshift=8cm]
\node at (-.2,0) {$x_{2}$};
\node at (.5,-.3) {$V_{5}$};
\draw[-] (0.0,0)--(1,0);
\draw[-] (.5,0)--(.5,1) node at (.5,1.2) {$x_{7}$};
\draw[-] (.5,0)--(1,0) node at (1.2,0) {$x_{9}$};
\draw[cyan] (1.5,.1) circle (.6 cm);
\draw[dashed,  orange]  (0,-.08) to[bend right=5] (.9,-.08);
\end{scope}
\begin{scope}[xshift=10cm]
\node at (-.2,0) {$x_{7}$};
\node at (.5,-.3) {$V_{6}$};
\draw[-] (0.0,0)--(1,0);
\draw[-] (.5,0)--(.5,1) node at (.5,1.2) {$x_{8}$};
\draw[-] (.5,0)--(1,0) node at (1.2,0) {$x_{9}$};
\draw[dashed,  orange]  (.1,.15) to[bend right=30] (.5,1.1);
\end{scope}
\end{tikzpicture}
: V_{5}V_{6}: v_{28}= {x_{28}^2 x_{79}^2 \over x_{29}^2 x_{78}^2  }
\end{align*}
\endgroup

By construction, the subscript labels of the $v_{rs}$ cross ratios identify the elements of the Mellin index set ${\cal V}$:
\eqn{}{
{\cal V} =\, & \{(14),(15),(16),(17),(18),(19),(25),(28),(35),(36),(37),(38),(39),(46),(47),(48),  \cr 
 & (56), (57),(58),(67), (68) \} \,.
}
With the set ${\cal V}$ in hand, we can use~\eno{PostMellinExternal} to work out the post-Mellin parameters $\ell_i$ associated to each external leg:
\begin{align}
\ell_{1}&= j_{14}+j_{15}+j_{16}+j_{17}+j_{18}+j_{19}
&
\ell_{2}&= j_{25}+j_{28}
\cr 
\ell_{3}&= j_{35}+j_{36}+j_{37}+j_{38}+j_{39}
&
\ell_{4}&= j_{14}+j_{46}+j_{47}+j_{48}
\cr 
\ell_{5}&=j_{15}+j_{25}+j_{35}+j_{56}+j_{57}+j_{58}
&
\ell_{6}&= j_{16}+j_{36}+j_{46}+j_{56}+j_{67}+ j_{68}
\cr 
\ell_{7}&= j_{17}+j_{37}+j_{47}+j_{57}+j_{67}
&
\ell_{8}&=j_{18}+j_{28}+j_{38}+j_{48}+j_{58}+j_{68}
\cr 
\ell_{9}&=j_{19}+j_{39} & &
\end{align}
We can now use~\eno{PostMellinInternal1}-\eno{PostMellinInternal2} to work out the post-Mellin parameters associated to the internal legs:
\begingroup
\allowdisplaybreaks
\begin{align*}
\begin{tikzpicture}[thick][r]
\begin{scope}
\node at (-.3,0) {${\cal{O}}_1$};
\node at (.5,-.3) {$V_7$};
\draw[-] (0.0,0)--(1,0);
\draw[-] (.5,0)--(.5,1) node at (.5,1.2) {${\cal{O}}_2$};
\draw[-] (.5,0)--(1,0) node at (1.4,0) {${\cal{O}}_{k_{1}}$};
\end{scope}
\end{tikzpicture}
 \ell_{k_{1}}&  \stackrel{2{\cal J}}{=} \ell_{1}+\ell_{2}= j_{14}+j_{15}+j_{16}+j_{17}+j_{18}+j_{19}+j_{25}+j_{28}
\\
\qquad 
\begin{tikzpicture}[thick][r]
\begin{scope}
\node at (-.3,0) {${\cal{O}}_8$};
\node at (.5,-.3) {$V_6$};
\draw[-] (0.0,0)--(1,0);
\draw[-] (.5,0)--(.5,1) node at (.5,1.2) {${\cal{O}}_9$};
\draw[-] (.5,0)--(1,0) node at (1.4,0) {${\cal{O}}_{k_{6}}$};
\end{scope}
\end{tikzpicture}
\ell_{k_{6}}& \stackrel{2{\cal J}}{=} \ell_{8}+\ell_{9}= j_{18}+j_{19}+j_{28}+j_{38}+j_{39}+j_{48}+j_{58}+j_{68}
\\ 
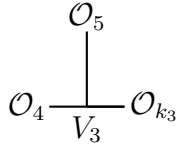
\begin{tikzpicture}[thick][r]
\begin{scope}
\node at (-.3,0) {${\cal{O}}_4$};
\node at (.5,-.3) {$V_3$};
\draw[-] (0.0,0)--(1,0);
\draw[-] (.5,0)--(.5,1) node at (.5,1.2) {${\cal{O}}_5$};
\draw[-] (.5,0)--(1,0) node at (1.4,0) {${\cal{O}}_{k_{3}}$};
\end{scope}
\end{tikzpicture}
\\ 
\ell_{k_{3}}& \stackrel{2{\cal J}}{=}\ell_{4}+\ell_{5}= j_{14}+j_{15}+j_{25}+j_{35}+j_{46}+j_{47}+j_{48}+j_{56}+j_{57}+j_{58}
\\
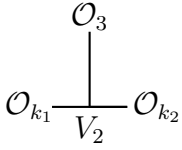
\begin{tikzpicture}[thick][r]
\begin{scope}
\node at (-.3,0) {${\cal{O}}_{k_{1}}$};
\node at (.5,-.3) {$V_2$};
\draw[-] (0.0,0)--(1,0);
\draw[-] (.5,0)--(.5,1) node at (.5,1.2) {${\cal{O}}_3$};
\draw[-] (.5,0)--(1,0) node at (1.4,0) {${\cal{O}}_{k_{2}}$};
\end{scope}
\end{tikzpicture}
\qquad
\ell_{k_{2}}& \stackrel{2{\cal J}}{=} \ell_{k_{1}}+\ell_{3} 
\\ 
& = j_{14}+j_{15}+j_{16}+j_{17}+j_{18}+j_{19}+j_{25}\cr
&\quad +j_{28}+j_{35}+j_{36}+j_{37}+j_{38}+j_{39}
\\ 
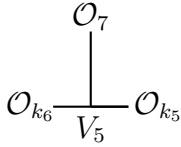
\begin{tikzpicture}[thick][r]
\begin{scope}
\node at (-.3,0) {${\cal{O}}_{k_{6}}$};
\node at (.5,-.3) {$V_5$};
\draw[-] (0.0,0)--(1,0);
\draw[-] (.5,0)--(.5,1) node at (.5,1.2) {${\cal{O}}_7$};
\draw[-] (.5,0)--(1,0) node at (1.4,0) {${\cal{O}}_{k_{5}}$};
\end{scope}
\end{tikzpicture}
\ell_{k_{5}}& \stackrel{2{\cal J}}{=} \ell_{k_{6}}+\ell_{7}
\\ 
& = j_{17}+j_{18}+j_{19}+j_{28}+j_{37}+j_{38}\cr
&\quad +j_{39}+j_{47}+j_{48}+j_{57}+j_{58}+j_{67}+j_{68}
\\ 
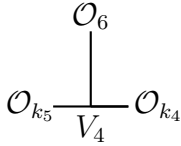
\begin{tikzpicture}[thick][r]
\begin{scope}
\node at (-.3,0) {${\cal{O}}_{k_{5}}$};
\node at (.5,-.3) {$V_4$};
\draw[-] (0.0,0)--(1,0);
\draw[-] (.5,0)--(.5,1) node at (.5,1.2) {${\cal{O}}_6$};
\draw[-] (.5,0)--(1,0) node at (1.4,0) {${\cal{O}}_{k_{4}}$};
\end{scope}
\end{tikzpicture}
\ell_{k_{4}}& \stackrel{2{\cal J}}{=} \ell_{k_{5}}+\ell_{6}
\\ 
&= j_{16}+j_{17}+j_{18}+j_{19}+j_{28}+j_{36}+j_{37}+j_{38}\cr
&\quad +j_{39}+j_{46}+j_{47}+j_{48}+j_{56}+j_{57}+j_{58}
\end{align*}
\endgroup

Having computed the post-Mellin parameters $\ell_{i}$ and $\ell_{k_i}$, we can use the Feynman rules to  explicitly write out all the vertex and edge factors.
For example, for the vertex $V_4$, we have
\begingroup\makeatletter\def\f@size{11.3}\check@mathfonts
   \eqn{ninepointV4}{
V_4 &=
     (\Delta_{k_{5}k_{4},6})_{m_{k_{4}k_{5},}
     +
j_{17}+j_{18}+j_{19}+j_{28}+j_{37}+j_{38}+j_{39}+j_{47}+j_{48}+j_{57}+j_{58}} (\Delta_{k_{5}6,k_{4}})_{m_{{5},{4}}+\ell_{j_{67}+j_{68}}} 
\cr 
&  \times (\Delta_{k_{4}6,k_{5}})_{m_{{4},{5}}+\ell_{j_{16}+j_{36}+j_{46}+j_{56}}}  \; F_A^{(2)}\!\left[\Delta_{k_{5}k_{4}6,}- h; 
\{-m_{{5}}, -m_{{4}}\}; \left\{\Delta_{k_{5}} -h+1, \Delta_{k_{4}} -h+1 \right\}; 1,1 \right].
     }
\endgroup
As an example of an edge factor, we write down the edge factor associated to the exchange of the operator ${\cal O}_{k_5}$ of dimension $\Delta_{k_5}$:
\eqn{}{
E_5 :=  {(\Delta_{k_5} - h +1)_{m_5} \over (\Delta_{k_5})_{2m_5 + j_{17}+j_{18}+j_{19}+j_{28}+j_{37}+j_{38}+j_{39}+j_{47}+j_{48}+j_{57}+j_{58}+j_{67}+j_{68}}}. 
}
The remaining vertex and edge factors are also trivially determined using~\eno{EdgeDef}-\eno{VertexDef}.

\subsection{Symmetries of the Block}

The nine-point block found above, associated to the topology of Figure \ref{fig:NineMixedCB}, must satisfy some identities related to the symmetries of its topology.

In the case at hand, the symmetry group is $H=\mathbb{Z}_2\times\mathbb{Z}_2\times\mathbb{Z}_2$ which corresponds to the three dendrite permutations $\mathcal{O}_1(x_1)\leftrightarrow\mathcal{O}_2(x_2)$, $\mathcal{O}_4(x_4)\leftrightarrow\mathcal{O}_5(x_5)$, and $\mathcal{O}_8(x_8)\leftrightarrow\mathcal{O}_9(x_9)$, respectively.

For $\mathcal{O}_1(x_1)\leftrightarrow\mathcal{O}_2(x_2)$, the identity is
\begin{align}
&g\left[\begin{array}{c}
u_1,u_2,u_3,u_4,u_5,u_6;v_{14},v_{15},v_{16},v_{17},v_{18},v_{19},v_{25},v_{28}\nonumber\\
v_{35},v_{36},v_{37},v_{38},v_{39},v_{46},v_{47},v_{48},v_{56},v_{57},v_{58},v_{67},v_{68}\end{array}\right]\\
&=v_{14}^{(\Delta_3-\Delta_{k_1}-\Delta_{k_3}+\Delta_{k_4})/2}v_{16}^{(-\Delta_6-\Delta_{k_4}+\Delta_{k_5})/2}v_{17}^{(-\Delta_7-\Delta_{k_5}+\Delta_{k_6})/2}v_{19}^{(\Delta_6+\Delta_7-\Delta_{k_2}+\Delta_{k_3}-\Delta_{k_6})/2}\label{eqn:12sym}\\
&\phantom{=}\times g_{\Delta_1\leftrightarrow\Delta_2}\left[\begin{array}{c}
\frac{u_1}{v_{14}},\frac{u_2}{v_{19}},\frac{u_3v_{19}}{v_{14}},\frac{u_4v_{14}}{v_{16}},\frac{u_5v_{16}}{v_{17}},\frac{u_6v_{17}}{v_{19}};\frac{1}{v_{14}},\frac{v_{25}}{v_{14}},\frac{1}{v_{16}},\frac{1}{v_{17}},\frac{v_{28}}{v_{19}},\frac{1}{v_{19}},\frac{v_{15}}{v_{14}},\frac{v_{18}}{v_{19}}\\
v_{35},\frac{v_{36}v_{14}}{v_{16}},\frac{v_{37}v_{14}}{v_{17}},\frac{v_{38}v_{14}}{v_{19}},\frac{v_{39}v_{14}}{v_{19}},\frac{v_{46}v_{19}}{v_{16}},\frac{v_{47}v_{19}}{v_{17}},v_{48},\frac{v_{56}v_{19}}{v_{16}},\frac{v_{57}v_{19}}{v_{17}},v_{58},\frac{v_{67}v_{19}}{v_{17}},v_{68}\end{array}\right],\nonumber
\end{align}
for $\mathcal{O}_4(x_4)\leftrightarrow\mathcal{O}_5(x_5)$, it takes the form
\begin{align}
&g\left[\begin{array}{c}
u_1,u_2,u_3,u_4,u_5,u_6;v_{14},v_{15},v_{16},v_{17},v_{18},v_{19},v_{25},v_{28}\nonumber\\
v_{35},v_{36},v_{37},v_{38},v_{39},v_{46},v_{47},v_{48},v_{56},v_{57},v_{58},v_{67},v_{68}\end{array}\right]\\
&=v_{25}^{(\Delta_3-\Delta_{k_1}-\Delta_{k_3}+\Delta_{k_4})/2}v_{35}^{(-\Delta_3+\Delta_{k_1}-\Delta_{k_2})/2}\label{eqn:45sym}\\
&\phantom{=}\times g_{\Delta_4\leftrightarrow\Delta_5}\left[\begin{array}{c}
\frac{u_1v_{35}}{v_{25}},\frac{u_2}{v_{35}},\frac{u_3}{v_{25}},u_4v_{25},u_5,u_6;\frac{v_{15}}{v_{25}},\frac{v_{14}}{v_{25}},v_{16},v_{17},v_{18},v_{19},\frac{1}{v_{25}},v_{28}\\
\frac{1}{v_{35}},\frac{v_{36}v_{25}}{v_{35}},\frac{v_{37}v_{25}}{v_{35}},\frac{v_{38}v_{25}}{v_{35}},\frac{v_{39}v_{25}}{v_{35}},v_{56},v_{57},v_{58},v_{46},v_{47},v_{48},v_{67},v_{68}\end{array}\right],\nonumber
\end{align}
while for $\mathcal{O}_8(x_8)\leftrightarrow\mathcal{O}_9(x_9)$, the identity becomes
\begin{align}
&g\left[\begin{array}{c}
u_1,u_2,u_3,u_4,u_5,u_6;v_{14},v_{15},v_{16},v_{17},v_{18},v_{19},v_{25},v_{28}\nonumber\\
v_{35},v_{36},v_{37},v_{38},v_{39},v_{46},v_{47},v_{48},v_{56},v_{57},v_{58},v_{67},v_{68}\end{array}\right]\\
&=v_{28}^{(\Delta_6+\Delta_7-\Delta_{k_2}+\Delta_{k_3}-\Delta_{k_6})/2}v_{48}^{(-\Delta_4+\Delta_5+\Delta_{k_2}-\Delta_{k_4})/2}v_{58}^{(\Delta_4-\Delta_5-\Delta_{k_3})/2}v_{68}^{(-\Delta_6+\Delta_{k_4}-\Delta_{k_5})/2}\label{eqn:89sym}\\
&\phantom{=}\times g_{\Delta_8\leftrightarrow\Delta_9}\left[\begin{array}{c}
u_1,\frac{u_2v_{48}}{v_{28}},\frac{u_3v_{28}}{v_{58}},\frac{u_4v_{68}}{v_{48}},\frac{u_5}{v_{68}},\frac{u_6}{v_{28}};v_{14},\frac{v_{15}v_{48}}{v_{58}},v_{16},v_{17},\frac{v_{19}}{v_{28}},\frac{v_{18}}{v_{28}},\frac{v_{25}v_{48}}{v_{58}},\frac{1}{v_{28}}\\
\frac{v_{35}v_{48}}{v_{58}},v_{36},v_{37},\frac{v_{39}}{v_{28}},\frac{v_{38}}{v_{28}},\frac{v_{46}v_{28}}{v_{48}},\frac{v_{47}v_{28}}{v_{48}},\frac{1}{v_{48}},\frac{v_{56}v_{28}}{v_{58}},\frac{v_{57}v_{28}}{v_{58}},\frac{1}{v_{58}},\frac{v_{67}v_{28}}{v_{68}},\frac{1}{v_{68}}\end{array}\right].\nonumber
\end{align}
Equations \eqref{eqn:12sym}, \eqref{eqn:45sym} and \eqref{eqn:89sym} are intricate identities that the scalar conformal block must satisfy due to the symmetry group of its associated topology.  We have only performed partial, though non-trivial checks to verify that they are consistent.  The checks were done by expanding both sides of the identities in powers of the cross ratios $u_i$ and $1-v_{rs}$ to finite order and comparing the coefficients of the expansions term by term.  Complete analytical proofs would presumably be extremely lengthy and complicated.

\section{Discussion}
\label{DISCUSSION}

In this paper we provided an inductive proof of the Feynman-like rules for scalar conformal blocks with scalar exchanges found in \cite{Hoback:2020pgj} using the flow diagrams developed in \cite{Fortin:2020zxw}.  These two results turned out to be complementary---their union was necessary to obtain a complete proof.  On the one hand, the Feynman-like rules led to an easy recipe for writing down conformal blocks, without however providing an efficient technique to generate the appropriate cross ratios. On the other hand, the flow diagrams by themselves did not directly lead to rules for conformal blocks but gave a straightforward and intuitive method for generating conformal cross ratios.

Indeed, with flow diagrams an $M$-point block's cross ratios can be written in terms of an associated $(M-1)$-point block's cross ratios.  Thus, assuming the Feynman-like rules and relying on the AdS interpretation of the conformal blocks, it is then possible to generate the appropriate conformal cross ratios that can be used to prove the rules for an arbitrary topology through repeated application of the First Barnes lemma along the topology's comb structures.  Consequently, with a simple demonstration for $M=4$ and the induction for $M$-point conformal blocks from $(M-1)$-conformal blocks, this paper provides a complete proof of the Feynman-like rules.

For completeness, we also showed the four- to five-point induction and we presented an explicit example for an asymmetric nine-point topology.  In the latter case, we also gave the identities the conformal blocks must satisfy from the symmetries of the associated topology, which we verified to low order in a power series expansion of the blocks.

Several interesting open questions ensue from our work.  First, the Feynman-like rules for conformal blocks only works for scalar conformal blocks with scalar exchanges.  It seems natural that such rules should exist for spinning exchanges as well as for spinning external operators. A possible route forward in finding Feynman-like rules for conformal blocks with spinning exchange arises from the search for the rules for Mellin Amplitudes with spinning exchange. Perhaps if one were able to find the Mellin Amplitudes rules for spinning exchange a similar proof could reveal the Feynman-like rules for conformal blocks with spin. The spinning case would also be of interest in AdS for graviton scattering and graviton exchange.  From the point of view of AdS, it would be of interest to generalize the induction of the single trace part of Witten diagrams presented here to the full Witten diagrams.  Further research is needed to generalize this prescription to higher-point blocks with spinning operators, either exchanged or external.

Another avenue would be to explore the analytic structure of the higher-point conformal blocks.  This could be done with the help of new cross ratios reminiscent of the Dolan-Osborn cross ratios that greatly simplify four-point conformal blocks in even spacetime dimensions \cite{Dolan:2000ut}.  From the bootstrap perspective, it would also be natural to analyse the domain of convergence of the various higher-point correlation functions to determine if there is any overlap of the regions of convergence between the different channels.  From our rules, blocks are expansions around $u_a\sim0$ and $v_{rs}\sim1$, and different channels lead to different cross ratios which must be compared.  For example, in the four-point crossing equations, the $s$- and $t$-channels cross ratios flip as in $(u,v)\to(v,u)$, suggesting an overlap of the regions of convergence.  Since there is only one topology, the expression for different channels are given in terms of the same blocks, and it is straightforward to determine the overlap.  The situation is significantly more complicated for higher-point correlation functions because there are many more cross ratios and different channels are not expressed in terms of the same blocks.  Indeed, although it can be argued that the region of convergence is finite, to determine its precise boundaries for an arbitrary block, expressed as a hypergeometric-like series, it is suitable to use the Horn method \cite{Horn1889}.  This usually leads to a complicated region described by a set of nonlinear equations between the cross ratios (see \textit{e.g.}\ \cite{Comeau:2019xco}).  For the six-point bootstrap, we did not attempt to obtain the regions of convergence for the comb and the snowflake topologies.  Instead, we directly compared the cross ratios from one channel in the comb topology to the cross ratios of the other channel in the snowflake topology.  We found that they are related as in the four-point case, with the $u$ cross ratios expressed in terms of the former $v$ cross ratios, and vice-versa, again suggesting an overlap of the regions of convergence.  We hope to return to these problems in the near future.

\section*{Acknowledgements}

The work of JFF is supported by NSERC.  The work of SH is supported by the National Science Foundation Graduate Research Fellowship under Grant No. 2021316516.  The work of WJM is supported by the China Scholarship Council and in part by NSERC and FRQNT. The work of SP is supported in part by the Young Faculty Incentive Fellowship from IIT Delhi. The work of WS is supported in part by U.S.~DOE HEP grant DE-SC00-17660.

\appendix 

\section{First Few Cases of Induction}
\label{BASECASE}
In this section we will instantiate the proof of the Feynman-like rules presented in Section~\ref{PROOF} by working out the first two non-trivial cases.
The methods provided here may seem overkill for these instances.
The strategies presented here, on the other hand, are critical in proving the general $M$-point case in Section~\ref{PROOF}.

\subsection{Three-point to Four-point}
\label{Base case 3}

\begin{figure}[!t]
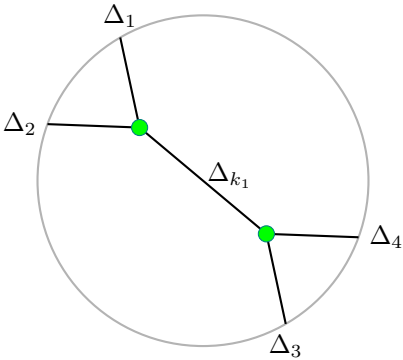

    \centering
    \[ \musepic{\figFourptAdSUnprime}  \]
    \caption{A canonical $4$-point AdS diagram.}
    \label{fig:4ptAdSUnprime}
\end{figure}
We consider the four-point scalar exchange Witten diagram shown in Figure~\ref{fig:4ptAdSUnprime}. 
The position space amplitude for the exchange diagram admits the following Mellin representation,
\eqn{A4}{
A_4 = {\cal N}_4 \left( \prod_{(rs) \in {\cal U} \bigcup {\cal V} \bigcup {\cal D}} \int {d\gamma_{rs} \over 2\pi i} {\Gamma(\gamma_{rs}) \over (x_{rs}^2)^{\gamma_{rs}}} \right) \left( \prod_{a=1}^4 2\pi i\, \delta(\sum_{b=1}^4 \gamma_{ab})  \right) {\cal M}_4(\gamma_{ab})
}
where the normalization constant is 
\eqn{}{
{\cal N}_4 =  { \pi^{2h}  \over \Gamma(\Delta_{k_1}) } \left( \prod_{a=1}^{4} {1\over \Gamma(\Delta_a)} \right) 
}
and the Mellin amplitude ${\cal M}_4$ is given by
\eqn{}{
{\cal M}_4 (\gamma_{ab}) = \sum_{m_1=0}^\infty E_1^{\rm Mellin} V_1^{\rm Mellin} V_2^{\rm Mellin} \,,
}
where the ``edge-'' and ``vertex-factors'' take the form
\eqn{}{
E_1^{\rm Mellin} &= {1 \over m_1!} {(\Delta_{k_1}-h+1)_{m_1} \over {\Delta_{k_1}-s_1 \over 2} + m_1 } \cr 
V_1^{\rm Mellin} &= {1\over 2} \Gamma(\Delta_{12k_1,}-h)\, F_A^{(1)}[\Delta_{12k_1,}-h; \{-m_1\};\{\Delta_{k_1}-h+1 \}; 1] \cr 
V_2^{\rm Mellin} &= {1\over 2} \Gamma(\Delta_{34k_1,}-h)\, F_A^{(1)}[\Delta_{34k_1,}-h; \{-m_1\};\{\Delta_{k_1}-h+1 \}; 1]
}
with the Mandelstam invariant $s_1 = \Delta_1 + \Delta_2 - 2\gamma_{12}$.\footnote{Here $\gamma_{ab} (= \gamma_{ba})$ are the Mellin variables satisfying $\sum_{b=1}^4\gamma_{ab} = 0$ for all $a$,
where we have defined $\gamma_{aa} := -\Delta_a$. These constraints can be solved in terms of auxiliary momenta variables $p_a$ such that they satisfy $p_a \cdot p_b = \gamma_{ab}$ for all $a,b$ with momentum conservation $\sum_{a=1}^4 p_a=0$. The Mandelstam invariant $s_1$ is then given by $s_1 = -(p_1 + p_2)^2 = \Delta_1+\Delta_2 - 2\gamma_{12}$. }
The functions $F_A^{(1)}$ appearing above are a special case of Lauricella functions defined in~\eno{LauricellaDef}.
The index sets in~\eno{A4} constitute entries of a $4 \times 4$ upper-triangular matrix
\eqn{}{
{\cal U} \bigcup {\cal V} \bigcup {\cal D} = \{ (12), (13), (14), (23), (24), (34) \} 
}
but due to the presence of the four delta functions in~\eno{A4}, there are only two independent Mellin variables. 
We follow the discussion of Section~\ref{INDEXCROSSRATIOS} to construct these sets and the corresponding cross ratios. Particularly, for the choice of OPE vertex depicted in the top line of~\eno{fourpointudiagrams} [equivalently, the top line of~\eno{fourVtypecross}], we have  
\eqn{}{
{\cal V} = \{(14) \} 
}
and the associated cross ratios are
\eqn{4uv}{
v_{14} = {x_{14}^2 x_{23}^2 \over x_{13}^2 x_{24}^2}  \qquad u_1 = {x_{12}^2 x_{34}^2 \over x_{13}^2 x_{24}^2} \,.
}
From the set ${\cal V}$ and the associated collection of independent cross-ratios~\eno{4uv}, we can construct the remaining sets
\eqn{}{
{\cal U} = \{ (12) \} \qquad {\cal D} = \{ (13), (23), (24), (34) \} \,.
}

We now define
\eqn{B4Def}{
B_4 := \left( \prod_{(rs) \in {\cal U} \bigcup {\cal V} \bigcup {\cal D}} \int {d\gamma_{rs} \over 2\pi i} {\Gamma(\gamma_{rs}) \over (x_{rs}^2)^{\gamma_{rs}}} \right) \left( \prod_{a=1}^4 2\pi i\, \delta(\sum_{b=1}^4 \gamma_{ab})  \right) {1 \over {\Delta_{k_1} -s_1 \over 2}+m_1}
}
so that
\eqn{}{
A_4 = {\cal N}_4 \sum_{m_1=0}^\infty V_1^{\rm Mellin} V_2^{\rm Mellin} {(\Delta_{k_1}-h+1)_{m_1} \over m_1!} B_4 \,.
}
Furthermore, we define 
\eqn{B3pDef}{
B_3 := \left( \prod_{(rs) \in {\cal U}^\prime \bigcup {\cal V}^\prime \bigcup {\cal D}^\prime} \int {d\gamma_{rs} \over 2\pi i} {\Gamma(\gamma_{rs}) \over (x_{rs}^2)^{\gamma_{rs}}} \right) \left( \prod_{a=2}^4 2\pi i\, \delta(\sum_{\substack{b=1 \\b\neq a}}^4 \gamma_{ab}-\Delta_a)  \right) 
}
where\footnote{While precise knowledge of the individual primed sets is not important in this subsection, to make connection with the general inductive proof to follow in Section~\ref{PROOF}, we note that ${\cal U}^\prime = \emptyset, {\cal V}^\prime = \emptyset$, and ${\cal D}^\prime = \{ (23), (24), (34) \}$.}  
\eqn{}{
{\cal U}^\prime \bigcup {\cal V}^\prime \bigcup {\cal D}^\prime = \{ (23), (24), (34) \} \,, 
}
so that
\eqn{B4B3}{
B_4 = \left(\prod_{a=2}^4 \int {d\gamma_{1a} \over 2\pi i} {\Gamma(\gamma_{1a}) \over (x_{1a}^2)^{\gamma_{1a}}} \right) 2\pi i\, \delta(\sum_{a=2}^4 \gamma_{1a}-\Delta_1) {1 \over {\Delta_{k_1}-s_1 \over 2}+m_1} B_3\,.
}
We can easily evaluate $B_3$ since the three delta functions eat up the three contour integrals, enforcing the following constraints
\eqn{}{
\gamma_{23} &= \Delta_{23,4} + {-\gamma_{12}-\gamma_{13}+\gamma_{14} \over 2} \cr 
\gamma_{24} &= \Delta_{24,3} + {-\gamma_{12}+\gamma_{13}-\gamma_{14} \over 2} \cr
\gamma_{34} &= \Delta_{34,2} + {\gamma_{12}-\gamma_{13}-\gamma_{14} \over 2} \,.
}
Consequently,
\eqn{B3peval}{
B_3 &= {\Gamma(\Delta_{23,4} + {-\gamma_{12}-\gamma_{13}+\gamma_{14} \over 2}) \Gamma(\Delta_{24,3} + {-\gamma_{12}+\gamma_{13}-\gamma_{14} \over 2}) \Gamma(\Delta_{34,2} + {\gamma_{12}-\gamma_{13}-\gamma_{14} \over 2}) \over (x_{23}^2)^{\Delta_{23,4} + {-\gamma_{12}-\gamma_{13}+\gamma_{14} \over 2}} (x_{24}^2)^{\Delta_{24,3} + {-\gamma_{12}+\gamma_{13}-\gamma_{14} \over 2}} (x_{34}^2)^{\Delta_{34,2} + {\gamma_{12}-\gamma_{13}-\gamma_{14} \over 2}} } \cr 
 &= L_2^\prime(\Delta_2^\prime) L_3^\prime(\Delta_3^\prime) L_4^\prime(\Delta_4^\prime)\, \Gamma(\Delta_{23,4}^\prime) \Gamma(\Delta_{24,3}^\prime )  \Gamma(\Delta_{34,2}^\prime) \,,
}
where we have defined the leg factors
\eqn{4LegFactors}{
L_2^\prime(\Delta_2^\prime) = \left({x_{34}^2 \over x_{23}^2 x_{24}^2} \right)^{\Delta_2^\prime \over 2} 
\quad 
L_3^\prime(\Delta_3^\prime) = \left({x_{24}^2 \over x_{23}^2 x_{34}^2} \right)^{\Delta_3^\prime \over 2} 
\quad 
L_4^\prime(\Delta_4^\prime) = \left({x_{23}^2 \over x_{24}^2 x_{34}^2} \right)^{\Delta_4^\prime \over 2} 
}
 with
\eqn{4ShiftedDims}{
\Delta_2^\prime := \Delta_2 - \gamma_{12} \qquad \Delta_3^\prime := \Delta_3 - \gamma_{13} \qquad \Delta_4^\prime := \Delta_4-\gamma_{14}\,.
}

To proceed, we substitute~\eno{B3peval} into~\eno{B4B3} and then calculate its single-trace projection $B_4^{\rm s.t.}$:
\eqn{B4st}{
B_4^{\rm s.t.} &= [ B_4]_{\rm s.t.} \cr 
&= \left[ \int {d\gamma_{12}\, d\gamma_{13}\, d\gamma_{14} \over (2\pi i)^3} {\Gamma(\gamma_{12}) \Gamma(\gamma_{13}) \Gamma(\gamma_{14}) \over (x_{12}^2)^{\gamma_{12}} (x_{13}^2)^{\gamma_{13}} (x_{14}^2)^{\gamma_{14}} } {2\pi i\, \delta(\gamma_{12}+\gamma_{13}+\gamma_{14} - \Delta_1) \over \Delta_{k_1,12}+\gamma_{12}+m_1 }  \right. \cr 
 &\quad \times   L_2^\prime(\Delta_2^\prime) L_3^\prime(\Delta_3^\prime) L_4^\prime(\Delta_4^\prime)\, \Gamma(\Delta_{23,4}^\prime) \Gamma(\Delta_{24,3}^\prime )  \Gamma(\Delta_{34,2}^\prime) \Bigg]_{\rm s.t.}\,.
}
Then the conformal block is obtained as
\eqn{CB-st}{
W_4(x_i) = {[A_4]_{\rm s.t.} \over C_{\Delta_1\Delta_2\Delta_{k_1}} C_{\Delta_3\Delta_4\Delta_{k_1}}}= {{\cal N}_4 \over  C_{\Delta_1\Delta_2\Delta_{k_1}} C_{\Delta_3\Delta_4\Delta_{k_1}}} \sum_{m_1=0}^\infty V_1^{\rm Mellin} V_2^{\rm Mellin} {(\Delta_{k_1}-h+1)_{m_1} \over m_1!} B_4^{\rm s.t.} 
}
where we have projected out the (known) theory-dependent mean field theory OPE coefficients
\eqn{OPEreal}{
C_{\Delta_i\Delta_j\Delta_k} &= \frac{\pi^h}{2} \, \Gamma\left(\Delta_{ijk,}-h\right)
\frac{\Gamma(\Delta_{ij,k})
\Gamma(\Delta_{jk,i})
\Gamma(\Delta_{ki,j})}{\Gamma(\Delta_i)\Gamma(\Delta_j)\Gamma(\Delta_k)}\,,
}
to extract the theory-independent conformal block.

The single-trace projection in~\eno{B4st} is obtained by evaluating the residue of the simple pole in the $\gamma_{12}$-plane, at $\gamma_{12} = \Delta_{12,k_1}-m_1$~\cite{Hoback:2020pgj}, which gives
\eqn{}{
B_4^{\rm s.t.} &=  \int {d\gamma_{13}\, d\gamma_{14} \over (2\pi i)^2} {\Gamma(\Delta_{12,k_1}-m_1) \Gamma(\gamma_{13}) \Gamma(\gamma_{14}) \over (x_{12}^2)^{\Delta_{12,k_1}-m_1} (x_{13}^2)^{\gamma_{13}} (x_{14}^2)^{\gamma_{14}} } 2\pi i\, \delta(\gamma_{13}+\gamma_{14} + \Delta_{2,1k_1} - m_1)  \cr 
 &\quad \times   \left({x_{34}^2 \over x_{23}^2 x_{24}^2} \right)^{\Delta_{2k_1,1}+m_1 \over 2}  L_3^\prime(\Delta_3^\prime) L_4^\prime(\Delta_4^\prime)\, \Gamma(\Delta_{23,4}-{1\over 2}\Delta_{12,k_1} +{m_1\over 2} + {\gamma_{14}-\gamma_{13}\over 2} )  \cr 
 &\quad \times \Gamma(\Delta_{24,3} -{1\over 2}\Delta_{12,k_1} + {m_1\over 2} + {\gamma_{13}-\gamma_{14} \over 2} )  \Gamma(\Delta_{34,2} +{1\over 2}\Delta_{12,k_1} - {m_1\over 2} - {\gamma_{13}+\gamma_{14} \over 2}) \,. 
}

Next, we evaluate the $\gamma_{13}$ integral by using the delta-function which enforces $\gamma_{13} = \Delta_{1k_1,2}+m_1-\gamma_{14}$:
\eqn{}{
B_4^{\rm s.t.} &= \int { d\gamma_{14} \over 2\pi i} {\Gamma(\Delta_{12,k_1}-m_1) \Gamma(\Delta_{1k_1,2}+m_1-\gamma_{14}) \Gamma(\gamma_{14}) \over (x_{12}^2)^{\Delta_{12,k_1}-m_1} (x_{13}^2)^{\Delta_{1k_1,2}+m_1-\gamma_{14}} (x_{14}^2)^{\gamma_{14}} } 
\Gamma(\Delta_{34,k_1}-m_1) \Gamma(\Delta_{23,14} + \gamma_{14})   \cr 
 &\quad \times  \Gamma(\Delta_{k_14,3} +m_1-\gamma_{14}) \left({x_{34}^2 \over x_{23}^2 x_{24}^2} \right)^{\Delta_{2k_1,1}+m_1 \over 2}  \left({x_{24}^2 \over x_{23}^2 x_{34}^2} \right)^{\Delta_3 - \Delta_{1k_1,2}-m_1+\gamma_{14} \over 2}  L_4^\prime(\Delta_4^\prime) \cr 
 &= \Gamma(\Delta_{12,k_1}-m_1) \Gamma(\Delta_{34,k_1}-m_1) L(\Delta_1,\Delta_2,\Delta_3,\Delta_4)\, u_1^{{\Delta_{k_1}\over 2} + m_1} \int {d\gamma_{14} \over 2\pi i} \, v_{14}^{-\gamma_{14}}  \cr 
 &\quad \times  \Gamma(\gamma_{14}) \Gamma(\Delta_{23,14} + \gamma_{14}) \Gamma(\Delta_{1k_1,2}+m_1-\gamma_{14}) \Gamma(\Delta_{k_14,3} +m_1-\gamma_{14}) \,,
}
where in the second equality, we have identified the cross ratios $u_1, v_{14}$ using~\eno{4uv} and defined the leg factor $L(\Delta_1,\Delta_2,\Delta_3,\Delta_4) = L_1(\Delta_1) L_2(\Delta_2) L_3(\Delta_3) L_4(\Delta_4)$ using~\eno{X}-\eno{2ILegs} and the choice of OPE vertex shown in the top line of~\eno{fourpointudiagrams}. More precisely,
\eqn{}{
L_1(\Delta_1) &= \left({x_{23}^2 \over x_{12}^2 x_{13}^2} \right)^{\Delta_1 \over 2} 
\qquad
L_2(\Delta_2) = \left({x_{13}^2 \over x_{12}^2 x_{23}^2} \right)^{\Delta_2 \over 2} 
\cr 
L_3(\Delta_3) &= \left({x_{24}^2 \over x_{23}^2 x_{34}^2} \right)^{\Delta_3 \over 2} 
\qquad 
L_4(\Delta_4) = \left({x_{23}^2 \over x_{24}^2 x_{34}^2} \right)^{\Delta_4 \over 2} \,.
}

Before evaluating the final $\gamma_{14}$ integral, we rewrite the factor of $v_{14}^{-\gamma_{14}}$ as
\eqn{}{
v_{14}^{-\gamma_{14}} = {1 \over \Gamma(\gamma_{14})} \int {d\widetilde{\gamma}_{14} \over 2\pi i} \Gamma(-\widetilde{\gamma}_{14}) \Gamma(\widetilde{\gamma}_{14} + \gamma_{14})  (1-v_{14})^{\widetilde{\gamma}_{14}}\,.
}
Substituting this in $B_4^{\rm s.t.}$, we get
\eqn{}{
B_4^{\rm s.t.} &=  \Gamma(\Delta_{12,k_1}-m_1) \Gamma(\Delta_{34,k_1}-m_1) L(\Delta_1,\Delta_2,\Delta_3,\Delta_4)\, u_1^{{\Delta_{k_1}\over 2} + m_1} \int {d\gamma_{14} \, d\widetilde{\gamma}_{14} \over (2\pi i)^2} \, (1-v_{14})^{\widetilde{\gamma}_{14}}  \cr 
 &\quad \times   \Gamma(-\widetilde{\gamma}_{14}) \Gamma(\widetilde{\gamma}_{14} + \gamma_{14})  \Gamma(\Delta_{23,14} + \gamma_{14}) \Gamma(\Delta_{1k_1,2}+m_1-\gamma_{14}) \Gamma(\Delta_{k_14,3} +m_1-\gamma_{14}) \,.
}
The utility of introducing an additional contour now becomes clear. We can evaluate the $\gamma_{14}$ contour integral using the first Barnes lemma \eno{FirstBarnes} which will subsequently leave us with a trivial-to-evaluate contour integral in the $\widetilde{\gamma}_{14}$-plane: 
\eqn{}{
B_4^{\rm s.t.} &= \Gamma(\Delta_{12,k_1}-m_1) \Gamma(\Delta_{34,k_1}-m_1) 
\Gamma(\Delta_{2k_1,1}+m_1) \Gamma(\Delta_{3k_1,4}+m_1)
L(\Delta_1,\Delta_2,\Delta_3,\Delta_4) \cr 
&\quad \times  u_1^{{\Delta_{k_1}\over 2} + m_1} \int { d\widetilde{\gamma}_{14} \over 2\pi i} \, (1-v_{14})^{\widetilde{\gamma}_{14}}   \Gamma(-\widetilde{\gamma}_{14}) 
  {\Gamma(\widetilde{\gamma}_{14} + \Delta_{1k_1,2}+m_1)  \Gamma(\widetilde{\gamma}_{14} +\Delta_{4k_1,3}+m_1)  \over \Gamma(\Delta_{k_1}+\widetilde{\gamma}_{14}+2m_1)}\,.
}
We close the contour in the $\widetilde{\gamma}_{14}$-plane to the right, picking up contributions from the simple poles at $\widetilde{\gamma}_{14} = j_{14}$ for all $j_{14} \in \mathbb{Z}^{\geq 0}$:
\eqn{}{
B_4^{\rm s.t.} &= L(\Delta_1,\Delta_2,\Delta_3,\Delta_4)  \Gamma(\Delta_{12,k_1}-m_1) \Gamma(\Delta_{34,k_1}-m_1) 
\Gamma(\Delta_{2k_1,1}+m_1) \Gamma(\Delta_{3k_1,4}+m_1)
\cr 
&\quad \times  u_1^{{\Delta_{k_1}\over 2} + m_1} \sum_{j_{14}=0}^{\infty} {(1-v_{14})^{j_{14}}  \over j_{14}!} 
  {\Gamma( \Delta_{1k_1,2}+m_1+j_{14})  \Gamma(\Delta_{4k_1,3}+m_1+j_{14})  \over \Gamma(\Delta_{k_1}+j_{14}+2m_1)}\,.
}
It is easily verified that the expression above reproduces the Feynman-like rule prescription given in~\eno{RulesBMst} for $M=4$.

We may go one step further. Substituting the final result in~\eno{CB-st} we can finally write down an explicit series representation for the four-point conformal block:
\eqn{}{
W_4(x_i) &=  L(\Delta_1,\Delta_2,\Delta_3,\Delta_4)  \sum_{m_1,j_{14}=0}^\infty { u_1^{{\Delta_{k_1}\over 2} + m_1}  \over m_1!} {(1-v_{14})^{j_{14}}  \over j_{14}!} { (\Delta_{k_1}-h+1)_{m_1} \over (\Delta_{k_1})_{j_{14}+2m_1} }  \cr 
 & \times (\Delta_{12,k_1})_{-m_1} (\Delta_{1k_1,2})_{m_1+j_{14}} 
(\Delta_{2k_1,1})_{m_1} F_A^{(1)}[\Delta_{12k_1,}-h; \{-m_1\};\{\Delta_{k_1}-h+1 \}; 1]
 \cr 
& \times (\Delta_{34,k_1})_{-m_1}   (\Delta_{3k_1,4})_{m_1}  (\Delta_{4k_1,3})_{m_1+j_{14}}  F_A^{(1)}[\Delta_{34k_1,}-h; \{-m_1\};\{\Delta_{k_1}-h+1 \}; 1]
}
which reproduces the well-known result from the literature~\cite{Dolan:2000ut},\footnote{Note that 
\eqn{}{
F_A^{(1)}[\Delta_{12k_1,-h};\{-m_1\};\{\Delta_{k_1}-h+1\};1] = (1-\Delta_{12,k_1})_{m_1}/(\Delta_{k_1}-h+1)_{m_1} \,.
} 
} consistent with the Feynman rules of Section~\ref{FEYNMAN}.

\subsection{Four-point to Five-point}
\label{Base case 4}

In this section we demonstrate that if $B_4^{\rm s.t.}$ defined by~\eno{BMst} (after setting $M=4$) evaluates to the series representation~\eno{RulesBMst},\footnote{Indeed, the expression we obtained for $B_4^{\rm s.t.}$ in Section~\ref{Base case 3} from first principles is precisely the one given in~\eno{RulesBMst}.}  then $B_5^{\rm s.t.}$ defined by~\eno{BMst} (after setting $M=5$) automatically satisfies~\eno{RulesBMst} as well.

\begin{figure}[!t]
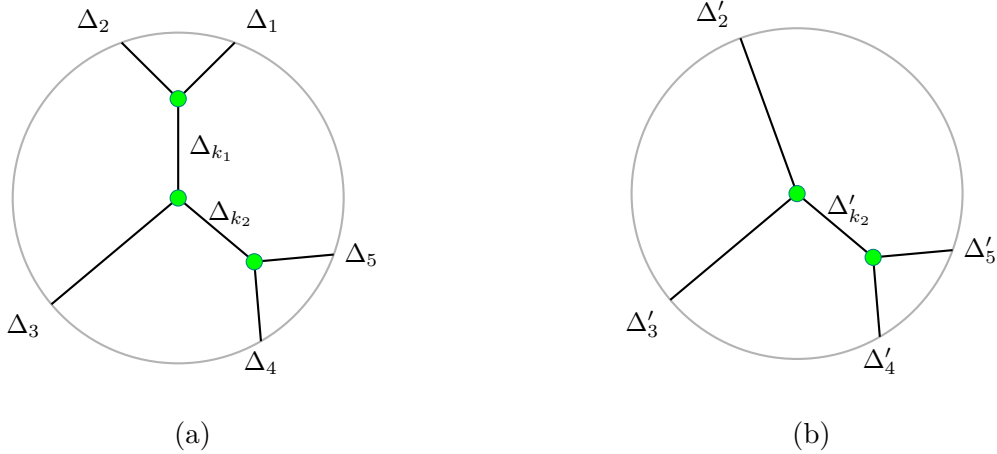

    \centering
    \begin{subfigure}[b]{0.49\textwidth}
    \[ \musepic{\figFiveptAdS}  \]
    \caption{}
    \label{fig:5ptAdS}
    \end{subfigure}
    \begin{subfigure}[b]{0.49\textwidth}
    \[ \musepic{\figFourptAdS}  \]
    \caption{}
    \label{fig:4ptAdS}
    \end{subfigure}
    \caption{(a) A canonical $5$-point AdS diagram, and (b) the associated $4$-point diagram.}
    \label{fig:45AdS}
\end{figure}

\begin{figure}[t!]
\centering
\resizebox{14cm}{!}{%
\begin{tikzpicture}[thick, scale=2.0]
    \coordinate (y1) at (-3/2,0);
    \coordinate (y2) at (-1,0);
    \coordinate (x3) at (-1,1);
    \coordinate (y3) at (0,0);
    \coordinate (z3) at (0,1/2);
    \coordinate (x4) at (0,1);
    \coordinate (y4) at (1,0);
    \coordinate (z4) at (1,1/2);
    \coordinate (x5) at (1,1);
    \coordinate (y5) at (3/2,0);
    \coordinate (x6) at (2,1);
    \coordinate (y6) at (9/2,0);
	\draw[thick] (y1)--(y2)--(x3);
	\draw[thick] (y3)--(x4);
	\draw (y1) node[anchor=east] { $\mathcal{O}_{1}(x_{1})$};
	\draw (x3) node[anchor=south] { $\mathcal{O}_{2}(x_{2})$};
	\draw (x4) node[anchor=south] { $\mathcal{O}_{3}(x_{3})$};
	\draw[thick] (y2)--(y3); 
	\draw ($(y2)!0.5!(y3)$) node[anchor=north] {\ $\mathcal{O}_{k_1}$};
	\draw[thick] (y3)--(y4); 
	\draw ($(y3)!0.5!(y4)$) node[anchor=north] { $\mathcal{O}_{k_2}$};
	\draw[thick] (x5)--(y4); 
	\draw (x5) node[anchor=south] { $\mathcal{O}_{4}(x_4)$};
	\draw[thick] (y5)--(y4); 
	\draw (y5) node[anchor=west] { $\mathcal{O}_{5}(x_5)$};
\end{tikzpicture} 
\begin{tikzpicture}[thick]
\begin{scope}[xshift=2cm]
\node at (0,0) {$\mathcal{O}_{2}^\prime(x_{2})$};
\draw[-] (1,0)--(4.5,0);
\draw[-] (2,0)--(2,1) node at (2,1.5) {$\mathcal{O}_{3}^\prime(x_{3})$};
\draw[-] (3.5,0)--(3.5,1) node at (3.5,1.5) {$\mathcal{O}_{4}^\prime(x_{4})$};
\draw[-] (3.5,0)--(4.5,0) node at (5.5,0)  {$\mathcal{O}_{5}^\prime(x_{5})$};
\node at (2.8,-0.5) {$\mathcal{O}_{k_{2}}^\prime$};
\end{scope}
\end{tikzpicture}
}
\caption{The $5$-point and $4$-point blocks, respectively. The $4$-point topology is obtained by removing the $\mathcal{O}_{1}(x_{1})$ leg. }
\label{FigU4ptOPEp}
\end{figure}
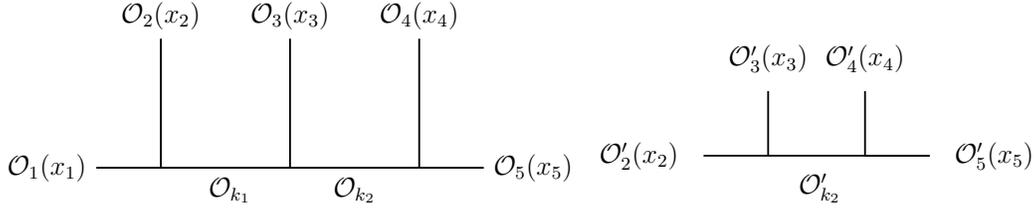

Our starting point is the integral representation for $B_5^{\rm s.t.}$ in terms of that for $B_4^{\rm s.t.}$,
\eqn{B5st}{
B_5^{\rm s.t.} =\left[\left(\prod_{a=2}^5 \int {d\gamma_{1a} \over 2\pi i} {\Gamma(\gamma_{1a}) \over (x_{1a}^2)^{\gamma_{1a}} } \right) 
2\pi i\delta(\sum_{a=2}^5 \gamma_{1a} - \Delta_1) 
 {1 \over {\Delta_{k_1}-s_1 \over 2} + m_1}B^{\text{s.t.}}_{4}\right]_{\text{s.t.}},
}
where 
\eqn{s1fivept}{
s_1 = \Delta_1 + \Delta_2 - 2\gamma_{12} \,,
}
and $B^{\text{s.t.}}_{4}$ is given by
\eqn{B4stNew}{
B^{\text{s.t.}}_{4} &= \left[\left(\prod_{(rs) \in {\cal U}^\prime \bigcup {\cal V}^\prime \bigcup {\cal D}^\prime} \int{ d \gamma_{rs} \over 2\pi i} {\Gamma(\gamma_{rs}) \over (x_{rs}^2)^{\gamma_{rs}} } \right) \!\!
\left(\prod_{a=2}^5 2\pi i\delta(\sum_{\substack{b=1\\b\neq a}}^5 \gamma_{ab} - \Delta_a) \right) \!\!
 {1 \over {\Delta_{k_2}-s_2 \over 2} +m_2} \right]_{\text{s.t.}}\!\!\!.
}
Here we have in mind the canonical diagrams shown in Figure~\ref{fig:45AdS}. The associated conformal blocks are those in Figure~\ref{FigU4ptOPEp}.
The primed Mellin index sets ${\cal U}^\prime, {\cal V}^\prime, {\cal D}^\prime$ are chosen as described in Section~\ref{SETINDUCTION}.
The unprimed Mellin index sets ${\cal U}, {\cal V}, {\cal D},$ will be determined after picking a convenient choice of the OPE vertex for the five-point block.

To proceed, we would like to re-express the Mandelstam invariants and conformal dimensions of the canonical five-point diagram appearing in~\eno{B4stNew} entirely in terms of the Mandelstam invariants and conformal dimensions of the reduced four-point diagram shown in Figure~\ref{fig:4ptAdS}. 
To this end, we define
\eqn{}{
\Delta_a^\prime = \Delta_a - \gamma_{1a}
}
for $a=2,\ldots,5$, so that the delta-functions in~\eno{B4stNew} can be rewritten as $\delta(\sum_{\substack{b=2\\b\neq a}}^5 \gamma_{ab} - \Delta_a^\prime)$, which is now entirely in terms of the Mellin variables of the reduced four-point diagram, corresponding to the primed index set ${\cal U}^\prime \bigcup {\cal V}^\prime \bigcup {\cal D}^\prime$ as necessary.
Furthermore, we define
\eqn{}{
s_2^\prime :=  \Delta_4^\prime + \Delta_5^\prime - 2\gamma_{45} 
}
such that the Mandelstam invariant of the reduced four-point diagram is written entirely in terms of four-point quantities.
Then, upon requiring
\eqn{}{
\Delta_{k_2} - s_2 = \Delta_{k_2}^\prime - s_2^\prime\,.
}
we obtain
\eqn{}{
\Delta_{k_2}^\prime = \Delta_{k_2} - \gamma_{14} - \gamma_{15} \,.
}
We can now rewrite~\eno{B4stNew} entirely in terms of physical quantities associated with the reduced four-point diagram,
\eqn{}{
B^{\text{s.t.}}_{4} &= \left[\left(\prod_{(rs) \in {\cal U}^\prime \bigcup {\cal V}^\prime \bigcup {\cal D}^\prime} \int{ d \gamma_{rs} \over 2\pi i} {\Gamma(\gamma_{rs}) \over (x_{rs}^2)^{\gamma_{rs}} } \right) \!\!
\left(\prod_{a=2}^5 2\pi i\delta(\sum_{\substack{b=2\\b\neq a}}^5 \gamma_{ab} - \Delta_a^\prime) \right) \!\!
 {1 \over {\Delta_{k_2}^\prime-s_2^\prime \over 2} +m_2} \right]_{\text{s.t.}}\!\!\!.
}
In Appendix~\ref{Base case 3} we showed that $B_4^{\rm s.t.}$ as written above satisfies the Feynman-like rules~\eno{RulesBMst}, and thus admits the following series representation,
\eqn{}{
B^{\text{s.t.}}_4 &=L^\prime(\Delta_2^\prime,\ldots,\Delta_5^\prime) \, (u_2^\prime)^{\frac{\Delta_{k_2}^\prime}{2}+m_2} \left(\prod_{(rs)\in\mathcal{V}^\prime} \sum_{j_{rs}=0}^{\infty} \frac{(1-v_{rs}^\prime)^{j_{rs}}}{j_{rs}!}\right) \widehat{V}_1^\prime \widehat{V}_2^\prime \widehat{E}_{2}^\prime\,,
}
where $\widehat{V}_a^\prime$ and $\widehat{E}_a^\prime$ are ``Gamma-vertex'' and ``Gamma-edge'' factors for the four-point conformal blocks, given by
\eqn{4factorbase}{
\widehat{V}_1^\prime &= \Gamma(\Delta^\prime_{23,k_2}-m_{2}+\frac{1}{2}\ell^\prime_{23,k_2}) \Gamma(\Delta^\prime_{3k_2,2}+m_{2}+\frac{1}{2}\ell^\prime_{3k_2,2}) \Gamma(\Delta^\prime_{2k_2,3}+m_{2}+\frac{1}{2}\ell^\prime_{2k_2,3})\cr
\widehat{V}_2^\prime &= \Gamma(\Delta^\prime_{45,k_2}-m_{2}+\frac{1}{2}\ell^\prime_{45,k_2}) \Gamma(\Delta^\prime_{5k_2,4}+m_{2}+\frac{1}{2}\ell^\prime_{5k_2,4}) \Gamma(\Delta^\prime_{4k_2,5}+m_{2}+\frac{1}{2}\ell^\prime_{4k_2,5})\cr
\widehat{E}_2^\prime &= \frac{1}{\Gamma(\Delta_{k_2}^\prime+2m_{2}+\ell_{k_2}^\prime)},
}
where $\ell_a^\prime$ are the post-Mellin parameters for the four-point topology.
To write down the post-Mellin parameters as well as the cross ratios explicitly, we need to make a choice of OPE vertex. We make the following convenient choice (refer to Sections~\ref{INDEXCROSSRATIOS} and~\ref{4POINTFLOWS}):
\eqn{OPEFlow4}{ 
\begin{tikzpicture}[thick]
\begin{scope}[xshift=2cm]
\node at (-.2,0) {$x_2$};
\node at (.5,-.3) {$V_{2}$};
\draw[-] (0.0,0)--(1,0);
\draw[-] (.5,0)--(.5,1) node at (.5,1.2) {$x_{3}$};
\draw[-] (.5,0)--(1,0) node at (1.2,0) {$x_4$};
\end{scope}
\begin{scope}[xshift=4cm]
\node at (-.2,0) {$x_3$};
\node at (.5,-.3) {$V_{3}$};
\draw[-] (0.0,0)--(1,0);
\draw[-] (.5,0)--(.5,1) node at (.5,1.2) {$x_{4}$};
\draw[-] (.5,0)--(1,0) node at (1.2,0) {$x_{5}$};
\end{scope}
\end{tikzpicture}
\qquad & :
\begin{tikzpicture}[thick]
\begin{scope}[xshift=2cm]
\node at (-.2,0) {$x_2$};
\node at (.5,-.3) {$V_{2}$};
\node at (1.6,-0.5) {$\mathcal{O}_{k_{2}}$};
\draw[-] (0.0,0)--(1,0);
\draw[-] (.5,0)--(.5,1) node at (.5,1.2) {$x_{3}$};
\draw[-] (.5,0)--(1,0) node at (1.2,0) {$x_4$};
\draw[black] (-.25,-.05) circle (.3 cm);
\draw[dashed,  red]  (0,-.08) to[bend right=5] (.9,-.08);
\draw[dashed,  teal]  (0,.08) to[bend right=5] (.45,1);
\end{scope}
\begin{scope}[xshift=4cm]
\node at (-.2,0) {$x_3$};
\node at (.5,-.3) {$V_{3}$};
\draw[-] (0.0,0)--(1,0);
\draw[-] (.5,0)--(.5,1) node at (.5,1.2) {$x_{4}$};
\draw[-] (.5,0)--(1,0) node at (1.2,0) {$x_{5}$};
\draw[black] (1.25,-.05) circle (.3 cm);
\draw[dashed,  red]  (0,-.08) to[bend right=5] (.9,-.08);
\draw[dashed,  teal]  (.55,1) to[bend right=5] (.8, .1);
\end{scope}
\end{tikzpicture}
\qquad : u_2' = {x_{23}^{2} x_{45}^2 \over x_{24}^2x_{35}^5 } 
\cr &
\begin{tikzpicture}[thick]
\begin{scope}[xshift=0cm]
\node at (-.2,0) {$x_{2}$};
\node at (.5,-.3) {$V_{2}$};
\draw[-] (0.0,0)--(1,0);
\draw[-] (.5,0)--(.5,1) node at (.5,1.2) {$x_{3}$};
\draw[-] (.5,0)--(1,0) node at (1.2,0) {$x_{4}$};
\draw[cyan] (1.5,.1) circle (.6 cm);
\draw[dashed,  orange]  (0,-.08) to[bend right=5] (.9,-.08);
\end{scope}
\begin{scope}[xshift=2cm]
\node at (-.2,0) {$x_{3}$};
\node at (.5,-.3) {$V_{3}$};
\draw[-] (0.0,0)--(1,0);
\draw[-] (.5,0)--(.5,1) node at (.5,1.2) {$x_{4}$};
\draw[-] (.5,0)--(1,0) node at (1.2,0) {$x_{5}$};
\draw[dashed,  orange]  (0,-.08) to[bend right=5] (.9,-.08);
\end{scope}
\end{tikzpicture} \qquad 
: V_{2}V_{3}:
v_{25}'= {x_{25}^2 x_{34}^2 \over x_{24}^2 x_{35}^2}
}
where we have also shown the four-point cross ratios associated with the choice of OPE flow.
The corresponding leg factor is given by 
\eqn{}{ 
L^\prime(\Delta_2^\prime,\ldots,\Delta_5^\prime)  = 
\left({x_{34}^2 \over x_{23}^2x_{24}^2}\right)^{\Delta_2^\prime \over 2}
\left({x_{24}^2 \over x_{23}^2x_{34}^2}\right)^{\Delta_3^\prime \over 2}
\left({x_{35}^2 \over x_{34}^2x_{45}^2}\right)^{\Delta_4^\prime \over 2}
\left({x_{34}^2 \over x_{45}^2x_{35}^2}\right)^{\Delta_5^\prime \over 2} .
}

The associated index sets are
\eqn{}{ 
{\cal{V}'} = \{ (25)\}  \qquad {\cal {U}'} = \{(45) \}  \qquad {\cal {D}}' = \{(23), (24), (34), (35) \}\,.
}
With the ${\cal V}^\prime$ index set in hand, we can use the rules described in Section~\ref{POSTMELLIN} to work out the post-Mellin parameters appearing in~\eno{4factorbase}:
\eqn{}{ 
\begin{gathered}
\ell_2'= j_{25}
\qquad
\ell_3'=0
\qquad 
\ell_{4}'=0
\qquad 
\ell_{5}'= j_{25}
\cr 
\ell_{k_2}'= \ell_4'+\ell_5'= \ell_{2}'+\ell_{3}'= j_{25}\,.
\end{gathered}
}

We can now plug $B_4^{\rm s.t.}$ back into the starting point~\eno{B5st} and proceed to evaluate the remaining four-dimensional contour integral,
\eqn{B5stNew}{
B_5^{\rm s.t.} &= \left[\int {d\gamma_{12}d\gamma_{13}d\gamma_{14}d\gamma_{15} \over (2\pi i)^4} 
{\Gamma(\gamma_{12}) \over (x_{12}^2)^{\gamma_{12}} }  
{\Gamma(\gamma_{13}) \over (x_{13}^2)^{\gamma_{13}} } 
{\Gamma(\gamma_{14}) \over (x_{14}^2)^{\gamma_{14}} } 
{\Gamma(\gamma_{15}) \over (x_{15}^2)^{\gamma_{15}} }
2\pi i\delta(\sum_{a=2}^5\gamma_{1a} - \Delta_1) 
 {1 \over {\Delta_{k_1}-s_1 \over 2} + m_1} \right.
\cr  
& \left. \times L^\prime(\Delta_2^\prime,\ldots,\Delta_5^\prime) \, (u_2^\prime)^{\frac{\Delta_{k_2}^\prime}{2}+m_2}  \sum_{j_{25}=0}^{\infty} \frac{(1-v_{25}^\prime)^{j_{25}}}{j_{25}!} \widehat{V}_1^\prime \widehat{V}_2^\prime \widehat{E}_{2}^\prime\right]_{\rm s.t.} ,
}
where the integrand is now explicitly specified.
Exactly like in the previous subsection, we first perform the single-trace projection by picking up the residue at the simple pole in the $\gamma_{12}$-plane [recall~\eno{s1fivept}], followed by the trivial integral in the $\gamma_{13}$-plane which simply eats up the delta-function.
This leads to
\eqn{B5st12}{
B_5^{\rm s.t.} &=
\int {d\gamma_{14}d\gamma_{15} \over (2\pi i)^2} 
{\Gamma(\Delta_{12,k_1}-m_1) \over (x_{12}^2)^{\Delta_{12,k_1}-m_1} }  
{\Gamma( \Delta_{1k_1,2}+m_1-\gamma_{14}-\gamma_{15}) \over (x_{13}^2)^{ \Delta_{1k_1,2}+m_1-\gamma_{14}-\gamma_{15} } }
{\Gamma(\gamma_{14}) \over (x_{14}^2)^{\gamma_{14}} } 
{\Gamma(\gamma_{15}) \over (x_{15}^2)^{\gamma_{15}} }
\cr 
 &\times  L^\prime(\Delta_2-\Delta_{12,k_1}-m_1,\Delta_3+\Delta_{2,1k_1}-m_1+\gamma_{14}+\gamma_{15},\Delta_4-\gamma_{14},\Delta_5-\gamma_{15}) 
 \cr 
 &\times \left(u_2^\prime\right)^{\frac{\Delta_{k_2}-\gamma_{14}-\gamma_{15}}{2}+m_2}
\sum_{j_{25}=0}^{\infty} \frac{
(1-v_{25}^\prime)^{j_{25}}}{j_{25}!}
\frac{1}{\Gamma(\Delta_{k_2}-\gamma_{14}-\gamma_{15}+2m_{2}+j_{25})}
\cr 
& \times  \Gamma(\Delta_{23,1k_2}+\gamma_{14}+\gamma_{15}-m_2)
\Gamma(\Delta_{45,k_2}-m_2)
\Gamma(\Delta_{k_1k_2,3}+m_{12,}-\gamma_{14}-\gamma_{15}+j_{25})
\cr
& \times \Gamma(\Delta_{3k_2,k_1}+m_{2,1})
\Gamma(\Delta_{4k_2,5}-\gamma_{14}+m_2)
\Gamma(\Delta_{5k_2,4}-\gamma_{15}+m_2+j_{25}) \,.
}

Before we proceed it is worthwhile to reorganize the position space factors in the integrand by collecting together factors with common exponents. This yields
\eqn{B5st12again}{
B_5^{\rm s.t.} &= 
\left({x_{23}^2\over x_{12}^2x_{13}^2}\right)^{{\Delta_{1}\over2}}
\left({x_{13}^2\over x_{23}^2x_{12}^2}\right)^{{\Delta_{2}\over 2}}
\left({x_{24}^2\over x_{23}^2x_{34}^2}\right)^{{\Delta_{3}\over 2}}
\left({x_{35}^2 \over x_{34}^2x_{45}^2 }   \right) ^{{\Delta_4 \over 2}}
\left({x_{34}^2 \over x_{35}^2x_{45}^2 }  \right)  ^{{\Delta_5 \over 2}} 
\cr 
&\times 
\left({x_{12}^2 x_{34}^2 \over x_{13}^2 x_{24}^2}\right)^{{\Delta_{k_1}\over 2}+m_1} \left(u_2^\prime\right)^{{\Delta_{k_2}\over 2}+m_2}
\Gamma(\Delta_{12,k_1}-m_1) \Gamma(\Delta_{45,k_2}-m_2) \Gamma(\Delta_{3k_2,k_1}+m_{2,1})
\cr 
&\times 
\int {d\gamma_{14}d\gamma_{15} \over (2\pi i)^2}  
\Gamma( \Delta_{1k_1,2}+m_1-\gamma_{14}-\gamma_{15})  
\Gamma(\gamma_{14}) 
\Gamma(\gamma_{15}) 
 \left({ x_{23}^2x_{14}^2 \over x_{24}^2x_{13}^2 }\right)^{-\gamma_{14}}
 \left( x_{23}^2x_{34}^2x_{15}^2 \over x_{24}^2x_{35}^2 x_{13}^2 \right)^{-\gamma_{15}}
 \cr 
 &\times 
\sum_{j_{25}=0}^{\infty} \frac{
(1-v_{25}^\prime)^{j_{25}}}{j_{25}!}
\frac{ \Gamma(\Delta_{23,1k_2}+\gamma_{14}+\gamma_{15}-m_2)
\Gamma(\Delta_{k_1k_2,3}+m_{12,}-\gamma_{14}-\gamma_{15}+j_{25})}{\Gamma(\Delta_{k_2}-\gamma_{14}-\gamma_{15}+2m_{2}+j_{25})}
\cr
& \times 
\Gamma(\Delta_{4k_2,5}-\gamma_{14}+m_2)
\Gamma(\Delta_{5k_2,4}-\gamma_{15}+m_2+j_{25}) \,.
}
The ratios of position space factors above take forms suggestive of cross ratios and leg factors. Indeed, the following extension of the four-point OPE flow~\eno{OPEFlow4} to the five-point topology shown in Figure~\ref{FigU4ptOPEp}, yields precisely these combinations as five-point cross ratios and leg factors [refer to Sections~\ref{INDEXCROSSRATIOS} and~\ref{5POINTFLOWS}]:
\eqn{ucross5A}{ 
&
\begin{tikzpicture}[thick]
\begin{scope}
\node at (-.2,0) {$x_{1}$};
\node at (.5,-.3) {$V_{1}$};
\draw[-] (0.0,0)--(1,0);
\draw[-] (.5,0)--(.5,1) node at (.5,1.2) {$x_{2}$};
\draw[-] (.5,0)--(1,0) node at (1.2,0) {$x_3$};
\end{scope}
\begin{scope}[xshift=2cm]
\node at (-.2,0) {$x_2$};
\node at (.5,-.3) {$V_{2}$};
\draw[-] (0.0,0)--(1,0);
\draw[-] (.5,0)--(.5,1) node at (.5,1.2) {$x_{3}$};
\draw[-] (.5,0)--(1,0) node at (1.2,0) {$x_4$};
\end{scope}
\begin{scope}[xshift=4cm]
\node at (-.2,0) {$x_3$};
\node at (.5,-.3) {$V_{3}$};
\draw[-] (0.0,0)--(1,0);
\draw[-] (.5,0)--(.5,1) node at (.5,1.2) {$x_{4}$};
\draw[-] (.5,0)--(1,0) node at (1.2,0) {$x_{5}$};
\end{scope}
\end{tikzpicture}
\cr 
&
\begin{tikzpicture}[thick]
\begin{scope}
\node at (-.2,0) {$x_{1}$};
\node at (.5,-.3) {$V_{1}$};
\node at (1.6,-0.5) {$\mathcal{O}_{k_{1}}$};
\draw[-] (0.0,0)--(1,0);
\draw[-] (.5,0)--(.5,1) node at (.5,1.2) {$x_{2}$};
\draw[-] (.5,0)--(1,0) node at (1.2,0) {$x_3$};
\draw[black] (-.25,-.05) circle (.3 cm);
\draw[dashed,  red]  (0,-.08) to[bend right=5] (.9,-.08);
\draw[dashed,  teal]  (0,.08) to[bend right=5] (.45,1);
\end{scope}
\begin{scope}[xshift=2cm]
\node at (-.2,0) {$x_2$};
\node at (.5,-.3) {$V_{2}$};
\draw[-] (0.0,0)--(1,0);
\draw[-] (.5,0)--(.5,1) node at (.5,1.2) {$x_{3}$};
\draw[-] (.5,0)--(1,0) node at (1.2,0) {$x_4$};
\draw[black] (1.25,-.05) circle (.3 cm);
\draw[dashed,  red]  (0,-.08) to[bend right=5] (.9,-.08);
\draw[dashed,  teal]  (.55,1) to[bend right=5] (.8, .1);
\end{scope}
\end{tikzpicture}: u_1 = \frac{x_{12}^2x_{43}^2}{x_{13}^2x_{42}^2}
\cr 
&\begin{tikzpicture}[thick]
\begin{scope}
\node at (-.2,0) {$x_{2}$};
\node at (.5,-.3) {$V_{2}$};
\node at (1.6,-0.5) {$\mathcal{O}_{k_{2}}$};
\draw[-] (0.0,0)--(1,0);
\draw[-] (.5,0)--(.5,1) node at (.5,1.2) {$x_{3}$};
\draw[-] (.5,0)--(1,0) node at (1.2,0) {$x_4$};
\draw[black] (-.25,-.05) circle (.3 cm);
\draw[dashed,  red]  (0,-.08) to[bend right=5] (.9,-.08);
\draw[dashed,  teal]  (0,.08) to[bend right=5] (.45,1);
\end{scope}
\begin{scope}[xshift=2cm]
\node at (-.2,0) {$x_3$};
\node at (.5,-.3) {$V_{3}$};
\draw[-] (0.0,0)--(1,0);
\draw[-] (.5,0)--(.5,1) node at (.5,1.2) {$x_{4}$};
\draw[-] (.5,0)--(1,0) node at (1.2,0) {$x_5$};
\draw[black] (1.25,-.05) circle (.3 cm);
\draw[dashed,  red]  (0,-.08) to[bend right=5] (.9,-.08);
\draw[dashed,  teal]  (.55,1) to[bend right=5] (.8, .1);
\end{scope}
\end{tikzpicture}
: u_{2}=\frac{x_{23}^2x_{54}^2}{x^2_{24}x^2_{53}}
}
\begingroup\makeatletter\def\f@size{10}\check@mathfonts 
\eqn{vcross5A}{ 
&\begin{tikzpicture}[thick]
\begin{scope}[xshift=0cm]
\node at (-.2,0) {$x_{1}$};
\node at (.5,-.3) {$V_{1}$};
\draw[-] (0.0,0)--(1,0);
\draw[-] (.5,0)--(.5,1) node at (.5,1.2) {$x_{2}$};
\draw[-] (.5,0)--(1,0) node at (1.2,0) {$x_{3}$};
\draw[cyan] (1.5,.1) circle (.6 cm);
\draw[dashed,  orange]  (0,-.08) to[bend right=5] (.9,-.08);
\end{scope}
\begin{scope}[xshift=2cm]
\node at (-.2,0) {$x_{2}$};
\node at (.5,-.3) {$V_{2}$};
\draw[-] (0.0,0)--(1,0);
\draw[-] (.5,0)--(.5,1) node at (.5,1.2) {$x_{3}$};
\draw[-] (.5,0)--(1,0) node at (1.2,0) {$x_{4}$};
\draw[dashed,  orange]  (0,-.08) to[bend right=5] (.9,-.08);
\end{scope}
\end{tikzpicture}
:V_{1}V_{2} :v_{14}= {x_{14}^2 x_{23}^2 \over x_{13}^2 x_{24}^2  }
\qquad 
\begin{tikzpicture}[thick]
\begin{scope}[xshift=2cm]
\node at (-.2,0) {$x_{2}$};
\node at (.5,-.3) {$V_{2}$};
\draw[-] (0.0,0)--(1,0);
\draw[-] (.5,0)--(.5,1) node at (.5,1.2) {$x_{3}$};
\draw[-] (.5,0)--(1,0) node at (1.2,0) {$x_{4}$};
\draw[cyan] (1.5,.1) circle (.6 cm);
\draw[dashed,  orange]  (0,-.08) to[bend right=5] (.9,-.08);
\end{scope}
\begin{scope}[xshift=4cm]
\node at (-.2,0) {$x_{3}$};
\node at (.5,-.3) {$V_{3}$};
\draw[-] (0.0,0)--(1,0);
\draw[-] (.5,0)--(.5,1) node at (.5,1.2) {$x_{4}$};
\draw[-] (.5,0)--(1,0) node at (1.2,0) {$x_{5}$};
\draw[dashed,  orange]  (0,-.08) to[bend right=5] (.9,-.08);
\end{scope}
\end{tikzpicture}
:V_{2}V_{3} :v_{25}= {x_{25}^2 x_{34}^2 \over x_{24}^2 x_{35}^2  }
\cr 
&\begin{tikzpicture}[thick]
\begin{scope}[xshift=0cm]
\node at (-.2,0) {$x_{1}$};
\node at (.5,-.3) {$V_{1}$};
\draw[-] (0.0,0)--(1,0);
\draw[-] (.5,0)--(.5,1) node at (.5,1.2) {$x_{2}$};
\draw[-] (.5,0)--(1,0) node at (1.2,0) {$x_{3}$};
\draw[cyan] (1.5,.1) circle (.6 cm);
\draw[dashed,  orange]  (0,-.08) to[bend right=5] (.9,-.08);
\end{scope}
\begin{scope}[xshift=2cm]
\node at (-.2,0) {$x_{2}$};
\node at (.5,-.3) {$V_{2}$};
\draw[-] (0.0,0)--(1,0);
\draw[-] (.5,0)--(.5,1) node at (.5,1.2) {$x_{3}$};
\draw[-] (.5,0)--(1,0) node at (1.2,0) {$x_{4}$};
\draw[cyan] (1.5,.1) circle (.6 cm);
\draw[dashed,  orange]  (0,-.08) to[bend right=5] (.9,-.08);
\end{scope}
\begin{scope}[xshift=4cm]
\node at (-.2,0) {$x_{3}$};
\node at (.5,-.3) {$V_{3}$};
\draw[-] (0.0,0)--(1,0);
\draw[-] (.5,0)--(.5,1) node at (.5,1.2) {$x_{4}$};
\draw[-] (.5,0)--(1,0) node at (1.2,0) {$x_{5}$};
\draw[dashed,  orange]  (0,-.08) to[bend right=5] (.9,-.08);
\end{scope}
\end{tikzpicture}
:V_{1}V_{3} :v_{15}= {x_{15}^2 x_{23}^2 x_{34}^2 \over x_{13}^2 x_{24}^2x_{35}^2  }
}
\endgroup
\eqn{}{
\begin{gathered}
L_1(\Delta_1)L_2(\Delta_2) = \left( { x_{23}^2 \over x_{12}^2 x_{13}^2 } \right)^{\Delta_1 \over 2} \left( { x_{13}^2 \over x_{12}^2 x_{23}^2 } \right)^{\Delta_2 \over 2}
\qquad
L_3(\Delta_3) = \left( {x_{24}^2 \over x_{23}^2 x_{34}^2} \right)^{\Delta_3 \over 2}
\cr
L_4(\Delta_4)L_5(\Delta_5) = \left( { x_{35}^2 \over x_{34}^2 x_{45}^2 } \right)^{\Delta_4 \over 2} \left( { x_{34}^2 \over x_{35}^2 x_{45}^2 } \right)^{\Delta_5 \over 2} .
\end{gathered}
}
Note that with this scheme of OPE flow, the four-point cross ratios get uplifted to five-point cross ratios:  $u_2^\prime = u_2$ and $v_{25}^\prime = v_{25}$.
With this, we can rewrite~\eno{B5st12again} concisely as
\eqn{B5st12again2}{
B_5^{\rm s.t.} &= 
L_1(\Delta_1)L_2(\Delta_2)L_3(\Delta_3)L_4(\Delta_4)L_5(\Delta_5)\,
u_1^{{\Delta_{k_1}\over 2}+m_1} u_2^{{\Delta_{k_2}\over 2}+m_2}
\cr 
&\times 
\Gamma(\Delta_{12,k_1}-m_1) \Gamma(\Delta_{45,k_2}-m_2) \Gamma(\Delta_{3k_2,k_1}+m_{2,1})
\cr 
&\times 
\int {d\gamma_{14}d\gamma_{15} \over (2\pi i)^2}  
\Gamma( \Delta_{1k_1,2}+m_1-\gamma_{14}-\gamma_{15})  
\Gamma(\gamma_{14}) 
\Gamma(\gamma_{15}) \,
 v_{14}^{-\gamma_{14}}
 v_{15}^{-\gamma_{15}}
 \cr 
 &\times 
\sum_{j_{25}=0}^{\infty} \frac{
(1-v_{25})^{j_{25}}}{j_{25}!}
\frac{ \Gamma(\Delta_{23,1k_2}+\gamma_{14}+\gamma_{15}-m_2)
\Gamma(\Delta_{k_1k_2,3}+m_{12,}-\gamma_{14}-\gamma_{15}+j_{25})}{\Gamma(\Delta_{k_2}-\gamma_{14}-\gamma_{15}+2m_{2}+j_{25})}
\cr
& \times 
\Gamma(\Delta_{4k_2,5}-\gamma_{14}+m_2)
\Gamma(\Delta_{5k_2,4}-\gamma_{15}+m_2+j_{25}) \,.
}

Now, making the following substitutions:
\eqn{vrsfivebase}{ 
v_{14}^{-\gamma_{14}} &=  {1 \over \Gamma{(\gamma_{14}})} \int {d \widetilde{\gamma}_{14} \over 2 \pi i} \Gamma(-\widetilde{\gamma}_{14})\Gamma(\widetilde{\gamma}_{14}+\gamma_{14})\left(v_{14}-1\right)^{\widetilde{\gamma}_{14}}
\cr 
v_{15}^{-\gamma_{15}} &= {1 \over \Gamma{(\gamma_{15}})} \int {d \widetilde{\gamma}_{15} \over  2 \pi i} \Gamma(-\widetilde{\gamma}_{15})\Gamma(\widetilde{\gamma}_{15}+\gamma_{15})\left(v_{15}-1\right)^{\widetilde{\gamma}_{15}} ,
}
we proceed to evaluate the $\gamma_{14}, \gamma_{15}$ integrals. 
First, sending $\gamma_{14} \to \gamma_{14} - \gamma_{15}$, we use the first Barnes lemma~\eno{FirstBarnes} to evaluate the $\gamma_{15}$ integral. 
Next, we employ the first Barnes lemma once again to evaluate the $\gamma_{14}$ integral.
When the dust settles, we obtain
\eqn{}{ 
B_5^{\rm s.t.} &= 
L_1(\Delta_1)L_2(\Delta_2)L_3(\Delta_3)L_4(\Delta_4)L_5(\Delta_5)\,
u_1^{{\Delta_{k_1}\over 2}+m_1} u_2^{{\Delta_{k_2}\over 2}+m_2}
\cr 
&\times  
\Gamma(\Delta_{12,k_1}-m_1)
\Gamma(\Delta_{45,k_2}-m_2)
\Gamma(\Delta_{3k_2,k_1}+m_{2,1}) 
 \Gamma( \Delta_{3k_1,k_2}+m_{1,2})
 \Gamma( \Delta_{2k_1,1}+m_{1}+j_{25})
\cr 
& \times \sum_{j_{25}=0}^{\infty} \frac{
(1-v_{25})^{j_{25}}}{j_{25}!}
\int {d \widetilde{\gamma}_{14} \over 2 \pi i} \int {d \widetilde{\gamma}_{15} \over 2 \pi i}  \Gamma(-\widetilde{\gamma}_{14}) \Gamma(-\widetilde{\gamma}_{15})  \left(v_{14}-1\right)^{\widetilde{\gamma}_{14}} \left(v_{15}-1\right)^{\widetilde{\gamma}_{15}}
\cr 
& \times 
{{
{\Gamma(\widetilde{\gamma}_{15}+\Delta_{5k_2,4}+m_2+j_{25})}
\Gamma(\Delta_{4k_2,5}+m_2+\widetilde{\gamma}_{14})} \over
{\Gamma{(\Delta_{k_2}+\widetilde{\gamma}_{15}+\widetilde{\gamma}_{14}+2m_2+j_{25})}}}
\cr 
&  \times
\frac{
 \Gamma{(\widetilde{\gamma}_{15}+\widetilde{\gamma}_{14}+\Delta_{1k_1,2}+{m_1})}
 \Gamma{(\widetilde{\gamma}_{15}+\widetilde{\gamma}_{14}+\Delta_{k_1k_2,3}+m_{12,}+j_{25})}}
 {\Gamma{(\Delta_{k_1}+\widetilde{\gamma}_{15}+\widetilde{\gamma}_{14}+2m_{1}+j_{25})}}\,.
}
The resulting $\widetilde{\gamma}_{14}$ and $\widetilde{\gamma}_{15}$ integrals are trivial to evaluate after closing the contours on the right, where they pick up simple poles at non-negative integers $\widetilde{\gamma}_{14} = j_{14} \in \{ 0,1,2,\ldots \}$ and $\widetilde{\gamma}_{15} = j_{15} \in \{ 0,1,2,\ldots \}$, yielding
\eqn{B5final}{ 
B_5^{\rm s.t.} &= 
L_1(\Delta_1)L_2(\Delta_2)L_3(\Delta_3)L_4(\Delta_4)L_5(\Delta_5)\,
u_1^{{\Delta_{k_1}\over 2}+m_1} u_2^{{\Delta_{k_2}\over 2}+m_2}
\left(\prod_{(rs) \in {\cal V} }    \sum_{j_{rs}=0}^{\infty} { (1-v_{rs})^{j_{rs}} \over j_{rs}!} \right) \cr
&\times 
{1 \over \Gamma(\Delta_{k_2}+2m_2+j_{14}+j_{15}+j_{25})}
{1 \over \Gamma(\Delta_{k_1}+2m_{1}+j_{14}+j_{15}+j_{25})}
\cr 
& \times 
\big( \Gamma(\Delta_{12,k_1}-m_1)  \Gamma( \Delta_{2k_1,1}+m_{1}+j_{25})  \Gamma(\Delta_{1k_1,2}+m_1 +j_{14}+j_{15}) \big) \cr 
&\times \big( \Gamma(\Delta_{3k_2,k_1}+m_{2,1})   \Gamma( \Delta_{3k_1,k_2}+m_{1,2})  \Gamma(\Delta_{k_1k_2,3}+m_{12,}+j_{14}+j_{15}+j_{25}) \big) \cr 
&\times  
\big( \Gamma(\Delta_{45,k_2}-m_2)
\Gamma(\Delta_{5k_2,4}+m_2+j_{15}+j_{25})
\Gamma(\Delta_{4k_2,5}+m_2+j_{14}) \big)
}
where ${\cal V} = \{(14),(15),(25) \}$.

It is straightforward to check that~\eno{B5final} precisely reproduces~\eno{RulesBMst}-\eno{GammaVertexEdge}, as we set out to show in this subsection.
Indeed, for the choice of OPE flow~\eno{ucross5A}, we have already verified that the conformal cross ratios and leg factors appearing out of the calculation are as expected according to the rules of Section~\ref{INDEXCROSSRATIOS}. 
Thus the only thing that needs to be checked is that the Gamma-vertex and Gamma-edge factors are reproduced as well. This is verified after noting that the post-Mellin parameters associated with the index set ${\cal V} = \{(14),(15),(25) \}$ for the choice of OPE flow~\eno{ucross5A} are [see Section~\ref{POSTMELLIN}]
\eqn{}{ 
\ell_{1}&= j_{14}+j_{15}  \qquad
\ell_{2}= j_{25} \qquad
\ell_{3}=0
\cr & 
\ell_{4}=j_{14} \qquad
\ell_{5}=j_{15}+j_{25}
}
and,
\eqn{}{ 
\ell_{k_1} = j_{14}+j_{15}+j_{25} \qquad
\ell_{k_2}=  j_{14}+j_{15}+j_{25}\,.
}
Thus the expected Gamma-edge and Gamma-vertex factors, according to the rules~\eno{GammaVertexEdge} are
\eqn{}{ 
\widehat{E}_1  =  {1 \over \Gamma(\Delta_{k_1}+2m_1+j_{14}+j_{15}+j_{25})}
\qquad
\widehat{E}_2 = {1 \over \Gamma(\Delta_{k_2}+2m_2+j_{14}+j_{15}+j_{25})}
}
and
\eqn{}{ 
\widehat{V}_1 &= \Gamma(\Delta _{12,k_1}-m _1) \Gamma(\Delta _{2k_1,1}+m_1+j_{25}) \Gamma(\Delta _{1k_1,2}+m _1+j_{14}+j_{15})
\cr 
\widehat{V}_2 &= \Gamma(\Delta _{3k_1,k_2}+m_{1,2}) \Gamma(\Delta _{3k_2,k_1}+m_{2,1}) \Gamma(\Delta _{k_1k_2,3}+m_{12,}+j_{14}+j_{15}+j_{25})
\cr 
\widehat{V}_3 & = \Gamma(\Delta _{45,k_2}-m _2) \Gamma(\Delta _{4k_2,5}+m_2+j_{14}) \Gamma(\Delta _{5k_2,4}+m_2+j_{15}+j_{25})\,.
}
These correspond precisely to the Gamma function factors appearing in~\eno{B5final}, completing the check and establishing the inductive step from the four-point case to the five-point case.

\bibliographystyle{ssg}
\bibliography{main}

\end{document}